\documentclass[aps,prd,reprint,superscriptaddress,twocolumn,preprintnumbers,floatfix,nofootinbib,amsmath,amssymb]{revtex4-1}
\usepackage{amsmath,amssymb,mathrsfs}
\usepackage{bm}
\usepackage{epsfig}
\usepackage{verbatim}
\usepackage{bm}
\usepackage[utf8]{inputenc}
\usepackage{graphics}
\usepackage{graphicx,epsfig,amssymb,amsmath,color, cancel}
\usepackage{soul}
\usepackage[normalem]{ulem}
\usepackage{slashed}

\usepackage[english]{babel}
\usepackage{amsfonts,amsmath,amssymb,amsthm,mathtools}
\usepackage{comment,enumerate,footnote,graphicx,subfloat,relsize}
\usepackage{array,tabularx,tabu,multirow,framed,afterpage}
\usepackage[usenames,dvipsnames]{xcolor}
\usepackage{hyperref}
\hypersetup{
pdfstartview=FitV,
colorlinks=true,
linkcolor=blue, 
citecolor=red, 
filecolor=black, 
urlcolor=blue}

\newcommand{\be}{\begin{equation}}
\newcommand{\ee}{\end{equation}}

\newcommand{\zp}{\text{$Z^\prime$}}

\begin{document}

\preprint{ULB-TH/19-06}
\preprint{CPHT-RR047.072019}
\preprint{LPT-Orsay-19-29}

\title{Forays into the dark side of the swamp}

\author{Iason Baldes}
\email{iason.baldes@ulb.ac.be}
\affiliation{Service de Physique Th\'eorique, Universit\'e Libre de Bruxelles, Boulevard du Triomphe, CP225, 1050 Brussels, Belgium}

\author{Debtosh Chowdhury}
\email{debtosh.chowdhury@polytechnique.edu}
\affiliation{Centre de Physique Th\'eorique, CNRS, \'Ecole Polytechnique, IP Paris, 91128 Palaiseau, France}
\affiliation{Laboratoire de Physique Th\'eorique (UMR8627), CNRS, Univ. Paris-Sud, Universit\'e Paris-Saclay, 91405 Orsay, France}

\author{Michel H.G.~Tytgat}
\email{mtytgat@ulb.ac.be}
\affiliation{Service de Physique Th\'eorique, Universit\'e Libre de Bruxelles, Boulevard du Triomphe, CP225, 1050 Brussels, Belgium}


\begin{abstract}
Motivated by the swampland conjectures, we study cosmological signatures of a quintessence potential which induces time variation in the low energy effective field theory. After deriving the evolution of the quintessence field, we illustrate its possible ramifications by exploring putative imprints in a number of directions of particle phenomenology. We first show that a dark matter self-interaction rate increasing with time gives a novel way of reconciling the large self interactions required to address small scale structure issues with the constraint coming from clusters. Next, we study  the effects of kinetic mixing variation during the radiation dominated era on freeze-in dark matter production. Last, we elucidate quintessence effects on the restoration of the electroweak symmetry at finite temperature and the lifetime of the electroweak vacuum through a modification of the effective Higgs mass and quartic coupling. 
\end{abstract}

\maketitle

\section{Introduction}

One aim of the swampland conjectures is to stimulate consideration of constraints coming from quantum gravity on low energy effective field theories~\cite{Vafa:2005ui,Palti:2019pca}. A crucial question is whether physics motivated by the conjectures themselves could have observable phenomenological signatures on physics today. Indeed the recently proposed swampland de Sitter conjecture~\cite{Obied:2018sgi,Garg:2018reu,Ooguri:2018wrx} has led to a number of interesting considerations in such a direction. 

The conjecture demands the observed dark energy is not a constant but rather originates from quintessence. Such a time varying dark energy is constrained from a number of observations~\cite{Aghanim:2018eyx,Scolnic:2017caz} as it departs from the usual equation-of-state for the cosmological constant~\cite{Agrawal:2018own,Heisenberg:2018yae,Akrami:2018ylq}. One is tempted to relate it to the tension in the determination of the Hubble parameter today between the Cosmic Microwave Background (CMB)~\cite{Ade:2015xua} and Supernovae measurements~\cite{Riess:2019cxk}. A priori, a slowly rolling quintessence field will have an equation of state $w_{\phi} > -1$, which does not fit with late-time resolutions of the discrepancy which would prefer $w_{\phi} < -1$~\cite{Salvatelli:2013wra,DiValentino:2017iww,Yang:2018euj,Colgain:2019joh}. Nevertheless, related frameworks of early dark energy, which then decays away sufficiently rapidly, can help resolve the tension~\cite{Poulin:2018cxd,Kaloper:2019lpl,Agrawal:2019lmo} (also see~\cite{Colgain:2018wgk}).\footnote{As this manuscript was in the final stages of preparation, it was shown a time dependent dark matter (DM) mass is able to alleviate the Hubble tension in a swampland inspired context~\cite{Agrawal:2019dlm}. For closely related earlier work see~\cite{vandeBruck:2019vzd}, for some historical references see~\cite{Casas:1991ky,GarciaBellido:1992de,Anderson:1997un}, and for an effective fluid description of energy transfer between DM and dark energy, see e.g.~\cite{Salvatelli:2013wra,DiValentino:2017iww,Yang:2018euj}.} The distance~\cite{Ooguri:2006in,Grimm:2018ohb,Heidenreich:2018kpg,Blumenhagen:2018hsh} and de Sitter conjectures have also been used to constrain models of inflation~\cite{Garg:2018reu,Kinney:2018nny}. 

That physical parameters, such as masses or couplings, may be varying is an ancient idea. Time variation of fundamental constants was famously considered by Dirac in the context of his large number hypothesis~\cite{Dirac:1937ti,Dirac:1973gk}. Somewhat more recently, quasar observations led to speculation that the fine structure constant was changing with time~\cite{Webb:1998cq,Webb:2000mn,Banks:2001qc,Chacko:2002mf}. Other possibilities that have been considered in the literature include variations of the DM mass~\cite{Agrawal:2019dlm,Casas:1991ky,GarciaBellido:1992de,Anderson:1997un,Coudarchet:2018ezq}, as mentioned above, or of the neutrino masses~\cite{Fardon:2003eh,Kaplan:2004dq}. In this paper, we will explore some novel directions of time variation\footnote{Spatial variation is also a possibility --- we focus here on the (arguably simpler) case of time variation only.} --- inspired by the distance and de Sitter conjectures --- which could leave imprints in other sectors of particle physics. As the possibilities are numerous, we focus, for the sake of illustration, on three possible phenomenological applications that are related to current topics in physics beyond the SM.

We first consider models of self-interacting Dark Matter (SIDM) coupled to quintessence. The motivation is the small scale structure issues which could be addressed with SIDM~\cite{Spergel:1999mh,Tulin:2017ara}. The aim here is to obtain a time dependent cross section for the SIDM, e.g.~through the time dependent mass of a scalar mediator (as opposed to a time dependent DM mass itself). We will speculate that the mediator is part of the tower of particles which, following the distance conjecture, becomes light once the quintessence field rolls a Planck scale distance in field space. 

Still on DM, we fancy the possibility that its production may be through a time-dependent portal to the SM.  A non-trivial momentum dependence in the kinetic mixing portal was recently shown to have interesting cosmological consequences for dark matter production in the early universe~\cite{Banerjee:2019asa}. We continue the exploration in this direction by considering a kinetic mixing portal controlled by the quintessence field. 

Following the proposal of the original de Sitter conjecture~\cite{Obied:2018sgi}, it was pointed out the SM Higgs and QCD potentials would violate it without drastic modifications~\cite{Denef:2018etk,Murayama:2018lie}. Hence, a coupling of quintessence to the Higgs was introduced in order to achieve compatibility with the conjecture~\cite{Denef:2018etk,Murayama:2018lie}. Such a coupling is, however, strongly constrained from fifth force experiments~\cite{Choi:2018rze}. Later, a refined de Sitter conjecture was introduced~\cite{Garg:2018reu,Ooguri:2018wrx}, which allows for de Sitter vacua provided they are sufficiently unstable, which removed the need for the quintessence-Higgs coupling (also see~\cite{Kobakhidze:2019ppv,Park:2019odd}). Nevertheless, we consider reintroducing such a coupling and explore its effects on the electroweak vacuum lifetime. Such a coupling can either be motivated from the original unrefined conjecture, or simply be kept as a general phenomenological possibility, as varying dark sector masses and couplings would lead one to speculate that the Higgs mass and/or coupling is also varying. 

The paper is organized in the following way. We first discuss the conjectures, introduce the quintessence field required by the de Sitter conjecture, and derive its motion in field space as a function of cosmic time. We include a derivation of the temperature dependent motion during the radiation dominated epoch in the context of a sum-of-exponentials type potential. We address the question of the mutual compatibility of the conjectures due to the back-reaction of the tower of states required by the distance conjecture on the quintessence potential required by the de Sitter conjecture. We then consider SIDM, studying both a weakly coupled model with a light mediator, and a model in which the interactions are governed by exotic strong dynamics. We next turn to a quintessence driven and hence temperature dependent kinetic mixing portal between the dark and visible sectors. Finally we discuss the restoration of EW symmetry and the decay of the EW vacuum. We then conclude.

\section{Swampland Avoidance}
\label{secII}
\subsection{The conjectures}

\underline{\emph{Swampland de Sitter Conjecture.}} The conjecture states that any self-consistent theory of quantum gravity requires the scalar potential to satisfy~\cite{Obied:2018sgi,Garg:2018reu,Ooguri:2018wrx}
	\begin{subequations}
	\label{eq:dSconjecture}
	\begin{align}
	M_{\rm Pl} \, |\nabla V| & \geq \xi \, V, 	\label{eq:conjecture} \\
	\mathrm{or} \nonumber \\
	M_{\rm Pl}^{2} \, \mathrm{Min}(\nabla_i \nabla_j V) & \leq  -\xi' \,  \, V, \label{eq:refinedconjecture}
	\end{align}
	\end{subequations}
for some $\xi, \xi^{'}$, which are positive $\mathcal{O}(1)$ constants, where $M_{\rm Pl}$ is the reduced Planck mass. Accepting the above means that the scalar potential, if positive, should be sufficiently unstable, as captured by the conditions on its first or second derivatives. The conjecture was introduced following the discussion in the literature regarding the possibility of constructing de Sitter vacua in string theory. On one side, constructions such as KKLT~\cite{Kachru:2003aw} claim de Sitter vacua are viable and lead to a string landscape or multiverse. On the other side, it has been argued that such mechanisms are not fully under calculable control, and there may be a fundamental reason why an example not liable to such criticism has yet been found~\cite{Danielsson:2018ztv}. The conjecture aims to provoke a closer examination of whether de Sitter vacua can indeed be constructed or alternatively the phenomenological consequences of an unstable potential.

\underline{\emph{ Swampland Distance Conjecture.}} The conjecture has been stated in a number of ways~\cite{Ooguri:2006in,Grimm:2018ohb,Heidenreich:2018kpg,Blumenhagen:2018hsh}. For our purposes, we adopt it as the following working hypothesis: if a field transverses a trans-Planckian field range, a tower of states becomes light
	\begin{equation}
	\label{eq:distance}
	\Delta \phi \gtrsim M_{\rm Pl} \implies V \supset \frac{1}{2}e^{-c_{S} \Delta \phi/M_{\rm Pl}}\sum_{i}m_{S_i}^{2}S_i^{2},
	\end{equation} 
where $c_{S}$ is a dimensionless quantity, ruining the original low-energy effective field theory description. The tower of states becoming light is associated with the decompactification of an extra-dimension in the full string theory. By speculating the tower is becoming light today, and coupling it to additional fields, we will attempt to derive some interesting phenomenological consequences below. 

\subsection{Introduction of quintessence and its evolution}

Supernovae light-curve data~\cite{Perlmutter:1998np,Riess:1998cb,Aghanim:2018eyx} established the accelerated expansion of the Universe associated with a component of dark energy. In the standard $\Lambda$CDM cosmology, the dark energy is taken to be a positive cosmological constant with a value $\Lambda \sim (M_{\rm Pl}/10^{30})$. More concretely, in the standard picture one takes this to be the value at the stable minimum of the Electroweak (EW) Higgs potential~\cite{Englert:1964et,Higgs:1964pj,Aad:2015zhl}, once QCD and other quantum corrections are taken into account (to which we add any additional effects from beyond-the-Standard Model fields). Clearly in the standard picture the conjecture~(\ref{eq:dSconjecture}) is violated at this point.

\underline{\emph{Exponential Potential.}}  To solve this issue, a quintessence potential for the dark energy is introduced~\cite{Ratra:1987rm,Ferreira:1997au,Ferreira:1997hj,Copeland:1997et,Tsujikawa:2013fta}
	\begin{equation}
	\label{eq:single}
	V(\phi) = \Lambda^{4}e^{-c_{\phi}\phi/M_{\rm Pl}},
	\end{equation}
where $c_{\phi}$ is a dimensionless constant, and we adopt the convention of setting $\phi=0$ today. Observations give a constraint $c_{\phi} \lesssim 0.6$~\cite{Agrawal:2018own,Heisenberg:2018yae,Akrami:2018ylq}. The conjecture~(\ref{eq:dSconjecture}) is then satisfied provided $c_{\phi} > \xi$. Furthermore, the additional stationary points in the QCD and Higgs potential satisfy the refined conjecture, i.e. with the inclusion of~(\ref{eq:refinedconjecture}), because the mass matrix at the stationary points at the origin (and at $\sim 10^{10}$ GeV~\cite{Buttazzo:2013uya} for the Higgs potential), have sufficiently large negative eigenvalues.

\begin{figure}[t]
\begin{center}
\includegraphics[width=200pt]{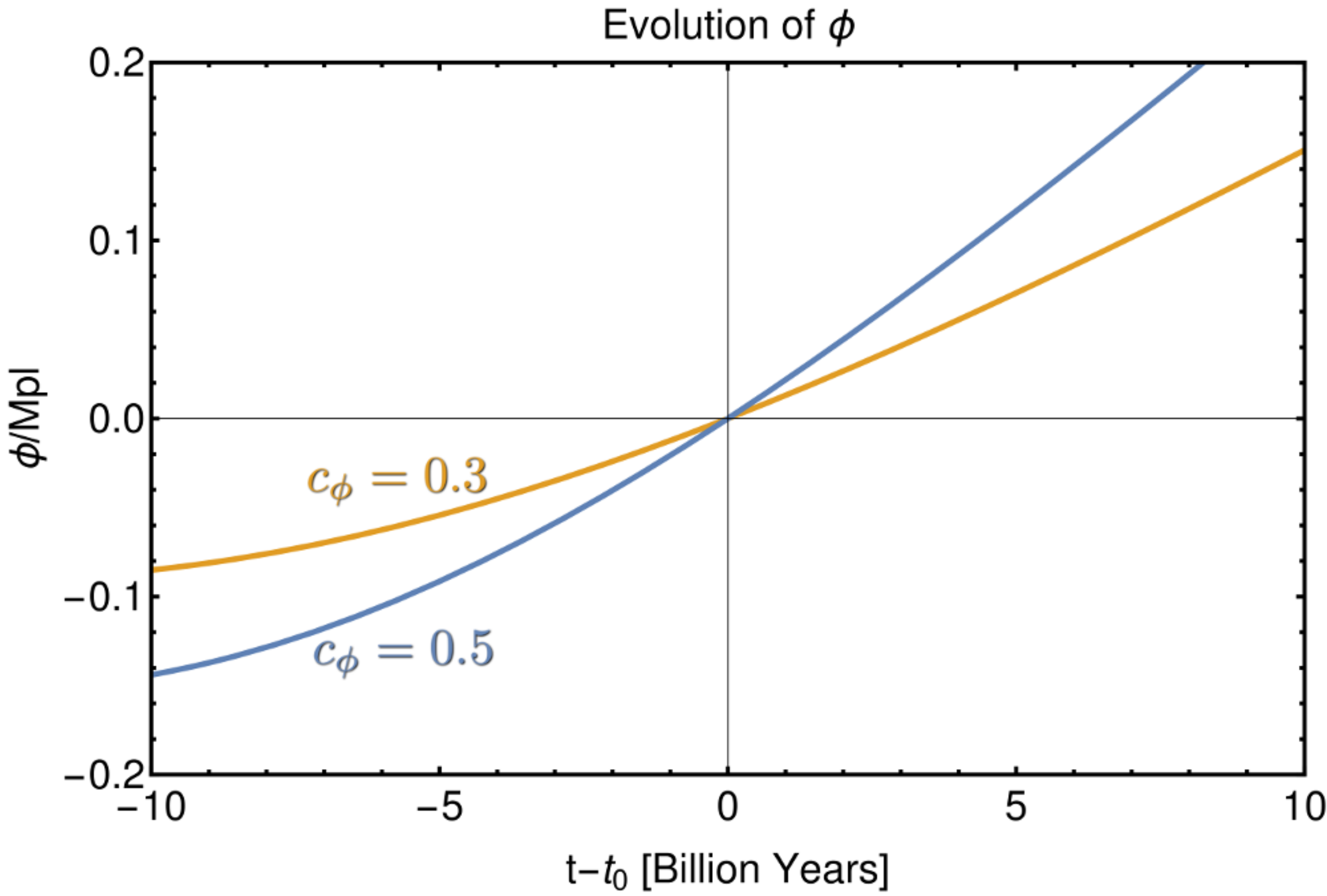}
\end{center}
\caption{The evolution of $\phi$ as a function of the cosmic time. Today corresponds to $t=t_0$. The change in $\phi$ is used as an input to our subsequent calculations below.
}
\label{fig:phievo}
\end{figure}

\begin{figure}[t]
\begin{center}
\includegraphics[width=200pt]{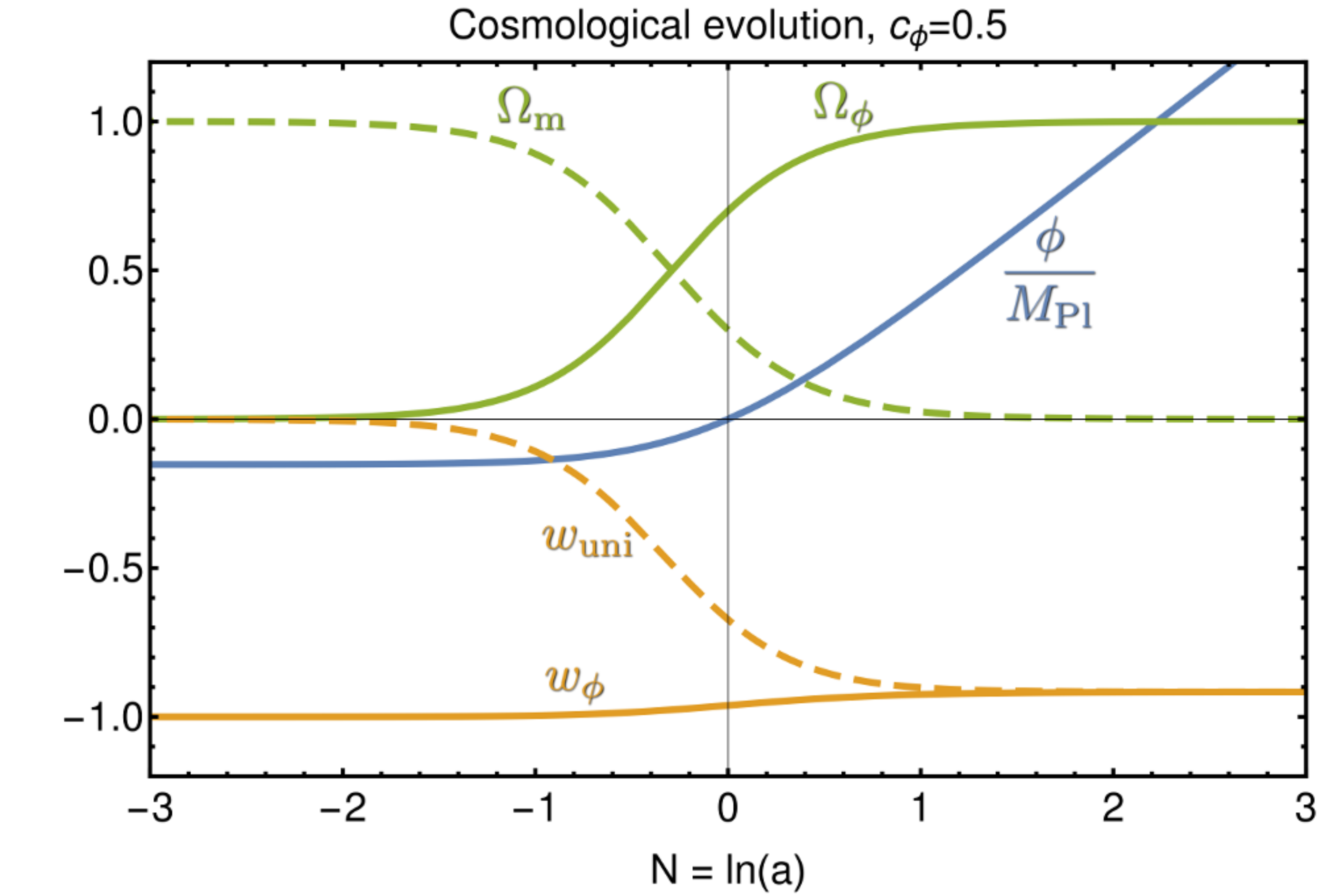}
\end{center}
\caption{The evolution of various quantities of cosmological interest as a function of the logarithmic scale factor. Today corresponds to $N=0$. The density in matter,  $\Omega_{m}$, and the density associated with the $\phi$ kinetic and potential energies, $\Omega_{\phi}$. Together with the equation of state of the Universe, $w_{\rm uni}$, and of the scalar field, $w_{\phi}$.
}
\label{fig:unievo}
\end{figure}

Next we solve for the evolution of $\phi$, starting at matter-radiation equality, using the by now standard methods which can, e.g. be found in~\cite{Copeland:1997et,Tsujikawa:2013fta}. We assume a flat Universe throughout. We start with the equation of motion
	\begin{equation}
	\ddot{\phi} + 3 H_{u} \dot{\phi} + \frac{dV}{d\phi} = 0, 
	\end{equation}
where $H_u$ is the Hubble parameter. We also require the two Friedmann equations
	\begin{align}
	3M_{\rm Pl}^{2}H_u^2 & = \, \, \frac{ \dot{\phi}^{2}}{2} + V(\phi)+\rho_b \\
	2M_{\rm Pl}^{2}\dot{H_u} & =  - \dot{\phi}^{2} - (1+w_b)\rho_b
	\end{align}
where $\rho_b$ is the background density and $w_b$ is its equation of state. The standard technique is now to write the equations in terms of the dimensionless kinetic energy and potential
	\begin{equation}
	\label{eq:xandy}
	x \equiv \frac{\dot{\phi}}{\sqrt{6}M_{\rm Pl}H_u}, \qquad \qquad 	y \equiv \frac{\sqrt{V(\phi)}}{\sqrt{3}M_{\rm Pl}H_u},
	\end{equation}
and then solve for the evolution in terms of $N=\mathrm{ln}(a)$, where $a$ is the scale factor~\cite{Copeland:1997et,Tsujikawa:2013fta} (we normalise to $a=1$ today). Note for $c_{\phi}^{2}<6$, there is a stable attractor in the corresponding phase space at
	\begin{equation}
	x^{2} = \frac{c_{\phi}^{2}}{6} , \qquad \qquad y^{2} =  1-\frac{ c_{\phi}^{2} }{ 6 },
	\end{equation}
towards which our system will evolve. Next we follow the method of~\cite{Agrawal:2018own}, in order to find suitable solutions which ensure agreement with cosmological observations. We find the initial conditions at matter-radiation equality ($N_{\rm eq} \approx -8.1$), $x_i = y_i$, such that the energy density in $\phi$ sector today matches the required $\Omega_{\phi} \approx 0.7$.\footnote{The results are not very sensitive to $x_{i}$, and we take $x_i = y_i$ for simplicity. We also find the kinetic energy in the scalar field is negligible today, typically $x^{2} \approx 0.02 \, y^{2}$.} By using the observed value of the Hubble parameter today $H_{0u} \approx 72$ km/s/Mpc~\cite{Riess:2019cxk}, and $\Lambda^{4} \simeq 0.7M_{\rm Pl}^{2}H_{0u}^{2}$, we can then convert the evolution of $x$ and $y$ into the required $H_u(t)$ and $\phi(t)$. The result of such a procedure, showing the time evolution of $\phi$, is shown in Fig.~\ref{fig:phievo}. In addition, we show in Fig.~\ref{fig:unievo} the concurrent evolution of various other cosmological quantities, which serve to illustrate the overall picture.

\underline{\emph{Sum-of-Exponentials Potential.}} More generally, we can consider more complicated forms of the potential. For example, rather than Eq.~(\ref{eq:single}), we may introduce~\cite{Barreiro:1999zs}
	\begin{equation}
	\label{eq:double}
	V(\phi) = \Lambda^{4}e^{-c_{\phi}\phi/M_{\rm Pl}}+\tilde{\Lambda}^{4}e^{-\tilde{c}_{\phi}\phi/M_{\rm Pl}},
	\end{equation}
which was also recently used in~\cite{Agrawal:2019dlm}. The idea here is to choose $\tilde{c}_{\phi} \gg c_{\phi}$ so the second term gives the dominant dark energy contribution at very early times. Indeed for $\tilde{c}_{\phi}^{2}>3(1+w_b)$ there is a stable fixed point in the phase space,
	\begin{equation}
	\label{eq:fixedpoint2}
	x^{2} = \frac{3}{2} \frac{(1+w_b)^{2}}{\tilde{c}_{\phi}^{2}} , \qquad \qquad y^{2} =  \frac{3}{2} \frac{(1-w_b^{2})}{\tilde{c}_{\phi}^{2}},
	\end{equation}
so that $\Omega_{\phi}=x^{2}+y^{2}$ remains constant~\cite{Ferreira:1997au,Ferreira:1997hj,Copeland:1997et}. This implies $\phi$ could already be moving significantly in the radiation and early matter dominated epochs. Constraints on additional radiation from Big Bang Nucleosynthesis (BBN) and the CMB~\cite{Aghanim:2018eyx} imply $\tilde{c}_{\phi} \gtrsim 10$ in the tracking regime. Detailed bounds from cosmological observables can be found in~\cite{Chiba:2012cb}, which give $\tilde{c}_{\phi} \gtrsim 16.3 \; (11.7)$ at 68\% (95\%) CL.

The crossover between $\phi$ evolution being described by the second and first terms in Eq.~(\ref{eq:double}) occurs when $\phi\equiv \phi_{\rm mc}$ and the two terms are equal. (The subscript mc stands for matching -- when we match the early time evolution to the late time one.) We will denote the scale factor when this occurs as $a_{\rm mc}$. Assuming the system has reached the fixed point in Eq.~(\ref{eq:fixedpoint2}) during the $\tilde{c}_{\phi}$ determined evolution, this fixed point then acts as the initial condition for the later $c_{\phi}$ determined evolution.

\begin{figure}[t]
\begin{center}
\includegraphics[width=200pt]{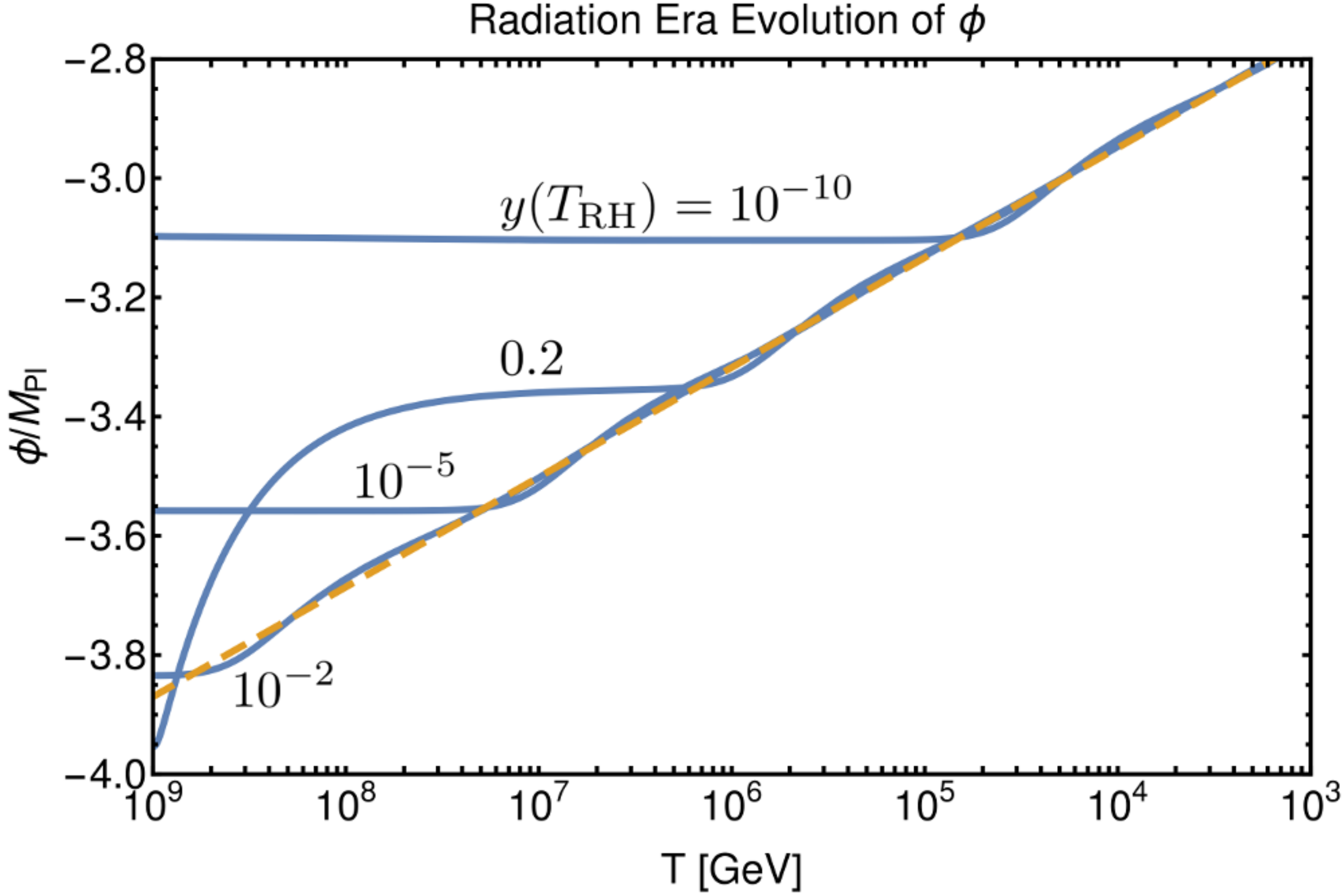}
\end{center}
\caption{Example of the movement of $\phi$ in the radiation dominated epoch using the sum-of-exponentials potential with initial conditions at reheating, $T_{\rm RH} = 10^{9}$ GeV, $x(T_{\rm RH}) =0$, and different choices for $y(T_{\rm RH})$. The solution is found for $\tilde{c}_{\phi} = 50$ and $c_{\rm \phi} = 0.5$. The dashed line shows the behaviour expected from Eq.~(\ref{eq:phitracking}) in the tracking regime. The initial oscillation is due to the evolution of $x$ and $y$ in phase space towards the fixed point in Eq.~(\ref{eq:fixedpoint2}).
}
\label{fig:phievo2}
\end{figure}

To recover the required late time behaviour of the $c_{\phi}$ term, discussed above, we find matching occurs in the matter dominated epoch provided $\tilde{c}_{\phi} \lesssim 10^{5}$. We then find
	\begin{equation}	
	a_{\rm mc} \approx \left( \frac{3}{ \zeta(c_{\phi}) \, \tilde{c}_{\phi}^{\, 2} } \right)^{1/3}a_{\rm eq},
	\end{equation}
where $a_{\rm eq} \approx 1/3370$ is the scale factor at matter-radiation equality, and $\zeta(c_{\phi}) \equiv x_{i}^{2}+y_{i}^{2}$ is a factor we determine numerically. Using the method below Eq.~(\ref{eq:xandy}) we find, e.g.~$\zeta(0.5) \approx 1.3\times 10^{-10}$ and $\zeta(0.3) \approx 1.2\times 10^{-10}$. One also has
	\begin{equation}
	\tilde{\Lambda}^{4} \approx \Lambda^{4} \, \mathrm{Exp} \left[ \frac{\phi_{\rm mc}}{M_{\rm Pl}} (\tilde{c}_{\phi} - c_{\phi})\right].
	\end{equation} 
This allows one to find the field value in the tracking regime, which is given by
	\begin{equation}
	\frac{\phi}{M_{\rm Pl}} \approx \frac{1}{\tilde{c}_{\phi}} \, \mathrm{ln}\left( \frac{ \tilde{\Lambda}^{4} }{ y^{2}\rho_{c}} \right).
	\end{equation}
Thus, in the radiation dominated epoch
	\begin{equation}
	\label{eq:phitracking}
	\frac{\phi(T_{2}) - \phi(T_{1})}{M_{\rm Pl}} = \frac{4}{\tilde{c}_{\phi}} \, \mathrm{ln} \left( \frac{T_{1}}{T_{2}} \right),
	\end{equation}
which we shall make use of below. An example of the early universe motion of $\phi$ is shown in Fig.~\ref{fig:phievo2}. Note the larger $\tilde{c}_{\phi}$ is, the smaller is $a_{\rm mc}$, and hence the more removed the late time implications on the 
cosmological observables such as the equation of state. A larger $\tilde{c}_{\phi}$, however, also corresponds to a suppression of the movement of $\phi$ in the very early universe. In the interval between the two extremes, we hope to find some implications for particle physics phenomenology below, in which early universe quantities have changed significantly to the present day.

\subsection{Compatibility of the conjectures}

A crucial issue is the compatibility of the de Sitter and distance conjectures. A priori, the quintessence field requires a doubly tuned potential; both its value and derivative must be incredibly small compared to their natural $\sim M_{\rm Pl}$ scale values. Conventionally, this can simply be achieved through a tuning of the various large contributions. The tower of states implied by the distance conjecture, however, will tend to lead to a detuning once $\phi$ begins to roll unless higher order terms in the series expansion of the quintessence potential are also tuned~\cite{Marsh:2018kub} (also see~\cite{Banks:2001qc}).

To see the tuning issue, expand the exponential in Eq.~(\ref{eq:distance}) to find the $\phi S_{i}^{2}$ coefficient. Evaluating the one-loop self energy of $\phi$, the finite part of the correction to the mass of the quintessence field scales as,
	\begin{equation}
	\label{eq:corr}
	\delta m_{\phi}^{2} \sim \sum_i \left( \frac{c_S^2 m_{Si}^{4} }{ 16\pi^{2} M_{\rm Pl}^{2} } \right) \, \mathrm{ln}\left( \frac{\mu^{2}}{m_{Si}^{2}} \right),
	\end{equation}
where $\mu$ is a renormalisation scale. For $m_{Si} \gtrsim \mathcal{O}$(meV), each contribution in the sum is large compared to the required quintessence mass $m_{\phi}^{2} \lesssim H_{0u}^{2}$~\cite{DAmico:2016jbm}. Even if tuned for one value of $\phi$ by introducing the requisite counterterms, the masses appearing in Eq.~(\ref{eq:corr}) vary as $\phi$ rolls, which will lead to an eventual detuning unless also higher order terms are tuned~\cite{Marsh:2018kub}.

Note from a bottom-up perspective, similar $\delta m_{\phi}^{2}$ corrections as to those elucidated above, also result when attempting to couple quintessence to DM~\cite{DAmico:2016jbm,Matsui:2018xwa} (or to other fields such as the SM Higgs as in~\cite{Denef:2018etk,Murayama:2018lie}). Of course, the induced corrections are small if the quintessence field couples only to sufficiently light fields, which means axion-like particles in the context of DM~\cite{DAmico:2016jbm,Matsui:2018xwa,Agrawal:2019lmo}.

As we study SIDM, however, axion-like particles appear to be unsuitable for our purposes. We therefore simply impose the required tuning for the quintessence potential. Note there may be a cancellation mechanism in a more complete picture. Although we do not explore this further here, it is interesting that cancellations in contributions from Kaluza-Klein towers in four-dimensional effective field theories have been found to occur in certain flux compactifications~\cite{Buchmuller:2016gib,Buchmuller:2018eog}.

In this paper we study exponential potentials inspired by the de Sitter conjecture, but in principle we could generalise our study to other quintessence potentials, such as the hilltop, $V(\phi)=\Lambda^{4}(1+\cos{[\phi/f]})$, or inverse power law potentials, $V(\phi)=\Lambda^{4+p}/\phi^{p}$~\cite{Chiba:2012cb}. These would also suffer the same tuning issues discussed above, however, once a coupling to massive fields is introduced.

\section{Self-Interacting Dark Matter}

Self interacting dark matter is one way of alleviating the purported tension between theory and observations of structure at small scales~\cite{Spergel:1999mh,Tulin:2017ara}. Interactions between DM particles can thermalise the dense inner regions of such structures leading to a reduction of their density. To achieve this $\sim \mathcal{O}(1)$ scattering events between DM particles are required in the age of the structure. The scattering rate is
 \begin{align}
\label{eq:SIDM}
\Gamma_{\rm scat.} & = \sigma v_{\rm rel} n_{\rm DM} =  \sigma v_{\rm rel} \frac{\rho_{\rm DM} }{m_{\rm DM}} \\
	     & \approx \Big( \frac{0.1}{{\rm Gyr}} \Big)  \Big( \frac{\rho_{\rm DM}}{0.1 \; M_{\rm sol}/{\rm pc}^3} \Big) \Big( \frac{v_{\rm rel}}{50 \; {\rm km/s}} \Big) \Big( \frac{\sigma /m_{\rm DM} }{1 \; {\rm cm^2/g}} \Big),   \nonumber
\end{align}
where $n_{\rm DM}$ ($\rho_{\rm DM}$) is the DM number (mass) density, and from which the canonical value of $\sigma /m_{\rm DM} \sim 1 \; {\rm cm^2/g}$ is obtained for an age $\sim 1/H_{0u}$. On the other hand, constraints coming from galaxy clusters point towards $\sigma/m_{\rm DM} \lesssim 0.1 \; \mathrm{cm}^{2}/\mathrm{g}$~\cite{Harvey:2015hha,Andrade:2019wzn}. The compatibility is conventionally ensured by invoking a velocity dependent $\sigma$, as the typical DM relative velocity in dwarf galaxies is $v_{\rm rel} \sim  30 \;\mathrm{km}/\mathrm{s}$, compared to $v_{\rm rel} \sim 1000 \; \mathrm{km}/\mathrm{s}$ on cluster scales~\cite{Kaplinghat:2015aga}. Note that there is not only a difference in $v_{\rm rel}$ between dwarf and galaxy cluster scales, but also a difference in the typical redshift and hence light-travel time between the observed structures and us. Observations of galaxy clusters therefore probe an earlier time of DM self interactions compared to dwarf galaxies. Hence a red-shift or time dependent cross section will allow us to achieve the required change in $\sigma$ between the two in a novel way.  As our SIDM will only obtain a large self interaction at late times, we expect the required cross section to be approximately an order-of-magnitude larger.

\begin{figure}[t]
\begin{center}
\includegraphics[width=200pt,height=135pt]{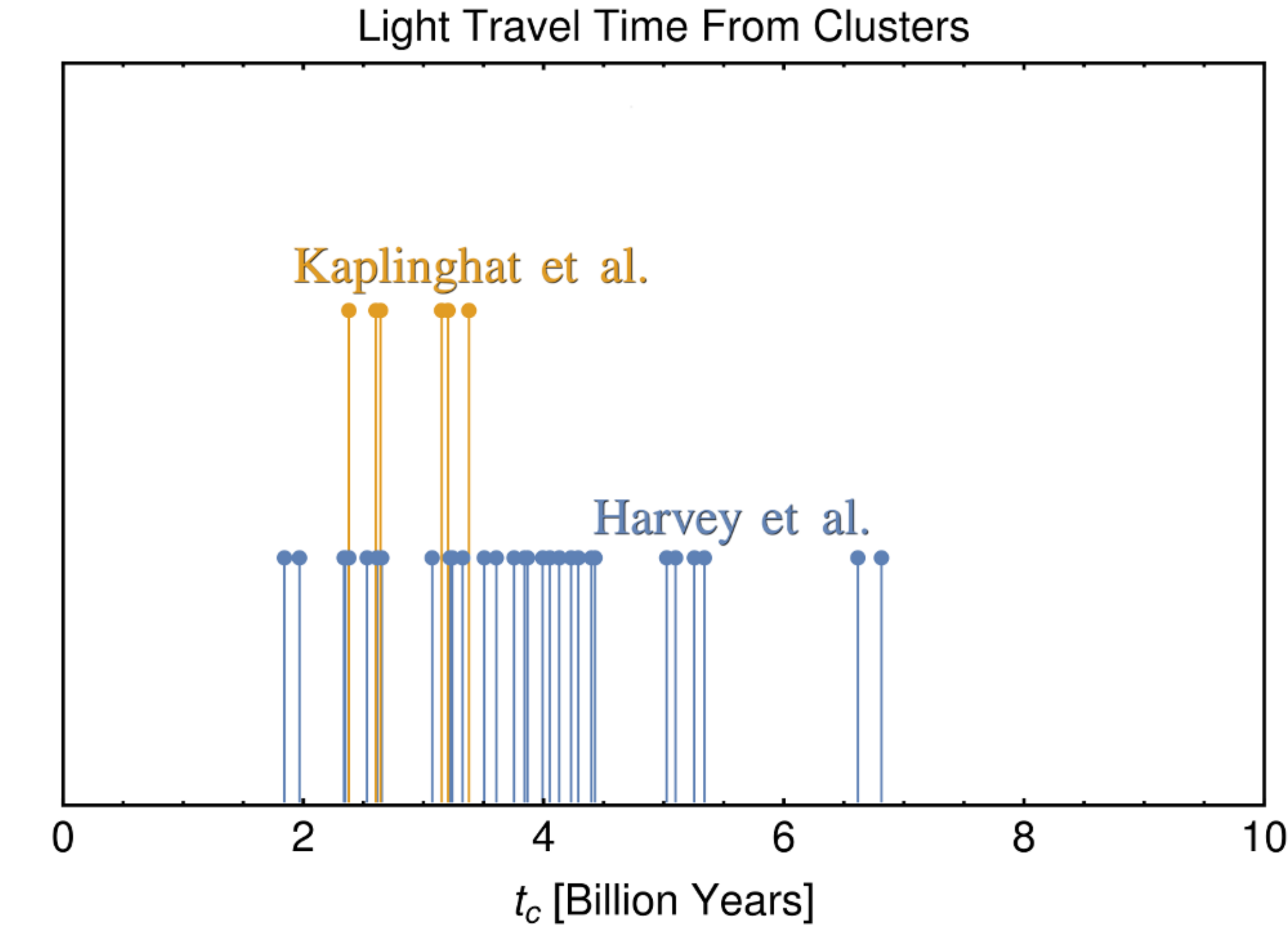}
\end{center}
\caption{The light travel time of the thirty merging complexes studied in~\cite{Harvey:2015hha}, and the six clusters analysed in~\cite{Kaplinghat:2015aga} in the context of our quintessence cosmology with $c_{\phi} = 0.5$. The median value for the former (latter) is 3.8 (2.9) billion years.
}
\label{fig:harvey}
\end{figure}

As mentioned above, limits on SIDM primarily come from the observation of galaxy clusters. The constraint from the bullet cluster at redshift $z=0.296$ was found to be $\sigma/m_{\rm DM} < 0.7  \; \mathrm{cm}^{2}/\mathrm{g}$~\cite{Randall:2007ph}. A stacked analysis of merging galaxy clusters, with redshifts in the range $0.154 < z < 0.87$, has placed a limit on the SIDM cross section $\sigma/m_{\rm DM} < 0.47  \; \mathrm{cm}^{2}/\mathrm{g}$~\cite{Harvey:2015hha} (although~\cite{Wittman:2017gxn} subsequently argued that the constraint should be relaxed to $\sigma/m_{\rm DM} < 2  \; \mathrm{cm}^{2}/\mathrm{g}$). Kaplinghat et al.~\cite{Kaplinghat:2015aga} adapted the analysis of~\cite{Newman:2012nv,Newman:2012nw} for six clusters in the redshift range $0.206 < z < 0.314$  to place a constraint $\sigma/m_{\rm DM} < 0.1  \; \mathrm{cm}^{2}/\mathrm{g}$. A recent analysis of Abell 611 at $z=0.288$ also points to $\sigma/m_{\rm DM} < 0.1  \; \mathrm{cm}^{2}/\mathrm{g}$~\cite{Andrade:2019wzn}. The key point is that limits coming from galaxy clusters originate at high redshifts, $z \gtrsim 0.2$, which correspond to light travel times
 	\begin{equation}
	t_{\rm c} = \int_{0}^{z} \frac{1}{(1+z')H_{u}(z')}dz',
	\end{equation}
of the order of a few billion years. Our conversion of the redshifts of the clusters used in~\cite{Harvey:2015hha,Kaplinghat:2015aga}, into the corresponding $t_{c}$ is shown in Fig.~\ref{fig:harvey}.

In contrast, due to observational limitations, small scale structures can only be observed at lower redshifts. Let us again follow the samples used in~\cite{Kaplinghat:2015aga}, which extracted a preferred $\sigma/m_{\rm DM} \sim 1-10 \; {\rm cm^2/g}$ using five dwarf galaxies from~\cite{Oh:2010ea} and seven low surface brightness galaxies from~\cite{KuziodeNaray:2007qi}. The former are at distances in the range $3.4 \; \mathrm{Mpc} < d < 5.3 \; \mathrm{Mpc}$~\cite{Walter:2008wy}, and the latter range from $10.1 \; \mathrm{Mpc} < d < 77 \; \mathrm{Mpc}$~\cite{KuziodeNaray:2007qi,deBlok:2002vgq,Swaters:2002rx,McGaugh:2001yc}. The corresponding light travel time from the dwarf and low surface brightness galaxies is therefore $\sim \mathcal{O}(10-100)$ million years.

To match onto observations, we therefore want to increase $\sigma/m_{\rm DM}\sim 0.1 \; {\rm cm^2/g}$ from a few billion years ago to $\sigma/m_{\rm DM}\sim 10 \; {\rm cm^2/g}$ today.  We wish to emphasise that given the uncertainties involved the overall picture for SIDM is still relatively unclear. Nevertheless, we take the values above as a working hypothesis for the purposes of the current discussion. Improvements in observations as well as simulations~\cite{Kummer:2019yrb} should eventually shed more light on this framework. Note we also expect an eventual accelerated gravothermal collapse of the dense DM cores, due to the ever increasing $\sigma$ in this picture~\cite{Nishikawa:2019lsc}. We now turn to the microphysical picture required to achieve such a cross section. We will sketch, in turn, a weakly coupled implementation and one relying on new strongly coupled dynamics.

\subsection{Weakly coupled model}

\begin{figure}[t]
\begin{center}
\includegraphics[width=200pt,height=135pt]{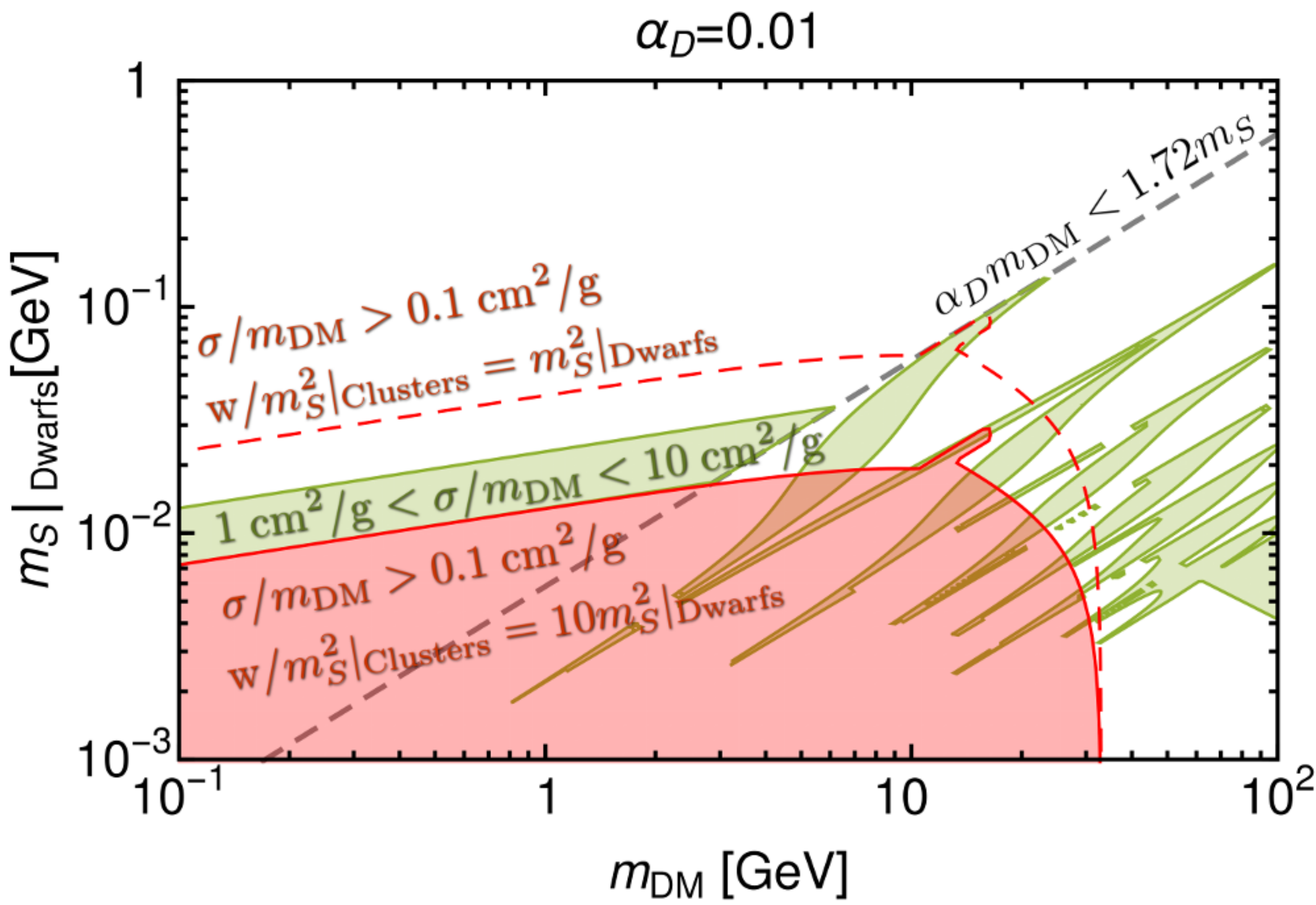}
\end{center}
\caption{The parameter space of the weakly coupled SIDM model with a light mediator. The green area shows the desired $\sigma$ at small scales and late times. The red shaded area is the cluster constraint with $m_{S}^{2}|_{\rm Clusters} = 10m_{S}^{2}|_{\rm Dwarfs}$. Also shown as a red dashed line is the cluster constraint assuming $m_{S}^{2}|_{\rm Clusters} = m_{S}^{2}|_{\rm Dwarfs}$. 
}
\label{fig:SIDM1}
\end{figure}

Consider a fermionic DM particle, $X$, coupled to a scalar mediator, $S$, through a Yukawa coupling
	\begin{equation}
	\mathcal{L} \supset y_{D}S\bar{X}X,
	\label{eq:DMyuk}	
	\end{equation}
where $y_{D}$ is a dimensionless coupling. We follow convention and write $\alpha_{D} \equiv y_{D}^{2}/(4\pi)$. One expects on naive dimensional grounds the self-interaction of the DM particles through $t$-channel $S$ exchange to be~\cite{Tulin:2013teo}
	\begin{equation}
	\label{eq:weaknaive}
	\sigma \approx 4 \pi\alpha_{D}^{2}\frac{ m_{\rm DM}^2 }{ m_{S}^4 }.
	\end{equation}
This is indeed valid for $\alpha_{D} m_{\rm DM} \ll m_{S}$ at low relative velocities. A more precise determination follows from identifying the momentum transfer cross section as the relevant quantity for SIDM. In the Born regime, $\alpha_{D} m_{\rm DM} \lesssim 1.72 m_{S}$, the SIDM transfer cross section following from Eq.~(\ref{eq:DMyuk}) is given by~\cite{Feng:2009hw,Tulin:2013teo}
	\begin{align}
	\sigma = \frac{ 8 \pi \alpha_{D}^2 }{ m_{\rm DM}^2v_{\rm rel}^4 } \left \{ \mathrm{ln} \left[1 + \frac{m_{\rm DM}^2 v_{\rm rel}^2}{m_{S}^2} \right] -  \frac{ m_{\rm DM}^2 v_{\rm rel}^2 }{ m_{S}^2 +m_{\rm DM}^2 v_{\rm rel}^2 }\right \}. \nonumber \\ \label{eq:weakself}
	\end{align}
Taking additionally the limit $m_{S}^{2} \gg m_{\rm DM}^{2} v_{\rm rel}^{2}$, one recovers Eq.~(\ref{eq:weaknaive}) from (\ref{eq:weakself}), i.e.~there is no velocity dependence.
Outside of the Born regime, the long range of the interaction must be taken into account, and one finds the velocity dependence which is usually invoked for SIDM. A plot of the parameter space showing the areas returning the desired $\sigma$ is shown in Fig.~\ref{fig:SIDM1}. Note in the plot we also show areas outside the Born regime, in which the long range of the interaction leads to parametric resonances~\cite{Tulin:2013teo}. As can be seen, we require an increase in $m_{S}^{2}$ by an order-of-magnitude in the last few billion years to obtain the desired time dependence for $\sigma$, which follows the naive expectation from Eq.~(\ref{eq:weaknaive}).

\begin{figure}[t]
\begin{center}
\includegraphics[width=200pt,height=135pt]{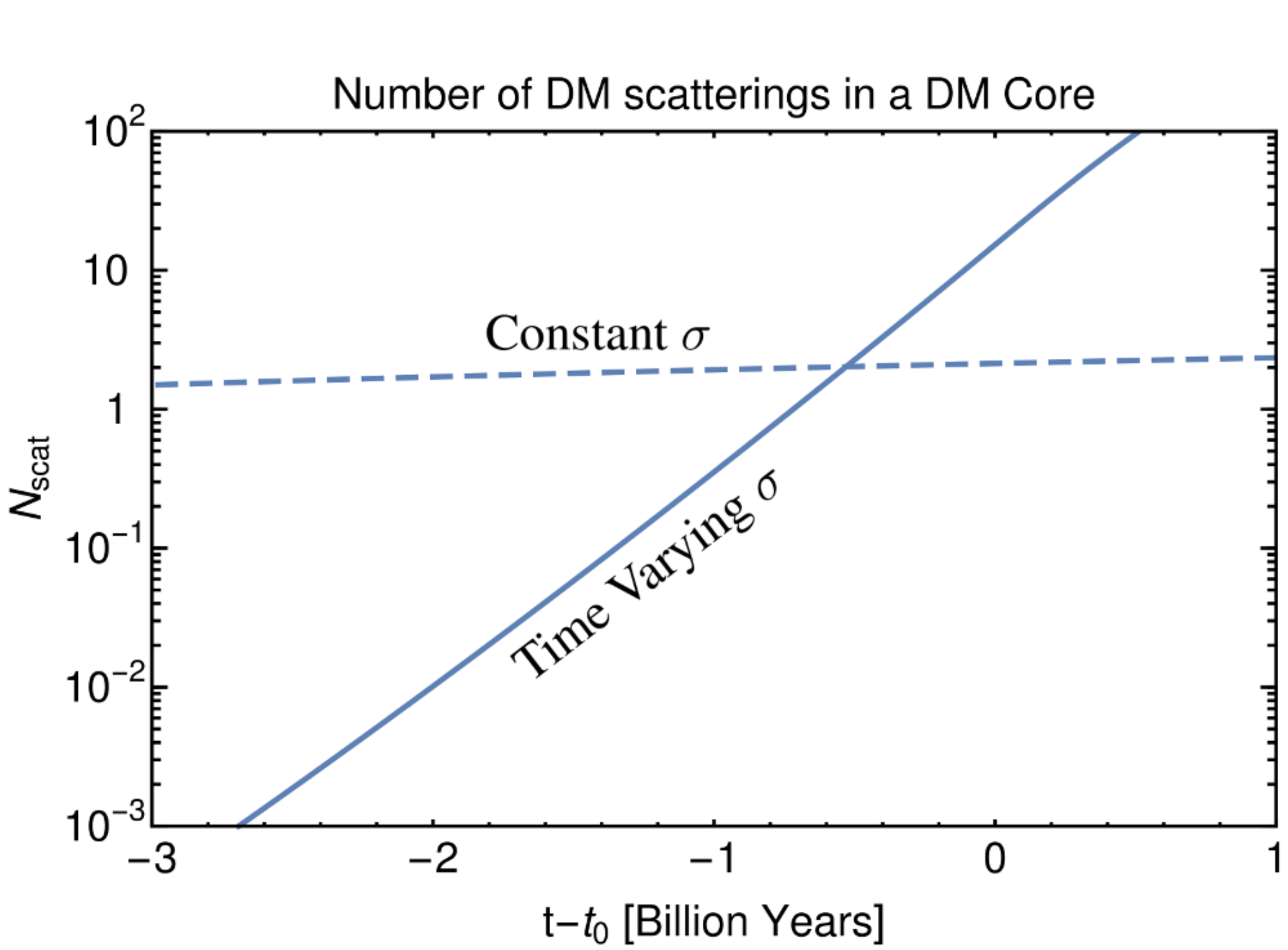}
\end{center}
\caption{The approximate number of DM-DM scatterings in a DM core with $v_{\rm rel} = 100 \; \mathrm{km}^{2}/\mathrm{s}$, $\rho_{\rm DM} = 0.1 \; \mathrm{M_{\rm sol}}/{\rm pc}^3$, and a DM core age of 10 billion years today. The solid line uses a cross section taken from the weakly coupled model with $\alpha_{D}=0.01$, $m_{\rm DM}=1$ GeV, $m_{S0}=10$ MeV, $c_{\phi}=0.5$ and $c_S=23$ which results in $\sigma/m_{\rm DM} = 28 \; {\rm cm^2/g}$ today. For comparison the dashed line shows the result for a time independent $\sigma/m_{\rm DM} = 0.1 \; {\rm cm^2/g}$.
}
\label{fig:SIDM1b}
\end{figure}

In order to achieve the desired variation, we adopt a $\phi$ dependent mass term of the form,
	\begin{equation}
	V \supset \frac{1}{2}e^{-c_S\phi/M_{\rm Pl}}m_{S0}^{2}S^{2},
	\label{eq:Smass}
	\end{equation}
where $c_S$ is a dimensionless constant, and $m_{S0}$ is the mass of $S$ today. We can think of the mediator as being the lightest particle in the tower of states arising from the distance conjecture. The required mass variation translates to $c_S = M_{\rm Pl}\mathrm{ln}(10)/\Delta \phi$, where $\Delta \phi$ is the change in $\phi$ over the past three billion years. Taking as our benchmark $c_{\phi}=0.5$, we see from Fig.~\ref{fig:phievo} that we require $c_S \approx 10 \mathrm{ln}(10) \approx  23$. Once these parameters have been specified, we may estimate the number of DM scattering events in cores, $N_{\rm scat.} = \int \Gamma_{\rm scat.}(dt) dt$, where the integral is performed over the lifetime of the structure, and we take into account the time dependence of $\sigma$. An evaluation is shown in Fig.~\ref{fig:SIDM1b}, showing we do indeed recover the desired behaviour of $N_{\rm scat.} \ll 1$ in the observed clusters, together with $N_{\rm scat.} \sim \mathcal{O}(1)$ at more recent times. Note that once the self interactions start, the estimated relaxation time for a structure with size $d \sim 1$ kpc and typical velocity $v \sim 10$ km/s is $t \sim 0.1$ billion years, and hence short enough to match our picture.

\begin{figure}[t]
\begin{center}
\includegraphics[width=200pt]{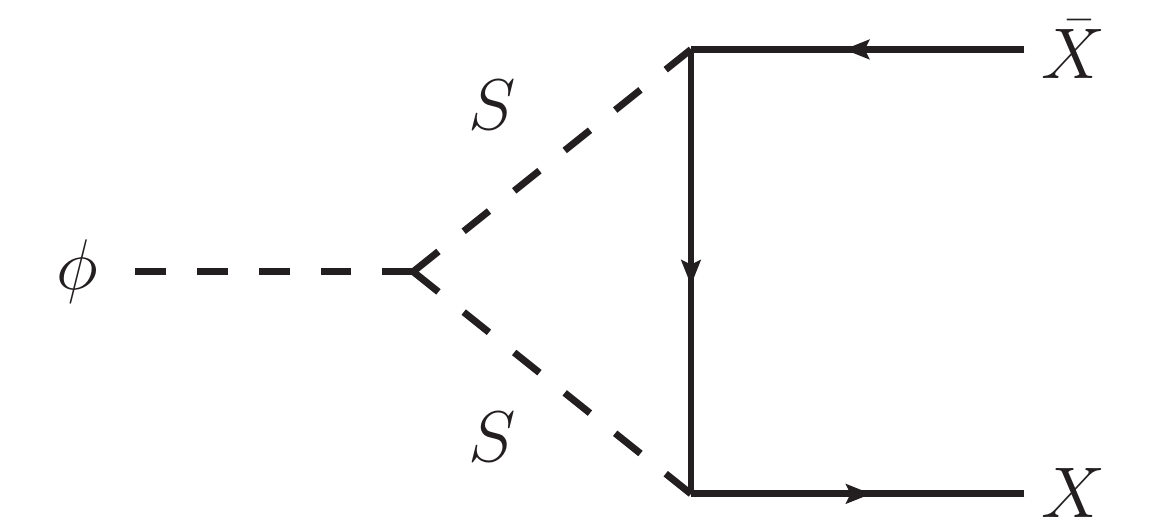}
\end{center}
\caption{The loop correction leading to the effection quintessence-DM interaction. 
}
\label{fig:triangle}
\end{figure}

The coupling in Eq.~(\ref{eq:DMyuk}), via the triangle diagram shown in Fig.~\ref{fig:triangle}, also leads to an effective quintessence-DM interaction $y_{\rm eff}\phi\bar{X}X$. We estimate
	\begin{equation}
	y_{\rm eff} \sim \frac{ c_{S}y_{D}^{2}m_{S}^{2} }{ 16\pi^{2}M_{\rm Pl}m_{\rm DM} }, \; \; \; \mathrm{for} \; m_{S}<m_{\rm DM}.
	\label{eq:yeff}
	\end{equation}
As the quintessence field has a mass $m_{\phi}^{2} < H_{u}^{2}$, this induces a long range fifth force between DM particles. Violations of the equivalence principle in the dark sector are constrained from observations of tidal streams of the Sagittarius dwarf spheroidal galaxy. This limits the fifth force normalised to gravity to $\beta^{2} \equiv 2y_{\rm eff}^{2}M_{\rm Pl}^{2}/m_{\rm DM}^{2} \lesssim 0.04$, i.e.~4\% of the gravitational strength~\cite{Kesden:2006zb,Kesden:2006vz} (also see~\cite{Friedman:1991dj,Bean:2001ys,Gubser:2004uh,Nusser:2004qu,Bean:2008ac,Bai:2015vca}, the last of which argues from CMB considerations that the constraint is much more stringent, $\beta^{2} \lesssim 10^{-4}$). Inserting the example parameters from Fig.~\ref{fig:SIDM1b}, we find $\beta^{2}\sim 10^{-11}$, i.e.~much weaker than the constraint. This may at first be surprising, given it is suppressed far below the typical loop factor squared, $\sim 1/(4\pi)^{4}$. But note there is an additional suppression in Eq.~(\ref{eq:yeff}) due to the hierarchy $m_{S} \ll m_{\rm DM}, \; M_{\rm Pl}$.

At early times in the single exponential potential, $\phi/M_{\rm Pl}$ is pinned due to Hubble friction. For our example with $c_{\phi}=0.5$, we find at early times (say prior to the CMB epoch),  $\phi/M_{\rm Pl} \approx -0.15$. Given $c_S \approx 23$ this implies $m_{S} \approx 50$ MeV at these and earlier times. Hence $S$ still provides an annihilation channel for the DM~\cite{Baldes:2017gzw}, and also the strong constraints on light mediators coming from the CMB and BBN can be somewhat relaxed~\cite{Hufnagel:2017dgo,Hufnagel:2018bjp,Forestell:2018txr} (here the CMB constraint on annihilating DM is avoided due to the p-wave annihilation, and BBN constraint and direct detection constraints are also satisfied, see Fig.~8 right of~\cite{Hufnagel:2018bjp}). Of course $m_{S}^{2}$ may have a more complicated functional dependence on $\phi$ than in Eq.~(\ref{eq:Smass}), which could lead to a much heavier $m_{S}^{2}$ in the early Universe and to different model building opportunities. Alternatively, $\phi$ itself may have a more complicated potential such as the sum-of-exponentials discussed above, leading to a larger field value at early times~\cite{Agrawal:2019dlm}. 

\subsection{Strongly coupled model}

\begin{figure}[t]
\begin{center}
\includegraphics[width=200pt]{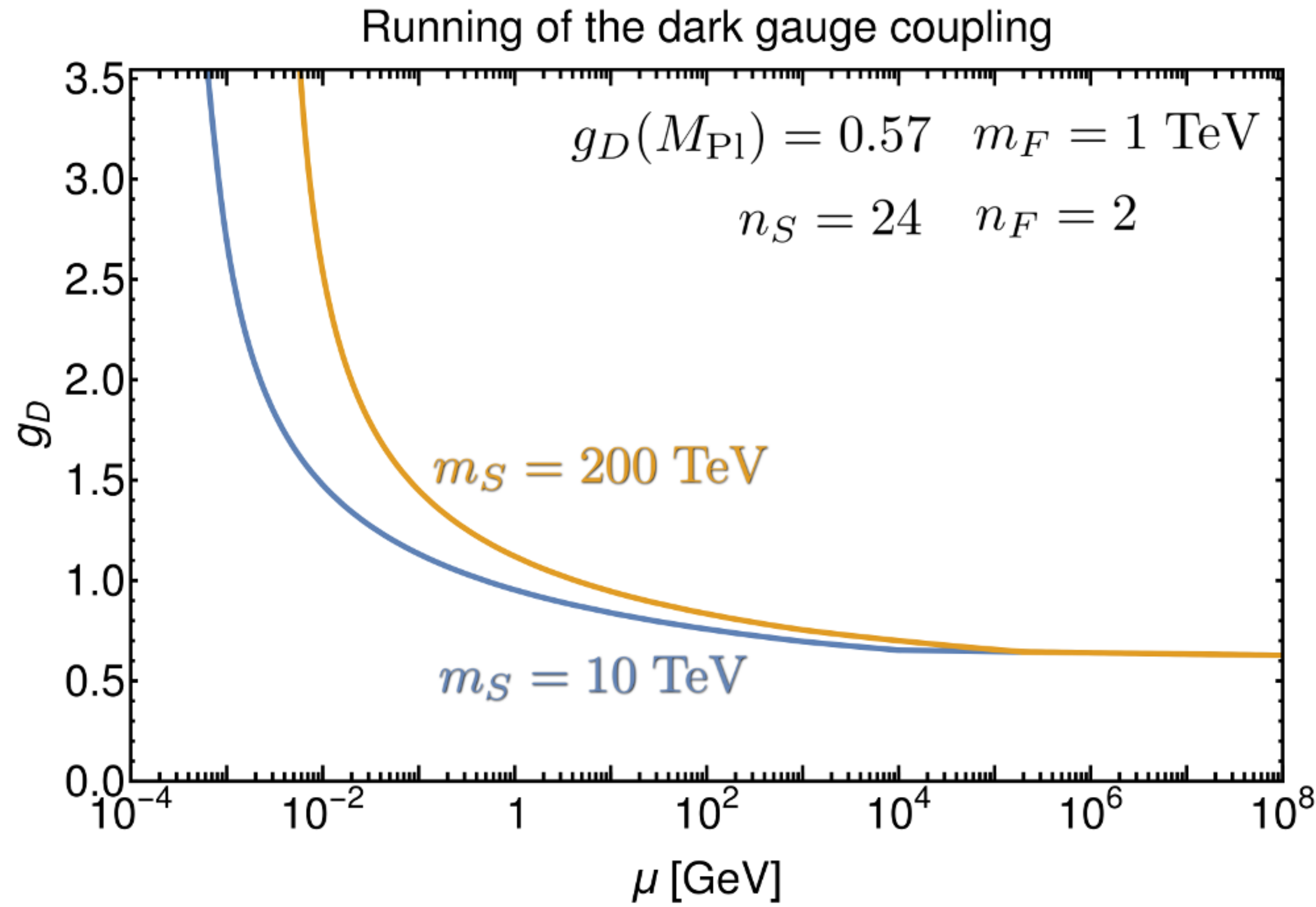}
\end{center}
\caption{Running of the dark gauge coupling in the strongly coupled SIDM model for different choices of $m_{S}$, with the same $g_{D}$ at $\mu = M_{\rm Pl}$. The confinement scale is changed by an order of magnitude between the two examples, for an increase of $m_{S}$ by a factor of twenty.
}
\label{fig:SIDM2}
\end{figure}

Consider DM interacting via strong dynamics. The interaction cross section scales on dimensional grounds as
	\begin{equation}
	\sigma \approx \frac{4\pi}{\Lambda_{\rm DM}^{2}},
	\end{equation}
where $\Lambda_{\rm DM}$ is the confinement scale of the dark gauge group. Clearly to increase $\sigma$ by a couple-of-orders of magnitude we need to decrease $\Lambda_{\rm DM}$ by a factor of ten. Note we do not want $m_{\rm DM}$ itself to change appreciably, hence we assume the constituent masses are much larger than $\Lambda_{\rm DM}$, leading to $m_{\rm DM} \sim 1$ TeV as in~\cite{Boddy:2014yra}. The scale of $\Lambda_{\rm DM}$ can be altered if the mass, $m_{S}$, of some scalar particles which enter the renormalisation group equations (RGEs) for the dark gauge coupling $g_{D}$ changes~\cite{Chacko:2002mf}. (As  the running of $g_{D}$ is different below and above $m_{S}$.) Here $S$ can again be part of the aforementioned tower of states. To pick a concrete example, let us study a dark $SU(3)_{D}$ for which the RGE can be written as
	\begin{align}
	\frac{d \, g_{D} }{ d \, \mathrm{ln} \mu } = \frac{ g_{D}^{3} }{ (4\pi)^{2} } \left(  \frac{n_{S}}{3}\Theta[\mu-m_{S}] + \frac{2n_{F}}{3}\Theta[\mu-m_{F}] - 11 \right), \nonumber \\
	\end{align} 
where $\Theta(x)$ is the Heaviside step function and we assume that in addition to the dark gluons, the field content comprises of $n_{S}$ complex scalars ($n_{F}$ Dirac fermions) transforming as the fundamental representation of $SU(3)_{D}$. An example showing the sensitivity to $m_{S}^{2}$ is shown in Fig.~\ref{fig:SIDM2}. Generally, a larger $n_{S}$ means we can achieve the required $\Delta \Lambda_{\rm DM}^{2}$ with a smaller $\Delta m_{S}/m_{S}$. In order to achieve the desired $\Delta m_{S}$, we can again invoke Eq.~(\ref{eq:Smass}). Typically we find $c' \sim \mathcal{O}(10)$, for $g_{D}(M_{\rm Pl}) \sim  \mathcal{O}(0.1)$, $n_{S}\sim \mathcal{O}(10)$, and $m_{S} \sim \mathcal{O}(10)$ TeV today. Note, similarly to the discussion above regarding the tuning of the quintessence potential, the induced change in the vacuum energy resulting from a change in $\Lambda_{\rm DM}$ must also be compensated for with a suitable tuning~\cite{Banks:2001qc}.

\section{Kinetic Mixing Portal}

Gauge kinetic mixing as a portal to DM production is widely used in the literature~\cite{Feldman:2006wd, Kang:2010mh,Chu:2013jja,Mambrini:2011dw}. Such a portal is also well motivated in various UV complete scenarios~\cite{Holdom:1985ag,Dienes:1996zr,Abel:2008ai}. The mixing strength between the dark U$(1)^\prime$ and the SM hypercharge U$(1)_\mathrm{Y}$ is often considered to be constant. Whereas, it has been shown~\cite{Banerjee:2019asa} that a varying gauge kinetic mixing naturally comes out while integrating out some heavy hybrid fermions charged under both gauge groups. As discussed in Sec.~\ref{secII}, one of the possible way to avoid the swampland is to introduce a nearly massless scalar field, $\phi$, slowly varying over time, attached to the potential of the theory. Introduction of such a quintessence field renders the parameters of the theory to be varying with time as well. In this section we shall concentrate on the phenomenological consequences of a dynamical gauge kinetic mixing due to the presence of a quintessence field $\phi$.

Let us the consider the presence of a massive vector mediator $Z^{\prime}_{\mu}$ coupled to a fermionic DM candidate $\chi$ while keeping the SM sector neutral with respect to the latter. The dark sector Lagrangian is then given by
\begin{equation}
\label{lag_dark}
\mathcal{L}_{\textrm{dark}}=-\dfrac{1}{4}\zp^{\mu\nu}Z^{\prime}_{\mu\nu}+\dfrac{1}{2}m_{Z^\prime}^2\zp^{\mu}Z^{\prime}_{\mu}+\bar{\chi}(i\slashed D-m_\chi)\chi\, ,
\end{equation}
where $\slashed D=\slashed\partial+ i g_{D}q_\chi\slashed Z^\prime$, $Z^{\prime}_{\mu \nu}= \partial_\mu Z^{\prime}_\nu-\partial_\nu Z^{\prime}_\mu$ is the field strength of $Z^{\prime}_\mu$, $g_D$ is the gauge coupling associated with U$(1)^{\prime}$ and $q_\chi$ is U$(1)^{\prime}$ charge of the DM $\chi$.  Due to gauge invariance, one can write a tree level kinetic mixing term between the dark  U$(1)^\prime$ and the SM hypercharge U$(1)_\mathrm{Y}$, given by
\begin{equation}
\label{lag_mix}
\mathcal{L}_{\textrm{mix}}=-\dfrac{\epsilon}{2}B^{\mu\nu}Z^{\prime}_{\mu\nu},
\end{equation}
\noindent
$B_\mu$ being the gauge field associated with the SM hypercharge. In the literature, the free parameter $\epsilon$ is generally taken to be small to avoid overproduction in freeze-out or freeze-in scenarios of DM. In what follows, we consider the kinetic mixing to be varying with time due to the presence of the quintessence field $\phi$. The ansatz for the varying gauge kinetic mixing we consider is
\begin{align}
\label{eq:kinmix}
\epsilon = \epsilon_0\, e^{-c_M \phi/M_\mathrm{Pl}}\ ,
\end{align} 
and $\phi(T)/M_\mathrm{Pl} = \ln(T_\mathrm{MAX}/T)$ as motivated by Eq.~(\ref{eq:phitracking}). Here $T_{\rm MAX}$ is the maximum temperature reached during the reheating process, which is completed at $T_{\rm RH}$. We absorb the prefactor appearing in Eq.~(\ref{eq:phitracking}) into $c_M$. Furthermore, as our purpose is to point out the possible effects of varying kinetic mixing through a simple example, we ignore the period in which $\phi$ is not close to its tracking value, and the departure from $w_{b} = 1/3$ during the reheating process. This is an approximation supported by the fact that in this setup, most of the DM production happens after reheating is complete, the Universe has entered the radiation dominated epoch, and the behaviour of $\phi$ can be well described by Eq.~(\ref{eq:phitracking}). We leave the inclusion of departures from this behaviour to future work.

\begin{figure}[t]
\centering
\includegraphics[width=0.45\textwidth]{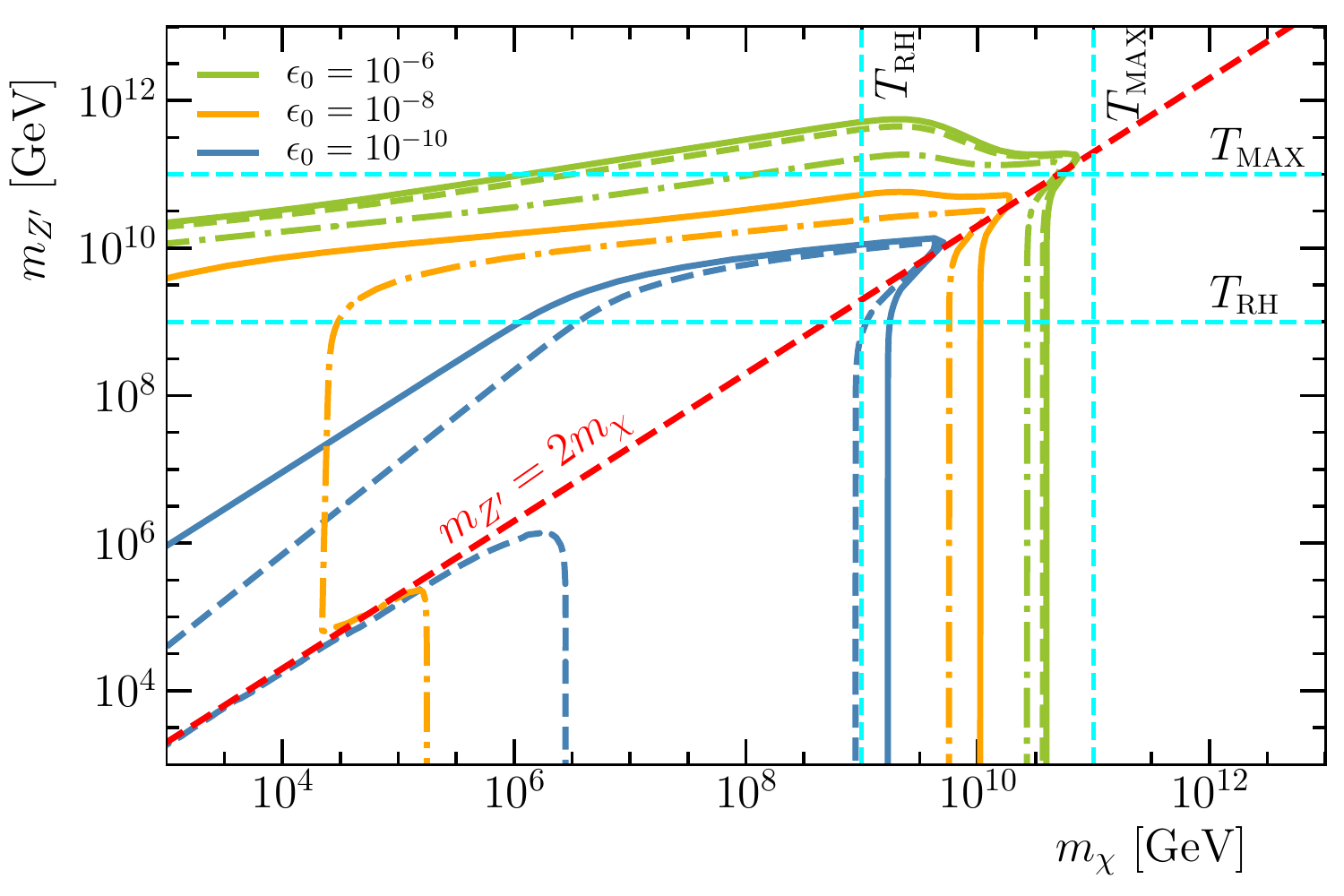}
\caption{
Contours of $\Omega h^2 = 0.12$ for both constant and varying $\epsilon$. The colours (green, yellow, blue) correspond to $\epsilon_0 = 10^{-6}, \, 10^{-8}, 10^{-10}$ (or equivalently top-to-bottom for the solid curves). The solid curves correspond to constant kinetic mixing, i.e.~$c_M=0$. The dashed curves correspond to $c_M = 0.1$ ($\epsilon_0 = 10^{-6}$ and  $10^{-10}$, top and bottom respectively) and the dash-dotted curves to $c_M = 0.5$ ($\epsilon_0 = 10^{-6}$ and $10^{-8}$, top and bottom respectively). As indicated, the diagonal red line shows the important kinematic threshold $m_{Z^\prime} =2 m_\chi$. For further details see the main text.}
\label{Fig:mdm-mzp}
\end{figure}

We consider DM production via the freeze-in mechanism~\cite{Hall:2009bx,Chu:2011be} (for more details on the setup, see~\cite{Banerjee:2019asa} and references therein). We assume that at $T_\mathrm{MAX}$ the gauge kinetic mixing is $\epsilon(T_\mathrm{MAX}) = \epsilon_0$. We consider $c_M > 0$ so the kinetic mixing is decreasing over time as the Universe cools down to the present temperature. In Fig.~\ref{Fig:mdm-mzp}, we compare the contours of $\Omega h^2 = 0.12$ for the constant mixing case with varying kinetic mixing scenario for different values of the coefficient $c_M$, for $T_\mathrm{RH} = 10^{9}$ GeV and $T_\mathrm{MAX} = 100\, T_\mathrm{RH}$.

With a larger $c_M$ one achieves a larger variation of the gauge kinetic mixing between $T_\mathrm{MAX}$ and the present day temperature, as can be understood from Eq.~\eqref{eq:kinmix}. On the other hand, larger value of $\epsilon_0$ makes the DM production rate large, which further pushes the $m_{Z^\prime}$ to larger values to tame the DM over production. As a result of this, green dashed contour with $\epsilon_0 = 10^{-6}$ and $c_M = 0.1$ does not considerably deviate from the green solid contour with constant kinetic mixing, whereas the slight deviation of the green dash-dotted contour from the latter is due to a larger variation of the kinetic mixing due to $c_M = 0.5$. The blue dashed contour with $\epsilon_0 = 10^{-10}$ and $c_M = 0.1$ starts deviating from the blue solid contour of constant mixing as the DM mass reaches below $T_\mathrm{RH}$. This is due to fact that, in the absence of any additional dependence of $T$ on the DM production rate, in this scenario DM production occurs mainly in the radiation dominated era (i.e. $T<T_\mathrm{RH}$). The larger the value of $c_M$, the smaller the kinetic mixing as the temperature of the Universe cools down, which further reduces the DM production rate. At a fixed DM mass, this reduction of DM production rate is compensated by moving to a smaller $m_{Z'}$ such that the DM relic density remains a constant. Thus, in Fig.~\ref{Fig:mdm-mzp} we observe the blue dashed contour with $c_M = 0.1$ significantly deviates from the contour with constant kinetic mixing as one approaches to a lower DM mass. The yellow dash-dotted contour is significantly different than the yellow solid contour with $\epsilon_0 = 10^{-8}$.  In this case, the large variation of the kinetic mixing makes the DM production rate small enough such that we get a lower limit on the DM mass for a given value of $m_{Z^\prime}$, beyond which the DM remains under abundant.

In the example above, we have seen the effects of the varying $\epsilon$ on the determination of the relic density. The DM and mediator masses are very high, however, and the couplings today so weak that direct or indirect detection signatures are completely out of experimental reach. The idea of varying kinetic mixing could, however, also be applied in alternative models which work at lower scales~\cite{Baldes:2017gzu,Hambye:2018dpi}, which have a myriad of cosmological, astrophysical and other experimental signatures. In such a case, the calculation of the variation of kinetic mixing could also be improved, e.g.~by including the precise variation of $\phi$ during the reheating era.

\section{Electroweak Sector}

\subsection{The potential}

We now consider adding a coupling of quintessence to the Higgs potential~\cite{Denef:2018etk,Murayama:2018lie,Han:2018yrk}
	\be 
	\label{Hphicoupling}
	V(\phi,H) \supset e^{-c_H \phi/M_{\rm Pl}}\,\bigg(\lambda_H\big[|H|^2-v^2\big]^2\bigg)\, ,
	\ee
where $c_H$ is a constant, $H$ is the EW Higgs doublet, $\lambda_H \approx 0.13$ is the SM Higgs quartic, and $v = 246$ GeV is the EW vacuum expectation value (VEV). Through a two-loop diagram, this potential induces an effective coupling of $\phi$ to nucleons at low energies, which results in a constraint from fifth force experiments, $c_{H} \lesssim 0.044$~\cite{Choi:2018rze}.

\subsection{Symmetry restoration}

Let us now ask what happens at finite temperature. Adding the leading temperature corrections to Eq.~(\ref{Hphicoupling}), we find
	\begin{align}  
	V(\phi,H) = & \; e^{-c_H \phi /M_{\rm Pl}}\,\bigg(\lambda_H\big[|H|^2-v^2\big]^2+\Lambda\bigg) \nonumber \\ & \; + c_T T^2 |H|^2 \, \label{finiteT} 
	\end{align}
where $c_T$ is the thermal mass coefficient of the Higgs~\cite{Carrington:1991hz}. We have implicitly used a high temperature expansion in writing the above potential, which breaks down for temperatures below the masses of the SM particles~\cite{Quiros:1999jp}. We deem this approximation sufficient for the present discussion as we are interested only in describing the qualitative behaviour of the system at high temperature. In this respect, Eq.~(\ref{finiteT}) mimics the SM case, where finite $T$ effects work to restore the EW symmetry. Because $\phi$ couples only to the Higgs, only the dependence on the Higgs quartic is modified compared to the SM case and
	\begin{equation}
	c_T = e^{-c_H \phi/M_{\rm Pl}} \frac{\lambda_H}{2} + \frac{3g_{2}^2}{16} + \frac{g_{Y}^2}{16} + \frac{y_{t}^2}{4} \approx 0.4,
	\end{equation}
remains SM like and dominated by the top Yukawa, $y_{t}$, with sub-leading contribution from the EW gauge couplings $g_{2}$ and $g_{Y}$~\cite{Carrington:1991hz}. Taking the first derivative with respect to $H$, we find the VEV depends on $T$ as
	\be
	\label{vevfiniteT}
	|H| = \mathrm{Re}\left[ \sqrt{v^{2} - e^{c_H \phi/M_{\rm Pl}}\frac{c_T T^2}{2\lambda_H}} \right] .
	\ee
During radiation domination $\phi$ is at some value $\phi_i \sim -\mathcal{O}(M_{\rm Pl})$. Thus at $T \gtrsim v$ the second term dominates and symmetry is restored, with some marginal, quantitatively small differences compared to the SM. We have only included the $T^{2}|H|^{2}$ correction in our analysis above, which does not allow us to study the order/strength of the phase transition. Nevertheless, we can reason that because the effective quartic term is strengthened through the quintessence coupling, we expect a smooth cross-over rather than a first-order phase transition in the present scenario, i.e.~a qualitatively same behaviour as in the SM~\cite{Carrington:1991hz,DOnofrio:2015gop}. This follows from the well-known behaviour of the SM phase diagram, which requires a smaller quartic coupling than $\lambda_{H} \approx 0.13$, inferred from the Higgs mass measurement, in order to return a strong first-order electroweak phase transition~\cite{Csikor:1998eu}.

The EW sphalerons are active at $T \gtrsim \mathcal{O}(100)$ GeV as the EW symmetry is restored. Thus allowing for leptogenesis if Majorana neutrinos are added to the field content~\cite{Fukugita:1986hr,Davidson:2008bu}, or alternatively EW baryogenesis is possible~\cite{Shaposhnikov:1987tw,Trodden:1998ym,Konstandin:2013caa}, provided additional modifications are made to the scalar potential in order to achieve a first order phase transition and an additional source of CP violation is present. Indeed, if coupling to quintessence more generally results in modified gauge or Yukawa couplings, the necessary ingredients could be attained~\cite{Berkooz:2004kx,Baldes:2016gaf,Ellis:2019flb}. Note that BBN is expected to proceed unaffected from the SM case~\cite{Ratra:1987rm}.

\subsection{Electroweak vacuum lifetime}

The effective potential $V_{\rm eff}$ in the SM famously contains a deeper minimum than the observed EW one. The deeper minimum corresponds to negative vacuum energy, i.e.~anti-de Sitter (AdS) space, which may be argued to be safer in the context of string compactification. In the current context, as $\phi$ evolves by rolling down its potential, the effective quartic coupling of the Higgs is suppressed, via Eq.~(\ref{Hphicoupling}). This implies, once quantum corrections are added, the Higgs instability scale is decreased and with it the implied lifetime of the EW vacuum. Here we will quantify this statement through an explicit calculation. An interesting question in light of the two conjectures, is whether the vacuum will first transition to the AdS space, or whether the light tower of states will appear due to the trans-Planckian motion of $\phi$? (Here we are assuming $\phi$ has not already moved $\sim M_{\rm Pl}$ by today and hence the light tower has not yet appeared in the hidden sector.)

\begin{figure}[t]
\begin{center}
\includegraphics[width=200pt]{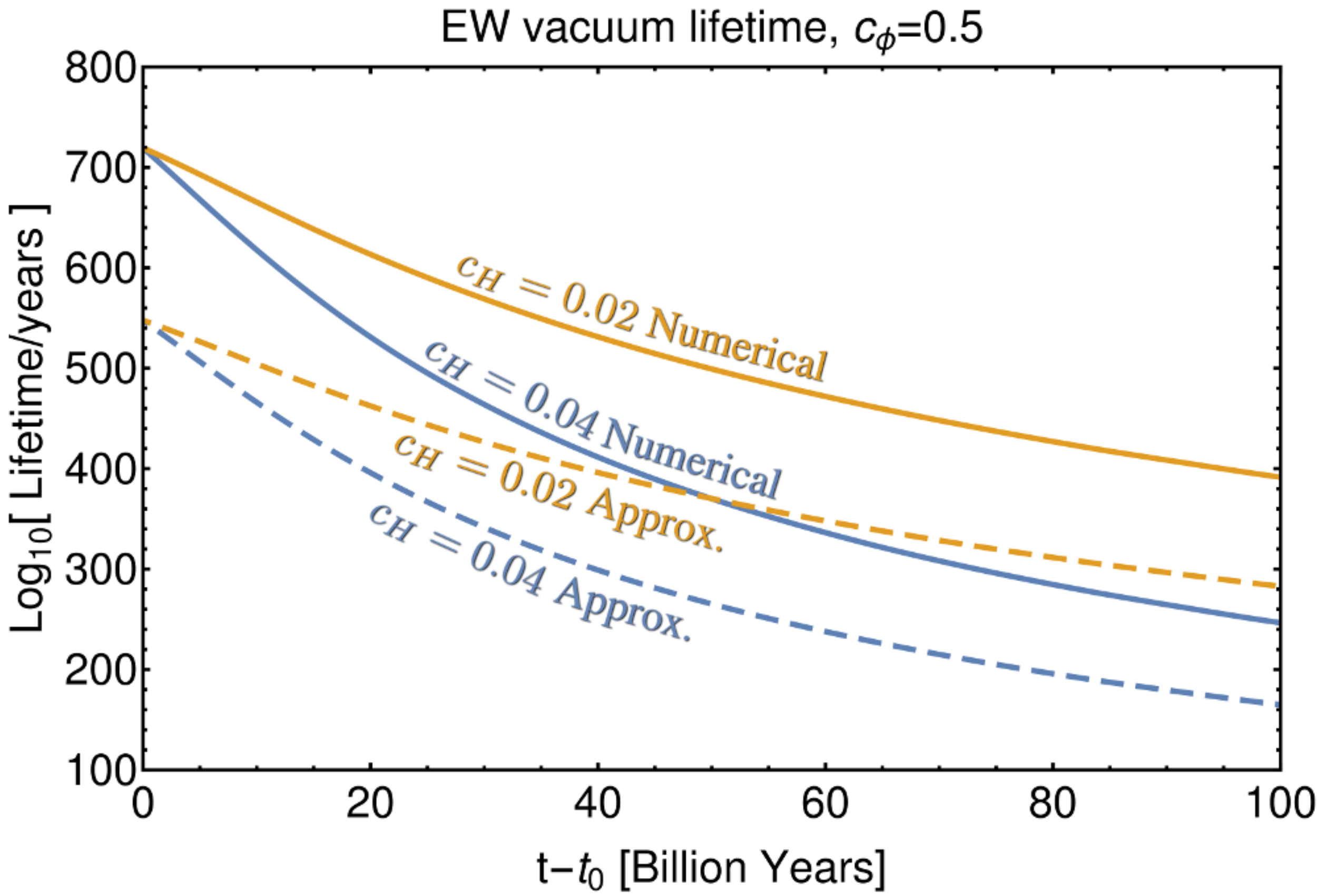}
\end{center}
\caption{The lifetime of the EW vacuum as a function of the cosmic time, today corresponds to $t=t_0$. The solid lines show the result using a numerical determination of $S_{4}$ and the lower line the approximation from Eq.~(\ref{eq:S4approx}). As time progresses, $\phi$ increases, $m_{H}$ decreases, and the EW vacuum becomes less stable.
}
\label{fig:lifetime}
\end{figure}

To proceed in answering this question, we consider the effective potential in the Higgs  field direction. We write the effective quartic at the EW scale as
	 \begin{equation}
	\tilde{\lambda}_{H}(v_{\rm EW})  \equiv e^{-c_H \phi/M_{\rm Pl}}\lambda_{H}(v_{\rm EW}).
	\label{eq:efflambda}
	\end{equation}
Once quantum corrections are included the effective potential in the Higgs field direction can be approximated as
	\begin{equation}
	\label{eq:veffapprox}
	V_{\rm eff}(h) \approx \frac{\tilde{\lambda}_{H}(\mu=h)}{4}h^{4},
	\end{equation}
where $\tilde{\lambda}_{H}(\mu)$ can be determined at the renormalisation scale $\mu$, using the renormalisation group equations (RGEs) and the boundary condition in Eq.~(\ref{eq:efflambda}). Note a more careful determination involves using the RGE improved effective potential, which takes into account the wave function renormalisation of the Higgs, together with corrections as SM particles gain masses proportional to the Higgs field value, instead of Eq.~(\ref{eq:veffapprox}).  (The difference in the two approaches is marginal for the accuracy in the desired final result of the current calculation.) In our numerical analysis we use the effective potential and RGEs provided in~\cite{Buttazzo:2013uya}.
	
The bubble nucleation rate per unit volume is given approximately by $\Gamma \sim R^{-4}e^{-S_4}$~\cite{Frampton:1976kf,Coleman:1977py,Callan:1977pt} where $R$ is the bubble size and 
	\be
	S_{4}=2\pi^{2}\int r^{3} \left\{\frac{1}{2}\left(\frac{d h}{dr}\right)^{2}+V_{\rm eff}(h)-V_{\rm eff}(v)\right\}dr,
	\label{eq:S4}
	\ee
is the O(4)-symmetric Euclidean action. Variation of the action yields the equation of motion
	\begin{equation}
	\frac{d^{2}h}{dr^{2}} + \frac{3}{r}\frac{d h}{dr} = \frac{\partial V_{eff}}{\partial h},
	\end{equation}
for which the associated boundary conditions are $(dh/dr)|_{r=0}=0$ and $h|_{r \to \infty}=v$. Given a close-to-conformal potential the action may be approximated as
	\begin{equation}
	\label{eq:S4approx}
	S_{4} \approx \frac{8\pi}{3|\mathrm{Min}[\lambda_{H}(\mu)]|}.
	\end{equation}
Furthermore, the bubble size is then set by $R^{-1} \sim \mu_{\rm Min}$, where $\mu_{\rm Min}$ is the scale at which $\mathrm{Min}[\lambda_{H}(\mu)]$ is reached.
Alternatively, we may  determine $S_{4}$ by finding the critical bubble profile numerically. Given a vacuum energy density dominated Universe the EW lifetime is~\cite{Buttazzo:2013uya}
	\begin{equation}
	\tau_{\rm EW} \approx \frac{3H_{u}^{3}}{4\pi\Gamma},
	\end{equation}
where $H_{u}$ is the Hubble parameter. Note we ignore gravitational and higher order quantum mechanical corrections to the tunneling rate, which still allows us to achieve a good approximation for our purposes. We have also assumed no higher dimensional operators enter which can act either to increase or decrease the $t_{\rm EW}$. Note that for the SM-like point, $\phi=0$, the field tunnels to $h=1.25M_{\rm Pl}$ in our determination, which means the tunneling rate can be very sensitive even to (non-reduced) Planck mass suppressed operators~\cite{Branchina:2013jra,Branchina:2014rva}. Nevertheless, as the instability worsens, the tunneling point moves in to lower field values, which reduces the sensitivity to Planck suppressed operators. Another question left for further work, is to consider two field tunneling, in which both $h$ and $\phi$ vary across the bubble wall. This poses a numerical challenge, but could decrease $t_{\rm EW}$ from Fig.~\ref{fig:lifetime}, which should therefore only be considered an upper bound. Nevertheless, assuming the calculated $t_{\rm EW}$, it is plausible that the light tower of states will appear before the EW vacuum decays. This is because, for $c_{\phi} =0.5$, we find that 100 billion years from today; $\Delta \phi \approx 2.5 M_{\rm Pl}$ which is certainly trans-Planckian, while $t_{\rm EW} \gtrsim \mathcal{O}(\mathrm{few} \; 100)$ billion years.   

\section{Conclusion}

Inspired by the swampland conjectures, we embarked on a tour of some selected topics in particle phenomenology, which could be hiding new physics associated with the de Sitter and distance conjectures. We speculated DM could be interacting via the tower of states associated with the distance conjecture and that these interactions are what solves the small scale structure issues. We found it plausible that time varying interactions could generically give a fit for the required thermalisation on dwarf galaxy scales while leaving observed galaxy cluster collisions unaffected. We presented two ways particles in the tower of states could influence SIDM, either by acting as the mediator directly, or by acting as a changing threshold in some RGEs which control the SIDM cross section.

We further explored the kinetic mixing portal. This served as an illustration, in a simple DM model, of the effects of varying couplings on DM production. In our example, the portal coupling is suppressed at late times and the DM is produced via freeze-in mechanism in the early universe. We showed that a larger variation of the kinetic mixing due to the quintessence field, gives a lower bound on the DM mass irrespective of the mediator mass. As the kinetic mixing is suppressed at late times, signatures of such portal at direct or indirect detection experiments are challenging. Instead, one could search for the effects of the quintessence field itself in cosmological observables. It would also be interesting to further explore this direction in models with light or massless dark photons which have implications for direct detection searches~\cite{Baldes:2017gzu,Hambye:2018dpi}.

Finally we studied the EW sector phase structure, in the light of a quintessence-Higgs coupling, showing symmetry restoration proceeds largely unaffected at high temperatures. We also calculated the modified lifetime of the EW vacuum, and showed the quintessence field will typically move a trans-Planckian distance in field space, prior to EW vacuum decay. Hence, observers in the far-future should be able to discover the effects of the descending tower of states, before the Universe transits into a deeper AdS phase.

\begin{acknowledgments}
We thank Riccardo Argurio, Emilian Dudas, Lucien Heurtier, Laura Lopez-Honorez, Yann Mambrini, Miguel~Montero, Filippo Sala, and Chang Sub Shin for useful discussions. This research was supported by the F.R.S.--FNRS under the Excellence of Science (EoS) project No. 30820817 -- be.h ``\emph{The H boson gateway to physics beyond the Standard Model}'' and by the (Indo-French) CEFIPRA/IFCPAR Project No.~5404-2. IB is a postdoctoral researcher of the F.R.S.--FNRS.
\end{acknowledgments}

\bibliographystyle{apsrev4-1}
\bibliography{swamp}

\begin{thebibliography}{126}%
\makeatletter
\providecommand \@ifxundefined [1]{%
 \@ifx{#1\undefined}
}%
\providecommand \@ifnum [1]{%
 \ifnum #1\expandafter \@firstoftwo
 \else \expandafter \@secondoftwo
 \fi
}%
\providecommand \@ifx [1]{%
 \ifx #1\expandafter \@firstoftwo
 \else \expandafter \@secondoftwo
 \fi
}%
\providecommand \natexlab [1]{#1}%
\providecommand \enquote  [1]{``#1''}%
\providecommand \bibnamefont  [1]{#1}%
\providecommand \bibfnamefont [1]{#1}%
\providecommand \citenamefont [1]{#1}%
\providecommand \href@noop [0]{\@secondoftwo}%
\providecommand \href [0]{\begingroup \@sanitize@url \@href}%
\providecommand \@href[1]{\@@startlink{#1}\@@href}%
\providecommand \@@href[1]{\endgroup#1\@@endlink}%
\providecommand \@sanitize@url [0]{\catcode `\\12\catcode `\$12\catcode
  `\&12\catcode `\#12\catcode `\^12\catcode `\_12\catcode `\%12\relax}%
\providecommand \@@startlink[1]{}%
\providecommand \@@endlink[0]{}%
\providecommand \url  [0]{\begingroup\@sanitize@url \@url }%
\providecommand \@url [1]{\endgroup\@href {#1}{\urlprefix }}%
\providecommand \urlprefix  [0]{URL }%
\providecommand \Eprint [0]{\href }%
\providecommand \doibase [0]{http://dx.doi.org/}%
\providecommand \selectlanguage [0]{\@gobble}%
\providecommand \bibinfo  [0]{\@secondoftwo}%
\providecommand \bibfield  [0]{\@secondoftwo}%
\providecommand \translation [1]{[#1]}%
\providecommand \BibitemOpen [0]{}%
\providecommand \bibitemStop [0]{}%
\providecommand \bibitemNoStop [0]{.\EOS\space}%
\providecommand \EOS [0]{\spacefactor3000\relax}%
\providecommand \BibitemShut  [1]{\csname bibitem#1\endcsname}%
\let\auto@bib@innerbib\@empty
\bibitem [{\citenamefont {Vafa}(2005)}]{Vafa:2005ui}%
  \BibitemOpen
  \bibfield  {author} {\bibinfo {author} {\bibfnamefont {C.}~\bibnamefont
  {Vafa}},\ }\href@noop {} {\  (\bibinfo {year} {2005})},\ \Eprint
  {http://arxiv.org/abs/hep-th/0509212} {arXiv:hep-th/0509212 [hep-th]}
  \BibitemShut {NoStop}%
\bibitem [{\citenamefont {Palti}(2019)}]{Palti:2019pca}%
  \BibitemOpen
  \bibfield  {author} {\bibinfo {author} {\bibfnamefont {E.}~\bibnamefont
  {Palti}}\ }(\bibinfo {year} {2019})\ \Eprint
  {http://arxiv.org/abs/1903.06239} {arXiv:1903.06239 [hep-th]} \BibitemShut
  {NoStop}%
\bibitem [{\citenamefont {Obied}\ \emph {et~al.}(2018)\citenamefont {Obied},
  \citenamefont {Ooguri}, \citenamefont {Spodyneiko},\ and\ \citenamefont
  {Vafa}}]{Obied:2018sgi}%
  \BibitemOpen
  \bibfield  {author} {\bibinfo {author} {\bibfnamefont {G.}~\bibnamefont
  {Obied}}, \bibinfo {author} {\bibfnamefont {H.}~\bibnamefont {Ooguri}},
  \bibinfo {author} {\bibfnamefont {L.}~\bibnamefont {Spodyneiko}}, \ and\
  \bibinfo {author} {\bibfnamefont {C.}~\bibnamefont {Vafa}},\ }\href@noop {}
  {\  (\bibinfo {year} {2018})},\ \Eprint {http://arxiv.org/abs/1806.08362}
  {arXiv:1806.08362 [hep-th]} \BibitemShut {NoStop}%
\bibitem [{\citenamefont {Garg}\ and\ \citenamefont
  {Krishnan}(2018)}]{Garg:2018reu}%
  \BibitemOpen
  \bibfield  {author} {\bibinfo {author} {\bibfnamefont {S.~K.}\ \bibnamefont
  {Garg}}\ and\ \bibinfo {author} {\bibfnamefont {C.}~\bibnamefont
  {Krishnan}},\ }\href@noop {} {\  (\bibinfo {year} {2018})},\ \Eprint
  {http://arxiv.org/abs/1807.05193} {arXiv:1807.05193 [hep-th]} \BibitemShut
  {NoStop}%
\bibitem [{\citenamefont {Ooguri}\ \emph {et~al.}(2019)\citenamefont {Ooguri},
  \citenamefont {Palti}, \citenamefont {Shiu},\ and\ \citenamefont
  {Vafa}}]{Ooguri:2018wrx}%
  \BibitemOpen
  \bibfield  {author} {\bibinfo {author} {\bibfnamefont {H.}~\bibnamefont
  {Ooguri}}, \bibinfo {author} {\bibfnamefont {E.}~\bibnamefont {Palti}},
  \bibinfo {author} {\bibfnamefont {G.}~\bibnamefont {Shiu}}, \ and\ \bibinfo
  {author} {\bibfnamefont {C.}~\bibnamefont {Vafa}},\ }\href {\doibase
  10.1016/j.physletb.2018.11.018} {\bibfield  {journal} {\bibinfo  {journal}
  {Phys. Lett.}\ }\textbf {\bibinfo {volume} {B788}},\ \bibinfo {pages} {180}
  (\bibinfo {year} {2019})},\ \Eprint {http://arxiv.org/abs/1810.05506}
  {arXiv:1810.05506 [hep-th]} \BibitemShut {NoStop}%
\bibitem [{\citenamefont {Aghanim}\ \emph {et~al.}(2018)\citenamefont {Aghanim}
  \emph {et~al.}}]{Aghanim:2018eyx}%
  \BibitemOpen
  \bibfield  {author} {\bibinfo {author} {\bibfnamefont {N.}~\bibnamefont
  {Aghanim}} \emph {et~al.} (\bibinfo {collaboration} {Planck}),\ }\href@noop
  {} {\  (\bibinfo {year} {2018})},\ \Eprint {http://arxiv.org/abs/1807.06209}
  {arXiv:1807.06209 [astro-ph.CO]} \BibitemShut {NoStop}%
\bibitem [{\citenamefont {Scolnic}\ \emph {et~al.}(2018)\citenamefont {Scolnic}
  \emph {et~al.}}]{Scolnic:2017caz}%
  \BibitemOpen
  \bibfield  {author} {\bibinfo {author} {\bibfnamefont {D.~M.}\ \bibnamefont
  {Scolnic}} \emph {et~al.},\ }\href {\doibase 10.3847/1538-4357/aab9bb}
  {\bibfield  {journal} {\bibinfo  {journal} {Astrophys. J.}\ }\textbf
  {\bibinfo {volume} {859}},\ \bibinfo {pages} {101} (\bibinfo {year}
  {2018})},\ \Eprint {http://arxiv.org/abs/1710.00845} {arXiv:1710.00845
  [astro-ph.CO]} \BibitemShut {NoStop}%
\bibitem [{\citenamefont {Agrawal}\ \emph {et~al.}(2018)\citenamefont
  {Agrawal}, \citenamefont {Obied}, \citenamefont {Steinhardt},\ and\
  \citenamefont {Vafa}}]{Agrawal:2018own}%
  \BibitemOpen
  \bibfield  {author} {\bibinfo {author} {\bibfnamefont {P.}~\bibnamefont
  {Agrawal}}, \bibinfo {author} {\bibfnamefont {G.}~\bibnamefont {Obied}},
  \bibinfo {author} {\bibfnamefont {P.~J.}\ \bibnamefont {Steinhardt}}, \ and\
  \bibinfo {author} {\bibfnamefont {C.}~\bibnamefont {Vafa}},\ }\href {\doibase
  10.1016/j.physletb.2018.07.040} {\bibfield  {journal} {\bibinfo  {journal}
  {Phys. Lett.}\ }\textbf {\bibinfo {volume} {B784}},\ \bibinfo {pages} {271}
  (\bibinfo {year} {2018})},\ \Eprint {http://arxiv.org/abs/1806.09718}
  {arXiv:1806.09718 [hep-th]} \BibitemShut {NoStop}%
\bibitem [{\citenamefont {Heisenberg}\ \emph {et~al.}(2018)\citenamefont
  {Heisenberg}, \citenamefont {Bartelmann}, \citenamefont {Brandenberger},\
  and\ \citenamefont {Refregier}}]{Heisenberg:2018yae}%
  \BibitemOpen
  \bibfield  {author} {\bibinfo {author} {\bibfnamefont {L.}~\bibnamefont
  {Heisenberg}}, \bibinfo {author} {\bibfnamefont {M.}~\bibnamefont
  {Bartelmann}}, \bibinfo {author} {\bibfnamefont {R.}~\bibnamefont
  {Brandenberger}}, \ and\ \bibinfo {author} {\bibfnamefont {A.}~\bibnamefont
  {Refregier}},\ }\href {\doibase 10.1103/PhysRevD.98.123502} {\bibfield
  {journal} {\bibinfo  {journal} {Phys. Rev.}\ }\textbf {\bibinfo {volume}
  {D98}},\ \bibinfo {pages} {123502} (\bibinfo {year} {2018})},\ \Eprint
  {http://arxiv.org/abs/1808.02877} {arXiv:1808.02877 [astro-ph.CO]}
  \BibitemShut {NoStop}%
\bibitem [{\citenamefont {Akrami}\ \emph {et~al.}(2019)\citenamefont {Akrami},
  \citenamefont {Kallosh}, \citenamefont {Linde},\ and\ \citenamefont
  {Vardanyan}}]{Akrami:2018ylq}%
  \BibitemOpen
  \bibfield  {author} {\bibinfo {author} {\bibfnamefont {Y.}~\bibnamefont
  {Akrami}}, \bibinfo {author} {\bibfnamefont {R.}~\bibnamefont {Kallosh}},
  \bibinfo {author} {\bibfnamefont {A.}~\bibnamefont {Linde}}, \ and\ \bibinfo
  {author} {\bibfnamefont {V.}~\bibnamefont {Vardanyan}},\ }\href {\doibase
  10.1002/prop.201800075} {\bibfield  {journal} {\bibinfo  {journal} {Fortsch.
  Phys.}\ }\textbf {\bibinfo {volume} {67}},\ \bibinfo {pages} {1800075}
  (\bibinfo {year} {2019})},\ \Eprint {http://arxiv.org/abs/1808.09440}
  {arXiv:1808.09440 [hep-th]} \BibitemShut {NoStop}%
\bibitem [{\citenamefont {Ade}\ \emph {et~al.}(2016)\citenamefont {Ade} \emph
  {et~al.}}]{Ade:2015xua}%
  \BibitemOpen
  \bibfield  {author} {\bibinfo {author} {\bibfnamefont {P.~A.~R.}\
  \bibnamefont {Ade}} \emph {et~al.} (\bibinfo {collaboration} {Planck}),\
  }\href {\doibase 10.1051/0004-6361/201525830} {\bibfield  {journal} {\bibinfo
   {journal} {Astron. Astrophys.}\ }\textbf {\bibinfo {volume} {594}},\
  \bibinfo {pages} {A13} (\bibinfo {year} {2016})},\ \Eprint
  {http://arxiv.org/abs/1502.01589} {arXiv:1502.01589 [astro-ph.CO]}
  \BibitemShut {NoStop}%
\bibitem [{\citenamefont {Riess}\ \emph {et~al.}(2019)\citenamefont {Riess},
  \citenamefont {Casertano}, \citenamefont {Yuan}, \citenamefont {Macri},\ and\
  \citenamefont {Scolnic}}]{Riess:2019cxk}%
  \BibitemOpen
  \bibfield  {author} {\bibinfo {author} {\bibfnamefont {A.~G.}\ \bibnamefont
  {Riess}}, \bibinfo {author} {\bibfnamefont {S.}~\bibnamefont {Casertano}},
  \bibinfo {author} {\bibfnamefont {W.}~\bibnamefont {Yuan}}, \bibinfo {author}
  {\bibfnamefont {L.~M.}\ \bibnamefont {Macri}}, \ and\ \bibinfo {author}
  {\bibfnamefont {D.}~\bibnamefont {Scolnic}},\ }\href {\doibase
  10.3847/1538-4357/ab1422} {\bibfield  {journal} {\bibinfo  {journal}
  {Astrophys. J.}\ }\textbf {\bibinfo {volume} {876}},\ \bibinfo {pages} {85}
  (\bibinfo {year} {2019})},\ \Eprint {http://arxiv.org/abs/1903.07603}
  {arXiv:1903.07603 [astro-ph.CO]} \BibitemShut {NoStop}%
\bibitem [{\citenamefont {Salvatelli}\ \emph {et~al.}(2013)\citenamefont
  {Salvatelli}, \citenamefont {Marchini}, \citenamefont {Lopez-Honorez},\ and\
  \citenamefont {Mena}}]{Salvatelli:2013wra}%
  \BibitemOpen
  \bibfield  {author} {\bibinfo {author} {\bibfnamefont {V.}~\bibnamefont
  {Salvatelli}}, \bibinfo {author} {\bibfnamefont {A.}~\bibnamefont
  {Marchini}}, \bibinfo {author} {\bibfnamefont {L.}~\bibnamefont
  {Lopez-Honorez}}, \ and\ \bibinfo {author} {\bibfnamefont {O.}~\bibnamefont
  {Mena}},\ }\href {\doibase 10.1103/PhysRevD.88.023531} {\bibfield  {journal}
  {\bibinfo  {journal} {Phys. Rev.}\ }\textbf {\bibinfo {volume} {D88}},\
  \bibinfo {pages} {023531} (\bibinfo {year} {2013})},\ \Eprint
  {http://arxiv.org/abs/1304.7119} {arXiv:1304.7119 [astro-ph.CO]} \BibitemShut
  {NoStop}%
\bibitem [{\citenamefont {Di~Valentino}\ \emph {et~al.}(2017)\citenamefont
  {Di~Valentino}, \citenamefont {Melchiorri},\ and\ \citenamefont
  {Mena}}]{DiValentino:2017iww}%
  \BibitemOpen
  \bibfield  {author} {\bibinfo {author} {\bibfnamefont {E.}~\bibnamefont
  {Di~Valentino}}, \bibinfo {author} {\bibfnamefont {A.}~\bibnamefont
  {Melchiorri}}, \ and\ \bibinfo {author} {\bibfnamefont {O.}~\bibnamefont
  {Mena}},\ }\href {\doibase 10.1103/PhysRevD.96.043503} {\bibfield  {journal}
  {\bibinfo  {journal} {Phys. Rev.}\ }\textbf {\bibinfo {volume} {D96}},\
  \bibinfo {pages} {043503} (\bibinfo {year} {2017})},\ \Eprint
  {http://arxiv.org/abs/1704.08342} {arXiv:1704.08342 [astro-ph.CO]}
  \BibitemShut {NoStop}%
\bibitem [{\citenamefont {Yang}\ \emph {et~al.}(2018)\citenamefont {Yang},
  \citenamefont {Pan}, \citenamefont {Di~Valentino}, \citenamefont {Nunes},
  \citenamefont {Vagnozzi},\ and\ \citenamefont {Mota}}]{Yang:2018euj}%
  \BibitemOpen
  \bibfield  {author} {\bibinfo {author} {\bibfnamefont {W.}~\bibnamefont
  {Yang}}, \bibinfo {author} {\bibfnamefont {S.}~\bibnamefont {Pan}}, \bibinfo
  {author} {\bibfnamefont {E.}~\bibnamefont {Di~Valentino}}, \bibinfo {author}
  {\bibfnamefont {R.~C.}\ \bibnamefont {Nunes}}, \bibinfo {author}
  {\bibfnamefont {S.}~\bibnamefont {Vagnozzi}}, \ and\ \bibinfo {author}
  {\bibfnamefont {D.~F.}\ \bibnamefont {Mota}},\ }\href {\doibase
  10.1088/1475-7516/2018/09/019} {\bibfield  {journal} {\bibinfo  {journal}
  {JCAP}\ }\textbf {\bibinfo {volume} {1809}},\ \bibinfo {pages} {019}
  (\bibinfo {year} {2018})},\ \Eprint {http://arxiv.org/abs/1805.08252}
  {arXiv:1805.08252 [astro-ph.CO]} \BibitemShut {NoStop}%
\bibitem [{\citenamefont {\'O~Colg\'ain}\ and\ \citenamefont
  {Yavartanoo}(2019)}]{Colgain:2019joh}%
  \BibitemOpen
  \bibfield  {author} {\bibinfo {author} {\bibfnamefont {E.}~\bibnamefont
  {\'O~Colg\'ain}}\ and\ \bibinfo {author} {\bibfnamefont {H.}~\bibnamefont
  {Yavartanoo}},\ }\href@noop {} {\  (\bibinfo {year} {2019})},\ \Eprint
  {http://arxiv.org/abs/1905.02555} {arXiv:1905.02555 [astro-ph.CO]}
  \BibitemShut {NoStop}%
\bibitem [{\citenamefont {Poulin}\ \emph {et~al.}(2019)\citenamefont {Poulin},
  \citenamefont {Smith}, \citenamefont {Karwal},\ and\ \citenamefont
  {Kamionkowski}}]{Poulin:2018cxd}%
  \BibitemOpen
  \bibfield  {author} {\bibinfo {author} {\bibfnamefont {V.}~\bibnamefont
  {Poulin}}, \bibinfo {author} {\bibfnamefont {T.~L.}\ \bibnamefont {Smith}},
  \bibinfo {author} {\bibfnamefont {T.}~\bibnamefont {Karwal}}, \ and\ \bibinfo
  {author} {\bibfnamefont {M.}~\bibnamefont {Kamionkowski}},\ }\href {\doibase
  10.1103/PhysRevLett.122.221301} {\bibfield  {journal} {\bibinfo  {journal}
  {Phys. Rev. Lett.}\ }\textbf {\bibinfo {volume} {122}},\ \bibinfo {pages}
  {221301} (\bibinfo {year} {2019})},\ \Eprint
  {http://arxiv.org/abs/1811.04083} {arXiv:1811.04083 [astro-ph.CO]}
  \BibitemShut {NoStop}%
\bibitem [{\citenamefont {Kaloper}(2019)}]{Kaloper:2019lpl}%
  \BibitemOpen
  \bibfield  {author} {\bibinfo {author} {\bibfnamefont {N.}~\bibnamefont
  {Kaloper}},\ }\href@noop {} {\  (\bibinfo {year} {2019})},\ \Eprint
  {http://arxiv.org/abs/1903.11676} {arXiv:1903.11676 [hep-th]} \BibitemShut
  {NoStop}%
\bibitem [{\citenamefont {Agrawal}\ \emph
  {et~al.}(2019{\natexlab{a}})\citenamefont {Agrawal}, \citenamefont
  {Cyr-Racine}, \citenamefont {Pinner},\ and\ \citenamefont
  {Randall}}]{Agrawal:2019lmo}%
  \BibitemOpen
  \bibfield  {author} {\bibinfo {author} {\bibfnamefont {P.}~\bibnamefont
  {Agrawal}}, \bibinfo {author} {\bibfnamefont {F.-Y.}\ \bibnamefont
  {Cyr-Racine}}, \bibinfo {author} {\bibfnamefont {D.}~\bibnamefont {Pinner}},
  \ and\ \bibinfo {author} {\bibfnamefont {L.}~\bibnamefont {Randall}},\
  }\href@noop {} {\  (\bibinfo {year} {2019}{\natexlab{a}})},\ \Eprint
  {http://arxiv.org/abs/1904.01016} {arXiv:1904.01016 [astro-ph.CO]}
  \BibitemShut {NoStop}%
\bibitem [{\citenamefont {\'O~Colg\'ain}\ \emph {et~al.}(2019)\citenamefont
  {\'O~Colg\'ain}, \citenamefont {van Putten},\ and\ \citenamefont
  {Yavartanoo}}]{Colgain:2018wgk}%
  \BibitemOpen
  \bibfield  {author} {\bibinfo {author} {\bibfnamefont {E.}~\bibnamefont
  {\'O~Colg\'ain}}, \bibinfo {author} {\bibfnamefont {M.~H. P.~M.}\
  \bibnamefont {van Putten}}, \ and\ \bibinfo {author} {\bibfnamefont
  {H.}~\bibnamefont {Yavartanoo}},\ }\href {\doibase
  10.1016/j.physletb.2019.04.032} {\bibfield  {journal} {\bibinfo  {journal}
  {Phys. Lett.}\ }\textbf {\bibinfo {volume} {B793}},\ \bibinfo {pages} {126}
  (\bibinfo {year} {2019})},\ \Eprint {http://arxiv.org/abs/1807.07451}
  {arXiv:1807.07451 [hep-th]} \BibitemShut {NoStop}%
\bibitem [{\citenamefont {Agrawal}\ \emph
  {et~al.}(2019{\natexlab{b}})\citenamefont {Agrawal}, \citenamefont {Obied},\
  and\ \citenamefont {Vafa}}]{Agrawal:2019dlm}%
  \BibitemOpen
  \bibfield  {author} {\bibinfo {author} {\bibfnamefont {P.}~\bibnamefont
  {Agrawal}}, \bibinfo {author} {\bibfnamefont {G.}~\bibnamefont {Obied}}, \
  and\ \bibinfo {author} {\bibfnamefont {C.}~\bibnamefont {Vafa}},\ }\href@noop
  {} {\  (\bibinfo {year} {2019}{\natexlab{b}})},\ \Eprint
  {http://arxiv.org/abs/1906.08261} {arXiv:1906.08261 [astro-ph.CO]}
  \BibitemShut {NoStop}%
\bibitem [{\citenamefont {van~de Bruck}\ and\ \citenamefont
  {Thomas}(2019)}]{vandeBruck:2019vzd}%
  \BibitemOpen
  \bibfield  {author} {\bibinfo {author} {\bibfnamefont {C.}~\bibnamefont
  {van~de Bruck}}\ and\ \bibinfo {author} {\bibfnamefont {C.~C.}\ \bibnamefont
  {Thomas}},\ }\href {\doibase 10.1103/PhysRevD.100.023515} {\bibfield
  {journal} {\bibinfo  {journal} {Phys. Rev.}\ }\textbf {\bibinfo {volume}
  {D100}},\ \bibinfo {pages} {023515} (\bibinfo {year} {2019})},\ \Eprint
  {http://arxiv.org/abs/1904.07082} {arXiv:1904.07082 [hep-th]} \BibitemShut
  {NoStop}%
\bibitem [{\citenamefont {Casas}\ \emph {et~al.}(1992)\citenamefont {Casas},
  \citenamefont {Garcia-Bellido},\ and\ \citenamefont {Quiros}}]{Casas:1991ky}%
  \BibitemOpen
  \bibfield  {author} {\bibinfo {author} {\bibfnamefont {J.~A.}\ \bibnamefont
  {Casas}}, \bibinfo {author} {\bibfnamefont {J.}~\bibnamefont
  {Garcia-Bellido}}, \ and\ \bibinfo {author} {\bibfnamefont {M.}~\bibnamefont
  {Quiros}},\ }\href {\doibase 10.1088/0264-9381/9/5/018} {\bibfield  {journal}
  {\bibinfo  {journal} {Class. Quant. Grav.}\ }\textbf {\bibinfo {volume}
  {9}},\ \bibinfo {pages} {1371} (\bibinfo {year} {1992})},\ \Eprint
  {http://arxiv.org/abs/hep-ph/9204213} {arXiv:hep-ph/9204213 [hep-ph]}
  \BibitemShut {NoStop}%
\bibitem [{\citenamefont {Garcia-Bellido}(1993)}]{GarciaBellido:1992de}%
  \BibitemOpen
  \bibfield  {author} {\bibinfo {author} {\bibfnamefont {J.}~\bibnamefont
  {Garcia-Bellido}},\ }\bibfield  {booktitle} {\emph {\bibinfo {booktitle}
  {{15th International Conference on Neutrino Physics and Astrophysics
  (Neutrino 92) Granada, Spain, June 7-12, 1992}}},\ }\href {\doibase
  10.1142/S0218271893000076} {\bibfield  {journal} {\bibinfo  {journal} {Int.
  J. Mod. Phys.}\ }\textbf {\bibinfo {volume} {D2}},\ \bibinfo {pages} {85}
  (\bibinfo {year} {1993})},\ \Eprint {http://arxiv.org/abs/hep-ph/9205216}
  {arXiv:hep-ph/9205216 [hep-ph]} \BibitemShut {NoStop}%
\bibitem [{\citenamefont {Anderson}\ and\ \citenamefont
  {Carroll}(1997)}]{Anderson:1997un}%
  \BibitemOpen
  \bibfield  {author} {\bibinfo {author} {\bibfnamefont {G.~W.}\ \bibnamefont
  {Anderson}}\ and\ \bibinfo {author} {\bibfnamefont {S.~M.}\ \bibnamefont
  {Carroll}},\ }in\ \href {\doibase 10.1142/9789814447263_0025} {\emph
  {\bibinfo {booktitle} {{Particle physics and the early universe. Proceedings,
  1st International Workshop, COSMO-97, Ambleside, UK, September 15-19,
  1997}}}}\ (\bibinfo {year} {1997})\ pp.\ \bibinfo {pages} {227--229},\
  \Eprint {http://arxiv.org/abs/astro-ph/9711288} {arXiv:astro-ph/9711288
  [astro-ph]} \BibitemShut {NoStop}%
\bibitem [{\citenamefont {Ooguri}\ and\ \citenamefont
  {Vafa}(2007)}]{Ooguri:2006in}%
  \BibitemOpen
  \bibfield  {author} {\bibinfo {author} {\bibfnamefont {H.}~\bibnamefont
  {Ooguri}}\ and\ \bibinfo {author} {\bibfnamefont {C.}~\bibnamefont {Vafa}},\
  }\href {\doibase 10.1016/j.nuclphysb.2006.10.033} {\bibfield  {journal}
  {\bibinfo  {journal} {Nucl. Phys.}\ }\textbf {\bibinfo {volume} {B766}},\
  \bibinfo {pages} {21} (\bibinfo {year} {2007})},\ \Eprint
  {http://arxiv.org/abs/hep-th/0605264} {arXiv:hep-th/0605264 [hep-th]}
  \BibitemShut {NoStop}%
\bibitem [{\citenamefont {Grimm}\ \emph {et~al.}(2018)\citenamefont {Grimm},
  \citenamefont {Palti},\ and\ \citenamefont {Valenzuela}}]{Grimm:2018ohb}%
  \BibitemOpen
  \bibfield  {author} {\bibinfo {author} {\bibfnamefont {T.~W.}\ \bibnamefont
  {Grimm}}, \bibinfo {author} {\bibfnamefont {E.}~\bibnamefont {Palti}}, \ and\
  \bibinfo {author} {\bibfnamefont {I.}~\bibnamefont {Valenzuela}},\ }\href
  {\doibase 10.1007/JHEP08(2018)143} {\bibfield  {journal} {\bibinfo  {journal}
  {JHEP}\ }\textbf {\bibinfo {volume} {08}},\ \bibinfo {pages} {143} (\bibinfo
  {year} {2018})},\ \Eprint {http://arxiv.org/abs/1802.08264} {arXiv:1802.08264
  [hep-th]} \BibitemShut {NoStop}%
\bibitem [{\citenamefont {Heidenreich}\ \emph {et~al.}(2018)\citenamefont
  {Heidenreich}, \citenamefont {Reece},\ and\ \citenamefont
  {Rudelius}}]{Heidenreich:2018kpg}%
  \BibitemOpen
  \bibfield  {author} {\bibinfo {author} {\bibfnamefont {B.}~\bibnamefont
  {Heidenreich}}, \bibinfo {author} {\bibfnamefont {M.}~\bibnamefont {Reece}},
  \ and\ \bibinfo {author} {\bibfnamefont {T.}~\bibnamefont {Rudelius}},\
  }\href {\doibase 10.1103/PhysRevLett.121.051601} {\bibfield  {journal}
  {\bibinfo  {journal} {Phys. Rev. Lett.}\ }\textbf {\bibinfo {volume} {121}},\
  \bibinfo {pages} {051601} (\bibinfo {year} {2018})},\ \Eprint
  {http://arxiv.org/abs/1802.08698} {arXiv:1802.08698 [hep-th]} \BibitemShut
  {NoStop}%
\bibitem [{\citenamefont {Blumenhagen}(2018)}]{Blumenhagen:2018hsh}%
  \BibitemOpen
  \bibfield  {author} {\bibinfo {author} {\bibfnamefont {R.}~\bibnamefont
  {Blumenhagen}},\ }\bibfield  {booktitle} {\emph {\bibinfo {booktitle}
  {{Proceedings, 17th Hellenic School and Workshops on Elementary Particle
  Physics and Gravity (CORFU2017): Corfu, Greece, September 2-28, 2017}}},\
  }\href {\doibase 10.22323/1.318.0175} {\bibfield  {journal} {\bibinfo
  {journal} {PoS}\ }\textbf {\bibinfo {volume} {CORFU2017}},\ \bibinfo {pages}
  {175} (\bibinfo {year} {2018})},\ \Eprint {http://arxiv.org/abs/1804.10504}
  {arXiv:1804.10504 [hep-th]} \BibitemShut {NoStop}%
\bibitem [{\citenamefont {Kinney}\ \emph {et~al.}(2019)\citenamefont {Kinney},
  \citenamefont {Vagnozzi},\ and\ \citenamefont {Visinelli}}]{Kinney:2018nny}%
  \BibitemOpen
  \bibfield  {author} {\bibinfo {author} {\bibfnamefont {W.~H.}\ \bibnamefont
  {Kinney}}, \bibinfo {author} {\bibfnamefont {S.}~\bibnamefont {Vagnozzi}}, \
  and\ \bibinfo {author} {\bibfnamefont {L.}~\bibnamefont {Visinelli}},\ }\href
  {\doibase 10.1088/1361-6382/ab1d87} {\bibfield  {journal} {\bibinfo
  {journal} {Class. Quant. Grav.}\ }\textbf {\bibinfo {volume} {36}},\ \bibinfo
  {pages} {117001} (\bibinfo {year} {2019})},\ \Eprint
  {http://arxiv.org/abs/1808.06424} {arXiv:1808.06424 [astro-ph.CO]}
  \BibitemShut {NoStop}%
\bibitem [{\citenamefont {Dirac}(1937)}]{Dirac:1937ti}%
  \BibitemOpen
  \bibfield  {author} {\bibinfo {author} {\bibfnamefont {P.~A.~M.}\
  \bibnamefont {Dirac}},\ }\href {\doibase 10.1038/139323a0} {\bibfield
  {journal} {\bibinfo  {journal} {Nature}\ }\textbf {\bibinfo {volume} {139}},\
  \bibinfo {pages} {323} (\bibinfo {year} {1937})}\BibitemShut {NoStop}%
\bibitem [{\citenamefont {Dirac}(1973)}]{Dirac:1973gk}%
  \BibitemOpen
  \bibfield  {author} {\bibinfo {author} {\bibfnamefont {P.~A.~M.}\
  \bibnamefont {Dirac}},\ }\href {\doibase 10.1098/rspa.1973.0070} {\bibfield
  {journal} {\bibinfo  {journal} {Proc. Roy. Soc. Lond.}\ }\textbf {\bibinfo
  {volume} {A333}},\ \bibinfo {pages} {403} (\bibinfo {year}
  {1973})}\BibitemShut {NoStop}%
\bibitem [{\citenamefont {Webb}\ \emph {et~al.}(1999)\citenamefont {Webb},
  \citenamefont {Flambaum}, \citenamefont {Churchill}, \citenamefont
  {Drinkwater},\ and\ \citenamefont {Barrow}}]{Webb:1998cq}%
  \BibitemOpen
  \bibfield  {author} {\bibinfo {author} {\bibfnamefont {J.~K.}\ \bibnamefont
  {Webb}}, \bibinfo {author} {\bibfnamefont {V.~V.}\ \bibnamefont {Flambaum}},
  \bibinfo {author} {\bibfnamefont {C.~W.}\ \bibnamefont {Churchill}}, \bibinfo
  {author} {\bibfnamefont {M.~J.}\ \bibnamefont {Drinkwater}}, \ and\ \bibinfo
  {author} {\bibfnamefont {J.~D.}\ \bibnamefont {Barrow}},\ }\href {\doibase
  10.1103/PhysRevLett.82.884} {\bibfield  {journal} {\bibinfo  {journal} {Phys.
  Rev. Lett.}\ }\textbf {\bibinfo {volume} {82}},\ \bibinfo {pages} {884}
  (\bibinfo {year} {1999})},\ \Eprint {http://arxiv.org/abs/astro-ph/9803165}
  {arXiv:astro-ph/9803165 [astro-ph]} \BibitemShut {NoStop}%
\bibitem [{\citenamefont {Webb}\ \emph {et~al.}(2001)\citenamefont {Webb},
  \citenamefont {Murphy}, \citenamefont {Flambaum}, \citenamefont {Dzuba},
  \citenamefont {Barrow}, \citenamefont {Churchill}, \citenamefont
  {Prochaska},\ and\ \citenamefont {Wolfe}}]{Webb:2000mn}%
  \BibitemOpen
  \bibfield  {author} {\bibinfo {author} {\bibfnamefont {J.~K.}\ \bibnamefont
  {Webb}}, \bibinfo {author} {\bibfnamefont {M.~T.}\ \bibnamefont {Murphy}},
  \bibinfo {author} {\bibfnamefont {V.~V.}\ \bibnamefont {Flambaum}}, \bibinfo
  {author} {\bibfnamefont {V.~A.}\ \bibnamefont {Dzuba}}, \bibinfo {author}
  {\bibfnamefont {J.~D.}\ \bibnamefont {Barrow}}, \bibinfo {author}
  {\bibfnamefont {C.~W.}\ \bibnamefont {Churchill}}, \bibinfo {author}
  {\bibfnamefont {J.~X.}\ \bibnamefont {Prochaska}}, \ and\ \bibinfo {author}
  {\bibfnamefont {A.~M.}\ \bibnamefont {Wolfe}},\ }\href {\doibase
  10.1103/PhysRevLett.87.091301} {\bibfield  {journal} {\bibinfo  {journal}
  {Phys. Rev. Lett.}\ }\textbf {\bibinfo {volume} {87}},\ \bibinfo {pages}
  {091301} (\bibinfo {year} {2001})},\ \Eprint
  {http://arxiv.org/abs/astro-ph/0012539} {arXiv:astro-ph/0012539 [astro-ph]}
  \BibitemShut {NoStop}%
\bibitem [{\citenamefont {Banks}\ \emph {et~al.}(2002)\citenamefont {Banks},
  \citenamefont {Dine},\ and\ \citenamefont {Douglas}}]{Banks:2001qc}%
  \BibitemOpen
  \bibfield  {author} {\bibinfo {author} {\bibfnamefont {T.}~\bibnamefont
  {Banks}}, \bibinfo {author} {\bibfnamefont {M.}~\bibnamefont {Dine}}, \ and\
  \bibinfo {author} {\bibfnamefont {M.~R.}\ \bibnamefont {Douglas}},\ }\href
  {\doibase 10.1103/PhysRevLett.88.131301} {\bibfield  {journal} {\bibinfo
  {journal} {Phys. Rev. Lett.}\ }\textbf {\bibinfo {volume} {88}},\ \bibinfo
  {pages} {131301} (\bibinfo {year} {2002})},\ \Eprint
  {http://arxiv.org/abs/hep-ph/0112059} {arXiv:hep-ph/0112059 [hep-ph]}
  \BibitemShut {NoStop}%
\bibitem [{\citenamefont {Chacko}\ \emph {et~al.}(2003)\citenamefont {Chacko},
  \citenamefont {Grojean},\ and\ \citenamefont {Perelstein}}]{Chacko:2002mf}%
  \BibitemOpen
  \bibfield  {author} {\bibinfo {author} {\bibfnamefont {Z.}~\bibnamefont
  {Chacko}}, \bibinfo {author} {\bibfnamefont {C.}~\bibnamefont {Grojean}}, \
  and\ \bibinfo {author} {\bibfnamefont {M.}~\bibnamefont {Perelstein}},\
  }\href {\doibase 10.1016/S0370-2693(03)00766-4} {\bibfield  {journal}
  {\bibinfo  {journal} {Phys. Lett.}\ }\textbf {\bibinfo {volume} {B565}},\
  \bibinfo {pages} {169} (\bibinfo {year} {2003})},\ \Eprint
  {http://arxiv.org/abs/hep-ph/0204142} {arXiv:hep-ph/0204142 [hep-ph]}
  \BibitemShut {NoStop}%
\bibitem [{\citenamefont {Coudarchet}\ \emph {et~al.}(2019)\citenamefont
  {Coudarchet}, \citenamefont {Heurtier},\ and\ \citenamefont
  {Partouche}}]{Coudarchet:2018ezq}%
  \BibitemOpen
  \bibfield  {author} {\bibinfo {author} {\bibfnamefont {T.}~\bibnamefont
  {Coudarchet}}, \bibinfo {author} {\bibfnamefont {L.}~\bibnamefont
  {Heurtier}}, \ and\ \bibinfo {author} {\bibfnamefont {H.}~\bibnamefont
  {Partouche}},\ }\href {\doibase 10.1007/JHEP03(2019)117} {\bibfield
  {journal} {\bibinfo  {journal} {JHEP}\ }\textbf {\bibinfo {volume} {03}},\
  \bibinfo {pages} {117} (\bibinfo {year} {2019})},\ \Eprint
  {http://arxiv.org/abs/1812.10134} {arXiv:1812.10134 [hep-th]} \BibitemShut
  {NoStop}%
\bibitem [{\citenamefont {Fardon}\ \emph {et~al.}(2004)\citenamefont {Fardon},
  \citenamefont {Nelson},\ and\ \citenamefont {Weiner}}]{Fardon:2003eh}%
  \BibitemOpen
  \bibfield  {author} {\bibinfo {author} {\bibfnamefont {R.}~\bibnamefont
  {Fardon}}, \bibinfo {author} {\bibfnamefont {A.~E.}\ \bibnamefont {Nelson}},
  \ and\ \bibinfo {author} {\bibfnamefont {N.}~\bibnamefont {Weiner}},\ }\href
  {\doibase 10.1088/1475-7516/2004/10/005} {\bibfield  {journal} {\bibinfo
  {journal} {JCAP}\ }\textbf {\bibinfo {volume} {0410}},\ \bibinfo {pages}
  {005} (\bibinfo {year} {2004})},\ \Eprint
  {http://arxiv.org/abs/astro-ph/0309800} {arXiv:astro-ph/0309800 [astro-ph]}
  \BibitemShut {NoStop}%
\bibitem [{\citenamefont {Kaplan}\ \emph {et~al.}(2004)\citenamefont {Kaplan},
  \citenamefont {Nelson},\ and\ \citenamefont {Weiner}}]{Kaplan:2004dq}%
  \BibitemOpen
  \bibfield  {author} {\bibinfo {author} {\bibfnamefont {D.~B.}\ \bibnamefont
  {Kaplan}}, \bibinfo {author} {\bibfnamefont {A.~E.}\ \bibnamefont {Nelson}},
  \ and\ \bibinfo {author} {\bibfnamefont {N.}~\bibnamefont {Weiner}},\ }\href
  {\doibase 10.1103/PhysRevLett.93.091801} {\bibfield  {journal} {\bibinfo
  {journal} {Phys. Rev. Lett.}\ }\textbf {\bibinfo {volume} {93}},\ \bibinfo
  {pages} {091801} (\bibinfo {year} {2004})},\ \Eprint
  {http://arxiv.org/abs/hep-ph/0401099} {arXiv:hep-ph/0401099 [hep-ph]}
  \BibitemShut {NoStop}%
\bibitem [{\citenamefont {Spergel}\ and\ \citenamefont
  {Steinhardt}(2000)}]{Spergel:1999mh}%
  \BibitemOpen
  \bibfield  {author} {\bibinfo {author} {\bibfnamefont {D.~N.}\ \bibnamefont
  {Spergel}}\ and\ \bibinfo {author} {\bibfnamefont {P.~J.}\ \bibnamefont
  {Steinhardt}},\ }\href {\doibase 10.1103/PhysRevLett.84.3760} {\bibfield
  {journal} {\bibinfo  {journal} {Phys. Rev. Lett.}\ }\textbf {\bibinfo
  {volume} {84}},\ \bibinfo {pages} {3760} (\bibinfo {year} {2000})},\ \Eprint
  {http://arxiv.org/abs/astro-ph/9909386} {arXiv:astro-ph/9909386 [astro-ph]}
  \BibitemShut {NoStop}%
\bibitem [{\citenamefont {Tulin}\ and\ \citenamefont
  {Yu}(2018)}]{Tulin:2017ara}%
  \BibitemOpen
  \bibfield  {author} {\bibinfo {author} {\bibfnamefont {S.}~\bibnamefont
  {Tulin}}\ and\ \bibinfo {author} {\bibfnamefont {H.-B.}\ \bibnamefont {Yu}},\
  }\href {\doibase 10.1016/j.physrep.2017.11.004} {\bibfield  {journal}
  {\bibinfo  {journal} {Phys. Rept.}\ }\textbf {\bibinfo {volume} {730}},\
  \bibinfo {pages} {1} (\bibinfo {year} {2018})},\ \Eprint
  {http://arxiv.org/abs/1705.02358} {arXiv:1705.02358 [hep-ph]} \BibitemShut
  {NoStop}%
\bibitem [{\citenamefont {Banerjee}\ \emph {et~al.}(2019)\citenamefont
  {Banerjee}, \citenamefont {Bhattacharyya}, \citenamefont {Chowdhury},\ and\
  \citenamefont {Mambrini}}]{Banerjee:2019asa}%
  \BibitemOpen
  \bibfield  {author} {\bibinfo {author} {\bibfnamefont {A.}~\bibnamefont
  {Banerjee}}, \bibinfo {author} {\bibfnamefont {G.}~\bibnamefont
  {Bhattacharyya}}, \bibinfo {author} {\bibfnamefont {D.}~\bibnamefont
  {Chowdhury}}, \ and\ \bibinfo {author} {\bibfnamefont {Y.}~\bibnamefont
  {Mambrini}},\ }\href@noop {} {\  (\bibinfo {year} {2019})},\ \Eprint
  {http://arxiv.org/abs/1905.11407} {arXiv:1905.11407 [hep-ph]} \BibitemShut
  {NoStop}%
\bibitem [{\citenamefont {Denef}\ \emph {et~al.}(2018)\citenamefont {Denef},
  \citenamefont {Hebecker},\ and\ \citenamefont {Wrase}}]{Denef:2018etk}%
  \BibitemOpen
  \bibfield  {author} {\bibinfo {author} {\bibfnamefont {F.}~\bibnamefont
  {Denef}}, \bibinfo {author} {\bibfnamefont {A.}~\bibnamefont {Hebecker}}, \
  and\ \bibinfo {author} {\bibfnamefont {T.}~\bibnamefont {Wrase}},\ }\href
  {\doibase 10.1103/PhysRevD.98.086004} {\bibfield  {journal} {\bibinfo
  {journal} {Phys. Rev.}\ }\textbf {\bibinfo {volume} {D98}},\ \bibinfo {pages}
  {086004} (\bibinfo {year} {2018})},\ \Eprint
  {http://arxiv.org/abs/1807.06581} {arXiv:1807.06581 [hep-th]} \BibitemShut
  {NoStop}%
\bibitem [{\citenamefont {Murayama}\ \emph {et~al.}(2018)\citenamefont
  {Murayama}, \citenamefont {Yamazaki},\ and\ \citenamefont
  {Yanagida}}]{Murayama:2018lie}%
  \BibitemOpen
  \bibfield  {author} {\bibinfo {author} {\bibfnamefont {H.}~\bibnamefont
  {Murayama}}, \bibinfo {author} {\bibfnamefont {M.}~\bibnamefont {Yamazaki}},
  \ and\ \bibinfo {author} {\bibfnamefont {T.~T.}\ \bibnamefont {Yanagida}},\
  }\href {\doibase 10.1007/JHEP12(2018)032} {\bibfield  {journal} {\bibinfo
  {journal} {JHEP}\ }\textbf {\bibinfo {volume} {12}},\ \bibinfo {pages} {032}
  (\bibinfo {year} {2018})},\ \Eprint {http://arxiv.org/abs/1809.00478}
  {arXiv:1809.00478 [hep-th]} \BibitemShut {NoStop}%
\bibitem [{\citenamefont {Choi}\ \emph {et~al.}(2018)\citenamefont {Choi},
  \citenamefont {Chway},\ and\ \citenamefont {Shin}}]{Choi:2018rze}%
  \BibitemOpen
  \bibfield  {author} {\bibinfo {author} {\bibfnamefont {K.}~\bibnamefont
  {Choi}}, \bibinfo {author} {\bibfnamefont {D.}~\bibnamefont {Chway}}, \ and\
  \bibinfo {author} {\bibfnamefont {C.~S.}\ \bibnamefont {Shin}},\ }\href
  {\doibase 10.1007/JHEP11(2018)142} {\bibfield  {journal} {\bibinfo  {journal}
  {JHEP}\ }\textbf {\bibinfo {volume} {11}},\ \bibinfo {pages} {142} (\bibinfo
  {year} {2018})},\ \Eprint {http://arxiv.org/abs/1809.01475} {arXiv:1809.01475
  [hep-th]} \BibitemShut {NoStop}%
\bibitem [{\citenamefont {Kobakhidze}(2019)}]{Kobakhidze:2019ppv}%
  \BibitemOpen
  \bibfield  {author} {\bibinfo {author} {\bibfnamefont {A.}~\bibnamefont
  {Kobakhidze}},\ }\href@noop {} {\  (\bibinfo {year} {2019})},\ \Eprint
  {http://arxiv.org/abs/1901.08137} {arXiv:1901.08137 [physics.gen-ph]}
  \BibitemShut {NoStop}%
\bibitem [{\citenamefont {Park}(2019)}]{Park:2019odd}%
  \BibitemOpen
  \bibfield  {author} {\bibinfo {author} {\bibfnamefont {J.-H.}\ \bibnamefont
  {Park}},\ }\href {\doibase 10.1103/PhysRevD.99.116013} {\bibfield  {journal}
  {\bibinfo  {journal} {Phys. Rev.}\ }\textbf {\bibinfo {volume} {D99}},\
  \bibinfo {pages} {116013} (\bibinfo {year} {2019})},\ \Eprint
  {http://arxiv.org/abs/1902.04559} {arXiv:1902.04559 [hep-ph]} \BibitemShut
  {NoStop}%
\bibitem [{\citenamefont {Kachru}\ \emph {et~al.}(2003)\citenamefont {Kachru},
  \citenamefont {Kallosh}, \citenamefont {Linde},\ and\ \citenamefont
  {Trivedi}}]{Kachru:2003aw}%
  \BibitemOpen
  \bibfield  {author} {\bibinfo {author} {\bibfnamefont {S.}~\bibnamefont
  {Kachru}}, \bibinfo {author} {\bibfnamefont {R.}~\bibnamefont {Kallosh}},
  \bibinfo {author} {\bibfnamefont {A.~D.}\ \bibnamefont {Linde}}, \ and\
  \bibinfo {author} {\bibfnamefont {S.~P.}\ \bibnamefont {Trivedi}},\ }\href
  {\doibase 10.1103/PhysRevD.68.046005} {\bibfield  {journal} {\bibinfo
  {journal} {Phys. Rev.}\ }\textbf {\bibinfo {volume} {D68}},\ \bibinfo {pages}
  {046005} (\bibinfo {year} {2003})},\ \Eprint
  {http://arxiv.org/abs/hep-th/0301240} {arXiv:hep-th/0301240 [hep-th]}
  \BibitemShut {NoStop}%
\bibitem [{\citenamefont {Danielsson}\ and\ \citenamefont
  {Van~Riet}(2018)}]{Danielsson:2018ztv}%
  \BibitemOpen
  \bibfield  {author} {\bibinfo {author} {\bibfnamefont {U.~H.}\ \bibnamefont
  {Danielsson}}\ and\ \bibinfo {author} {\bibfnamefont {T.}~\bibnamefont
  {Van~Riet}},\ }\href {\doibase 10.1142/S0218271818300070} {\bibfield
  {journal} {\bibinfo  {journal} {Int. J. Mod. Phys.}\ }\textbf {\bibinfo
  {volume} {D27}},\ \bibinfo {pages} {1830007} (\bibinfo {year} {2018})},\
  \Eprint {http://arxiv.org/abs/1804.01120} {arXiv:1804.01120 [hep-th]}
  \BibitemShut {NoStop}%
\bibitem [{\citenamefont {Perlmutter}\ \emph {et~al.}(1999)\citenamefont
  {Perlmutter} \emph {et~al.}}]{Perlmutter:1998np}%
  \BibitemOpen
  \bibfield  {author} {\bibinfo {author} {\bibfnamefont {S.}~\bibnamefont
  {Perlmutter}} \emph {et~al.} (\bibinfo {collaboration} {Supernova Cosmology
  Project}),\ }\href {\doibase 10.1086/307221} {\bibfield  {journal} {\bibinfo
  {journal} {Astrophys. J.}\ }\textbf {\bibinfo {volume} {517}},\ \bibinfo
  {pages} {565} (\bibinfo {year} {1999})},\ \Eprint
  {http://arxiv.org/abs/astro-ph/9812133} {arXiv:astro-ph/9812133 [astro-ph]}
  \BibitemShut {NoStop}%
\bibitem [{\citenamefont {Riess}\ \emph {et~al.}(1998)\citenamefont {Riess}
  \emph {et~al.}}]{Riess:1998cb}%
  \BibitemOpen
  \bibfield  {author} {\bibinfo {author} {\bibfnamefont {A.~G.}\ \bibnamefont
  {Riess}} \emph {et~al.} (\bibinfo {collaboration} {Supernova Search Team}),\
  }\href {\doibase 10.1086/300499} {\bibfield  {journal} {\bibinfo  {journal}
  {Astron. J.}\ }\textbf {\bibinfo {volume} {116}},\ \bibinfo {pages} {1009}
  (\bibinfo {year} {1998})},\ \Eprint {http://arxiv.org/abs/astro-ph/9805201}
  {arXiv:astro-ph/9805201 [astro-ph]} \BibitemShut {NoStop}%
\bibitem [{\citenamefont {Englert}\ and\ \citenamefont
  {Brout}(1964)}]{Englert:1964et}%
  \BibitemOpen
  \bibfield  {author} {\bibinfo {author} {\bibfnamefont {F.}~\bibnamefont
  {Englert}}\ and\ \bibinfo {author} {\bibfnamefont {R.}~\bibnamefont
  {Brout}},\ }\href {\doibase 10.1103/PhysRevLett.13.321} {\bibfield  {journal}
  {\bibinfo  {journal} {Phys. Rev. Lett.}\ }\textbf {\bibinfo {volume} {13}},\
  \bibinfo {pages} {321} (\bibinfo {year} {1964})},\ \bibinfo {note}
  {[,157(1964)]}\BibitemShut {NoStop}%
\bibitem [{\citenamefont {Higgs}(1964)}]{Higgs:1964pj}%
  \BibitemOpen
  \bibfield  {author} {\bibinfo {author} {\bibfnamefont {P.~W.}\ \bibnamefont
  {Higgs}},\ }\href {\doibase 10.1103/PhysRevLett.13.508} {\bibfield  {journal}
  {\bibinfo  {journal} {Phys. Rev. Lett.}\ }\textbf {\bibinfo {volume} {13}},\
  \bibinfo {pages} {508} (\bibinfo {year} {1964})},\ \bibinfo {note}
  {[,160(1964)]}\BibitemShut {NoStop}%
\bibitem [{\citenamefont {Aad}\ \emph {et~al.}(2015)\citenamefont {Aad} \emph
  {et~al.}}]{Aad:2015zhl}%
  \BibitemOpen
  \bibfield  {author} {\bibinfo {author} {\bibfnamefont {G.}~\bibnamefont
  {Aad}} \emph {et~al.} (\bibinfo {collaboration} {ATLAS, CMS}),\ }\href
  {\doibase 10.1103/PhysRevLett.114.191803} {\bibfield  {journal} {\bibinfo
  {journal} {Phys. Rev. Lett.}\ }\textbf {\bibinfo {volume} {114}},\ \bibinfo
  {pages} {191803} (\bibinfo {year} {2015})},\ \Eprint
  {http://arxiv.org/abs/1503.07589} {arXiv:1503.07589 [hep-ex]} \BibitemShut
  {NoStop}%
\bibitem [{\citenamefont {Ratra}\ and\ \citenamefont
  {Peebles}(1988)}]{Ratra:1987rm}%
  \BibitemOpen
  \bibfield  {author} {\bibinfo {author} {\bibfnamefont {B.}~\bibnamefont
  {Ratra}}\ and\ \bibinfo {author} {\bibfnamefont {P.~J.~E.}\ \bibnamefont
  {Peebles}},\ }\href {\doibase 10.1103/PhysRevD.37.3406} {\bibfield  {journal}
  {\bibinfo  {journal} {Phys. Rev.}\ }\textbf {\bibinfo {volume} {D37}},\
  \bibinfo {pages} {3406} (\bibinfo {year} {1988})}\BibitemShut {NoStop}%
\bibitem [{\citenamefont {Ferreira}\ and\ \citenamefont
  {Joyce}(1997)}]{Ferreira:1997au}%
  \BibitemOpen
  \bibfield  {author} {\bibinfo {author} {\bibfnamefont {P.~G.}\ \bibnamefont
  {Ferreira}}\ and\ \bibinfo {author} {\bibfnamefont {M.}~\bibnamefont
  {Joyce}},\ }\href {\doibase 10.1103/PhysRevLett.79.4740} {\bibfield
  {journal} {\bibinfo  {journal} {Phys. Rev. Lett.}\ }\textbf {\bibinfo
  {volume} {79}},\ \bibinfo {pages} {4740} (\bibinfo {year} {1997})},\ \Eprint
  {http://arxiv.org/abs/astro-ph/9707286} {arXiv:astro-ph/9707286 [astro-ph]}
  \BibitemShut {NoStop}%
\bibitem [{\citenamefont {Ferreira}\ and\ \citenamefont
  {Joyce}(1998)}]{Ferreira:1997hj}%
  \BibitemOpen
  \bibfield  {author} {\bibinfo {author} {\bibfnamefont {P.~G.}\ \bibnamefont
  {Ferreira}}\ and\ \bibinfo {author} {\bibfnamefont {M.}~\bibnamefont
  {Joyce}},\ }\href {\doibase 10.1103/PhysRevD.58.023503} {\bibfield  {journal}
  {\bibinfo  {journal} {Phys. Rev.}\ }\textbf {\bibinfo {volume} {D58}},\
  \bibinfo {pages} {023503} (\bibinfo {year} {1998})},\ \Eprint
  {http://arxiv.org/abs/astro-ph/9711102} {arXiv:astro-ph/9711102 [astro-ph]}
  \BibitemShut {NoStop}%
\bibitem [{\citenamefont {Copeland}\ \emph {et~al.}(1998)\citenamefont
  {Copeland}, \citenamefont {Liddle},\ and\ \citenamefont
  {Wands}}]{Copeland:1997et}%
  \BibitemOpen
  \bibfield  {author} {\bibinfo {author} {\bibfnamefont {E.~J.}\ \bibnamefont
  {Copeland}}, \bibinfo {author} {\bibfnamefont {A.~R.}\ \bibnamefont
  {Liddle}}, \ and\ \bibinfo {author} {\bibfnamefont {D.}~\bibnamefont
  {Wands}},\ }\href {\doibase 10.1103/PhysRevD.57.4686} {\bibfield  {journal}
  {\bibinfo  {journal} {Phys. Rev.}\ }\textbf {\bibinfo {volume} {D57}},\
  \bibinfo {pages} {4686} (\bibinfo {year} {1998})},\ \Eprint
  {http://arxiv.org/abs/gr-qc/9711068} {arXiv:gr-qc/9711068 [gr-qc]}
  \BibitemShut {NoStop}%
\bibitem [{\citenamefont {Tsujikawa}(2013)}]{Tsujikawa:2013fta}%
  \BibitemOpen
  \bibfield  {author} {\bibinfo {author} {\bibfnamefont {S.}~\bibnamefont
  {Tsujikawa}},\ }\href {\doibase 10.1088/0264-9381/30/21/214003} {\bibfield
  {journal} {\bibinfo  {journal} {Class. Quant. Grav.}\ }\textbf {\bibinfo
  {volume} {30}},\ \bibinfo {pages} {214003} (\bibinfo {year} {2013})},\
  \Eprint {http://arxiv.org/abs/1304.1961} {arXiv:1304.1961 [gr-qc]}
  \BibitemShut {NoStop}%
\bibitem [{\citenamefont {Buttazzo}\ \emph {et~al.}(2013)\citenamefont
  {Buttazzo}, \citenamefont {Degrassi}, \citenamefont {Giardino}, \citenamefont
  {Giudice}, \citenamefont {Sala}, \citenamefont {Salvio},\ and\ \citenamefont
  {Strumia}}]{Buttazzo:2013uya}%
  \BibitemOpen
  \bibfield  {author} {\bibinfo {author} {\bibfnamefont {D.}~\bibnamefont
  {Buttazzo}}, \bibinfo {author} {\bibfnamefont {G.}~\bibnamefont {Degrassi}},
  \bibinfo {author} {\bibfnamefont {P.~P.}\ \bibnamefont {Giardino}}, \bibinfo
  {author} {\bibfnamefont {G.~F.}\ \bibnamefont {Giudice}}, \bibinfo {author}
  {\bibfnamefont {F.}~\bibnamefont {Sala}}, \bibinfo {author} {\bibfnamefont
  {A.}~\bibnamefont {Salvio}}, \ and\ \bibinfo {author} {\bibfnamefont
  {A.}~\bibnamefont {Strumia}},\ }\href {\doibase 10.1007/JHEP12(2013)089}
  {\bibfield  {journal} {\bibinfo  {journal} {JHEP}\ }\textbf {\bibinfo
  {volume} {12}},\ \bibinfo {pages} {089} (\bibinfo {year} {2013})},\ \Eprint
  {http://arxiv.org/abs/1307.3536} {arXiv:1307.3536 [hep-ph]} \BibitemShut
  {NoStop}%
\bibitem [{\citenamefont {Barreiro}\ \emph {et~al.}(2000)\citenamefont
  {Barreiro}, \citenamefont {Copeland},\ and\ \citenamefont
  {Nunes}}]{Barreiro:1999zs}%
  \BibitemOpen
  \bibfield  {author} {\bibinfo {author} {\bibfnamefont {T.}~\bibnamefont
  {Barreiro}}, \bibinfo {author} {\bibfnamefont {E.~J.}\ \bibnamefont
  {Copeland}}, \ and\ \bibinfo {author} {\bibfnamefont {N.~J.}\ \bibnamefont
  {Nunes}},\ }\href {\doibase 10.1103/PhysRevD.61.127301} {\bibfield  {journal}
  {\bibinfo  {journal} {Phys. Rev.}\ }\textbf {\bibinfo {volume} {D61}},\
  \bibinfo {pages} {127301} (\bibinfo {year} {2000})},\ \Eprint
  {http://arxiv.org/abs/astro-ph/9910214} {arXiv:astro-ph/9910214 [astro-ph]}
  \BibitemShut {NoStop}%
\bibitem [{\citenamefont {Chiba}\ \emph {et~al.}(2013)\citenamefont {Chiba},
  \citenamefont {De~Felice},\ and\ \citenamefont {Tsujikawa}}]{Chiba:2012cb}%
  \BibitemOpen
  \bibfield  {author} {\bibinfo {author} {\bibfnamefont {T.}~\bibnamefont
  {Chiba}}, \bibinfo {author} {\bibfnamefont {A.}~\bibnamefont {De~Felice}}, \
  and\ \bibinfo {author} {\bibfnamefont {S.}~\bibnamefont {Tsujikawa}},\ }\href
  {\doibase 10.1103/PhysRevD.87.083505} {\bibfield  {journal} {\bibinfo
  {journal} {Phys. Rev.}\ }\textbf {\bibinfo {volume} {D87}},\ \bibinfo {pages}
  {083505} (\bibinfo {year} {2013})},\ \Eprint {http://arxiv.org/abs/1210.3859}
  {arXiv:1210.3859 [astro-ph.CO]} \BibitemShut {NoStop}%
\bibitem [{\citenamefont {David~Marsh}(2019)}]{Marsh:2018kub}%
  \BibitemOpen
  \bibfield  {author} {\bibinfo {author} {\bibfnamefont {M.~C.}\ \bibnamefont
  {David~Marsh}},\ }\href {\doibase 10.1016/j.physletb.2018.11.001} {\bibfield
  {journal} {\bibinfo  {journal} {Phys. Lett.}\ }\textbf {\bibinfo {volume}
  {B789}},\ \bibinfo {pages} {639} (\bibinfo {year} {2019})},\ \Eprint
  {http://arxiv.org/abs/1809.00726} {arXiv:1809.00726 [hep-th]} \BibitemShut
  {NoStop}%
\bibitem [{\citenamefont {D'Amico}\ \emph {et~al.}(2016)\citenamefont
  {D'Amico}, \citenamefont {Hamill},\ and\ \citenamefont
  {Kaloper}}]{DAmico:2016jbm}%
  \BibitemOpen
  \bibfield  {author} {\bibinfo {author} {\bibfnamefont {G.}~\bibnamefont
  {D'Amico}}, \bibinfo {author} {\bibfnamefont {T.}~\bibnamefont {Hamill}}, \
  and\ \bibinfo {author} {\bibfnamefont {N.}~\bibnamefont {Kaloper}},\ }\href
  {\doibase 10.1103/PhysRevD.94.103526} {\bibfield  {journal} {\bibinfo
  {journal} {Phys. Rev.}\ }\textbf {\bibinfo {volume} {D94}},\ \bibinfo {pages}
  {103526} (\bibinfo {year} {2016})},\ \Eprint
  {http://arxiv.org/abs/1605.00996} {arXiv:1605.00996 [hep-th]} \BibitemShut
  {NoStop}%
\bibitem [{\citenamefont {Matsui}\ \emph {et~al.}(2019)\citenamefont {Matsui},
  \citenamefont {Takahashi},\ and\ \citenamefont {Yamada}}]{Matsui:2018xwa}%
  \BibitemOpen
  \bibfield  {author} {\bibinfo {author} {\bibfnamefont {H.}~\bibnamefont
  {Matsui}}, \bibinfo {author} {\bibfnamefont {F.}~\bibnamefont {Takahashi}}, \
  and\ \bibinfo {author} {\bibfnamefont {M.}~\bibnamefont {Yamada}},\ }\href
  {\doibase 10.1016/j.physletb.2018.12.055} {\bibfield  {journal} {\bibinfo
  {journal} {Phys. Lett.}\ }\textbf {\bibinfo {volume} {B789}},\ \bibinfo
  {pages} {387} (\bibinfo {year} {2019})},\ \Eprint
  {http://arxiv.org/abs/1809.07286} {arXiv:1809.07286 [astro-ph.CO]}
  \BibitemShut {NoStop}%
\bibitem [{\citenamefont {Buchmuller}\ \emph {et~al.}(2017)\citenamefont
  {Buchmuller}, \citenamefont {Dierigl}, \citenamefont {Dudas},\ and\
  \citenamefont {Schweizer}}]{Buchmuller:2016gib}%
  \BibitemOpen
  \bibfield  {author} {\bibinfo {author} {\bibfnamefont {W.}~\bibnamefont
  {Buchmuller}}, \bibinfo {author} {\bibfnamefont {M.}~\bibnamefont {Dierigl}},
  \bibinfo {author} {\bibfnamefont {E.}~\bibnamefont {Dudas}}, \ and\ \bibinfo
  {author} {\bibfnamefont {J.}~\bibnamefont {Schweizer}},\ }\href {\doibase
  10.1007/JHEP04(2017)052} {\bibfield  {journal} {\bibinfo  {journal} {JHEP}\
  }\textbf {\bibinfo {volume} {04}},\ \bibinfo {pages} {052} (\bibinfo {year}
  {2017})},\ \Eprint {http://arxiv.org/abs/1611.03798} {arXiv:1611.03798
  [hep-th]} \BibitemShut {NoStop}%
\bibitem [{\citenamefont {Buchmuller}\ \emph {et~al.}(2018)\citenamefont
  {Buchmuller}, \citenamefont {Dierigl},\ and\ \citenamefont
  {Dudas}}]{Buchmuller:2018eog}%
  \BibitemOpen
  \bibfield  {author} {\bibinfo {author} {\bibfnamefont {W.}~\bibnamefont
  {Buchmuller}}, \bibinfo {author} {\bibfnamefont {M.}~\bibnamefont {Dierigl}},
  \ and\ \bibinfo {author} {\bibfnamefont {E.}~\bibnamefont {Dudas}},\ }\href
  {\doibase 10.1007/JHEP08(2018)151} {\bibfield  {journal} {\bibinfo  {journal}
  {JHEP}\ }\textbf {\bibinfo {volume} {08}},\ \bibinfo {pages} {151} (\bibinfo
  {year} {2018})},\ \Eprint {http://arxiv.org/abs/1804.07497} {arXiv:1804.07497
  [hep-th]} \BibitemShut {NoStop}%
\bibitem [{\citenamefont {Harvey}\ \emph {et~al.}(2015)\citenamefont {Harvey},
  \citenamefont {Massey}, \citenamefont {Kitching}, \citenamefont {Taylor},\
  and\ \citenamefont {Tittley}}]{Harvey:2015hha}%
  \BibitemOpen
  \bibfield  {author} {\bibinfo {author} {\bibfnamefont {D.}~\bibnamefont
  {Harvey}}, \bibinfo {author} {\bibfnamefont {R.}~\bibnamefont {Massey}},
  \bibinfo {author} {\bibfnamefont {T.}~\bibnamefont {Kitching}}, \bibinfo
  {author} {\bibfnamefont {A.}~\bibnamefont {Taylor}}, \ and\ \bibinfo {author}
  {\bibfnamefont {E.}~\bibnamefont {Tittley}},\ }\href {\doibase
  10.1126/science.1261381} {\bibfield  {journal} {\bibinfo  {journal}
  {Science}\ }\textbf {\bibinfo {volume} {347}},\ \bibinfo {pages} {1462}
  (\bibinfo {year} {2015})},\ \Eprint {http://arxiv.org/abs/1503.07675}
  {arXiv:1503.07675 [astro-ph.CO]} \BibitemShut {NoStop}%
\bibitem [{\citenamefont {Andrade}\ \emph {et~al.}(2019)\citenamefont
  {Andrade}, \citenamefont {Minor}, \citenamefont {Nierenberg},\ and\
  \citenamefont {Kaplinghat}}]{Andrade:2019wzn}%
  \BibitemOpen
  \bibfield  {author} {\bibinfo {author} {\bibfnamefont {K.~E.}\ \bibnamefont
  {Andrade}}, \bibinfo {author} {\bibfnamefont {Q.}~\bibnamefont {Minor}},
  \bibinfo {author} {\bibfnamefont {A.}~\bibnamefont {Nierenberg}}, \ and\
  \bibinfo {author} {\bibfnamefont {M.}~\bibnamefont {Kaplinghat}},\
  }\href@noop {} {\  (\bibinfo {year} {2019})},\ \Eprint
  {http://arxiv.org/abs/1901.00507} {arXiv:1901.00507 [astro-ph.GA]}
  \BibitemShut {NoStop}%
\bibitem [{\citenamefont {Kaplinghat}\ \emph {et~al.}(2016)\citenamefont
  {Kaplinghat}, \citenamefont {Tulin},\ and\ \citenamefont
  {Yu}}]{Kaplinghat:2015aga}%
  \BibitemOpen
  \bibfield  {author} {\bibinfo {author} {\bibfnamefont {M.}~\bibnamefont
  {Kaplinghat}}, \bibinfo {author} {\bibfnamefont {S.}~\bibnamefont {Tulin}}, \
  and\ \bibinfo {author} {\bibfnamefont {H.-B.}\ \bibnamefont {Yu}},\ }\href
  {\doibase 10.1103/PhysRevLett.116.041302} {\bibfield  {journal} {\bibinfo
  {journal} {Phys. Rev. Lett.}\ }\textbf {\bibinfo {volume} {116}},\ \bibinfo
  {pages} {041302} (\bibinfo {year} {2016})},\ \Eprint
  {http://arxiv.org/abs/1508.03339} {arXiv:1508.03339 [astro-ph.CO]}
  \BibitemShut {NoStop}%
\bibitem [{\citenamefont {Randall}\ \emph {et~al.}(2008)\citenamefont
  {Randall}, \citenamefont {Markevitch}, \citenamefont {Clowe}, \citenamefont
  {Gonzalez},\ and\ \citenamefont {Bradac}}]{Randall:2007ph}%
  \BibitemOpen
  \bibfield  {author} {\bibinfo {author} {\bibfnamefont {S.~W.}\ \bibnamefont
  {Randall}}, \bibinfo {author} {\bibfnamefont {M.}~\bibnamefont {Markevitch}},
  \bibinfo {author} {\bibfnamefont {D.}~\bibnamefont {Clowe}}, \bibinfo
  {author} {\bibfnamefont {A.~H.}\ \bibnamefont {Gonzalez}}, \ and\ \bibinfo
  {author} {\bibfnamefont {M.}~\bibnamefont {Bradac}},\ }\href {\doibase
  10.1086/587859} {\bibfield  {journal} {\bibinfo  {journal} {Astrophys. J.}\
  }\textbf {\bibinfo {volume} {679}},\ \bibinfo {pages} {1173} (\bibinfo {year}
  {2008})},\ \Eprint {http://arxiv.org/abs/0704.0261} {arXiv:0704.0261
  [astro-ph]} \BibitemShut {NoStop}%
\bibitem [{\citenamefont {Wittman}\ \emph {et~al.}(2018)\citenamefont
  {Wittman}, \citenamefont {Golovich},\ and\ \citenamefont
  {Dawson}}]{Wittman:2017gxn}%
  \BibitemOpen
  \bibfield  {author} {\bibinfo {author} {\bibfnamefont {D.}~\bibnamefont
  {Wittman}}, \bibinfo {author} {\bibfnamefont {N.}~\bibnamefont {Golovich}}, \
  and\ \bibinfo {author} {\bibfnamefont {W.~A.}\ \bibnamefont {Dawson}},\
  }\href {\doibase 10.3847/1538-4357/aaee77} {\bibfield  {journal} {\bibinfo
  {journal} {Astrophys. J.}\ }\textbf {\bibinfo {volume} {869}},\ \bibinfo
  {pages} {104} (\bibinfo {year} {2018})},\ \Eprint
  {http://arxiv.org/abs/1701.05877} {arXiv:1701.05877 [astro-ph.CO]}
  \BibitemShut {NoStop}%
\bibitem [{\citenamefont {Newman}\ \emph
  {et~al.}(2013{\natexlab{a}})\citenamefont {Newman}, \citenamefont {Treu},
  \citenamefont {Ellis}, \citenamefont {Sand}, \citenamefont {Nipoti},
  \citenamefont {Richard},\ and\ \citenamefont {Jullo}}]{Newman:2012nv}%
  \BibitemOpen
  \bibfield  {author} {\bibinfo {author} {\bibfnamefont {A.~B.}\ \bibnamefont
  {Newman}}, \bibinfo {author} {\bibfnamefont {T.}~\bibnamefont {Treu}},
  \bibinfo {author} {\bibfnamefont {R.~S.}\ \bibnamefont {Ellis}}, \bibinfo
  {author} {\bibfnamefont {D.~J.}\ \bibnamefont {Sand}}, \bibinfo {author}
  {\bibfnamefont {C.}~\bibnamefont {Nipoti}}, \bibinfo {author} {\bibfnamefont
  {J.}~\bibnamefont {Richard}}, \ and\ \bibinfo {author} {\bibfnamefont
  {E.}~\bibnamefont {Jullo}},\ }\href {\doibase 10.1088/0004-637X/765/1/24}
  {\bibfield  {journal} {\bibinfo  {journal} {Astrophys. J.}\ }\textbf
  {\bibinfo {volume} {765}},\ \bibinfo {pages} {24} (\bibinfo {year}
  {2013}{\natexlab{a}})},\ \Eprint {http://arxiv.org/abs/1209.1391}
  {arXiv:1209.1391 [astro-ph.CO]} \BibitemShut {NoStop}%
\bibitem [{\citenamefont {Newman}\ \emph
  {et~al.}(2013{\natexlab{b}})\citenamefont {Newman}, \citenamefont {Treu},
  \citenamefont {Ellis},\ and\ \citenamefont {Sand}}]{Newman:2012nw}%
  \BibitemOpen
  \bibfield  {author} {\bibinfo {author} {\bibfnamefont {A.~B.}\ \bibnamefont
  {Newman}}, \bibinfo {author} {\bibfnamefont {T.}~\bibnamefont {Treu}},
  \bibinfo {author} {\bibfnamefont {R.~S.}\ \bibnamefont {Ellis}}, \ and\
  \bibinfo {author} {\bibfnamefont {D.~J.}\ \bibnamefont {Sand}},\ }\href
  {\doibase 10.1088/0004-637X/765/1/25} {\bibfield  {journal} {\bibinfo
  {journal} {Astrophys. J.}\ }\textbf {\bibinfo {volume} {765}},\ \bibinfo
  {pages} {25} (\bibinfo {year} {2013}{\natexlab{b}})},\ \Eprint
  {http://arxiv.org/abs/1209.1392} {arXiv:1209.1392 [astro-ph.CO]} \BibitemShut
  {NoStop}%
\bibitem [{\citenamefont {Oh}\ \emph {et~al.}(2011)\citenamefont {Oh},
  \citenamefont {de~Blok}, \citenamefont {Brinks}, \citenamefont {Walter},\
  and\ \citenamefont {Kennicutt}}]{Oh:2010ea}%
  \BibitemOpen
  \bibfield  {author} {\bibinfo {author} {\bibfnamefont {S.-H.}\ \bibnamefont
  {Oh}}, \bibinfo {author} {\bibfnamefont {W.~J.~G.}\ \bibnamefont {de~Blok}},
  \bibinfo {author} {\bibfnamefont {E.}~\bibnamefont {Brinks}}, \bibinfo
  {author} {\bibfnamefont {F.}~\bibnamefont {Walter}}, \ and\ \bibinfo {author}
  {\bibfnamefont {R.~C.}\ \bibnamefont {Kennicutt}, \bibfnamefont {Jr}},\
  }\href {\doibase 10.1088/0004-6256/141/6/193} {\bibfield  {journal} {\bibinfo
   {journal} {Astron. J.}\ }\textbf {\bibinfo {volume} {141}},\ \bibinfo
  {pages} {193} (\bibinfo {year} {2011})},\ \Eprint
  {http://arxiv.org/abs/1011.0899} {arXiv:1011.0899 [astro-ph.CO]} \BibitemShut
  {NoStop}%
\bibitem [{\citenamefont {Kuzio~de Naray}\ \emph {et~al.}(2008)\citenamefont
  {Kuzio~de Naray}, \citenamefont {McGaugh},\ and\ \citenamefont
  {de~Blok}}]{KuziodeNaray:2007qi}%
  \BibitemOpen
  \bibfield  {author} {\bibinfo {author} {\bibfnamefont {R.}~\bibnamefont
  {Kuzio~de Naray}}, \bibinfo {author} {\bibfnamefont {S.~S.}\ \bibnamefont
  {McGaugh}}, \ and\ \bibinfo {author} {\bibfnamefont {W.~J.~G.}\ \bibnamefont
  {de~Blok}},\ }\href {\doibase 10.1086/527543} {\bibfield  {journal} {\bibinfo
   {journal} {Astrophys. J.}\ }\textbf {\bibinfo {volume} {676}},\ \bibinfo
  {pages} {920} (\bibinfo {year} {2008})},\ \Eprint
  {http://arxiv.org/abs/0712.0860} {arXiv:0712.0860 [astro-ph]} \BibitemShut
  {NoStop}%
\bibitem [{\citenamefont {Walter}\ \emph {et~al.}(1985)\citenamefont {Walter},
  \citenamefont {Brinks}, \citenamefont {de~Blok}, \citenamefont {Bigiel},
  \citenamefont {Kennicutt}, \citenamefont {Jr.}, \citenamefont {Thornley},\
  and\ \citenamefont {Leroy}}]{Walter:2008wy}%
  \BibitemOpen
  \bibfield  {author} {\bibinfo {author} {\bibfnamefont {F.}~\bibnamefont
  {Walter}}, \bibinfo {author} {\bibfnamefont {E.}~\bibnamefont {Brinks}},
  \bibinfo {author} {\bibfnamefont {W.~J.~G.}\ \bibnamefont {de~Blok}},
  \bibinfo {author} {\bibfnamefont {F.}~\bibnamefont {Bigiel}}, \bibinfo
  {author} {\bibfnamefont {R.~C.}\ \bibnamefont {Kennicutt}}, \bibinfo {author}
  {\bibnamefont {Jr.}}, \bibinfo {author} {\bibfnamefont {M.~D.}\ \bibnamefont
  {Thornley}}, \ and\ \bibinfo {author} {\bibfnamefont {A.~K.}\ \bibnamefont
  {Leroy}},\ }\href {\doibase 10.1088/0004-6256/136/6/2563} {\bibfield
  {journal} {\bibinfo  {journal} {Astron. J.}\ }\textbf {\bibinfo {volume}
  {136}},\ \bibinfo {pages} {2563} (\bibinfo {year} {1985})},\ \Eprint
  {http://arxiv.org/abs/0810.2125} {arXiv:0810.2125 [astro-ph]} \BibitemShut
  {NoStop}%
\bibitem [{\citenamefont {de~Blok}\ and\ \citenamefont
  {Bosma}(2002)}]{deBlok:2002vgq}%
  \BibitemOpen
  \bibfield  {author} {\bibinfo {author} {\bibfnamefont {W.~J.~G.}\
  \bibnamefont {de~Blok}}\ and\ \bibinfo {author} {\bibfnamefont
  {A.}~\bibnamefont {Bosma}},\ }\href {\doibase 10.1051/0004-6361:20020080}
  {\bibfield  {journal} {\bibinfo  {journal} {Astron. Astrophys.}\ }\textbf
  {\bibinfo {volume} {385}},\ \bibinfo {pages} {816} (\bibinfo {year}
  {2002})},\ \Eprint {http://arxiv.org/abs/astro-ph/0201276}
  {arXiv:astro-ph/0201276 [astro-ph]} \BibitemShut {NoStop}%
\bibitem [{\citenamefont {Swaters}\ \emph {et~al.}(2003)\citenamefont
  {Swaters}, \citenamefont {Madore}, \citenamefont {van~den Bosch},\ and\
  \citenamefont {Balcells}}]{Swaters:2002rx}%
  \BibitemOpen
  \bibfield  {author} {\bibinfo {author} {\bibfnamefont {R.~A.}\ \bibnamefont
  {Swaters}}, \bibinfo {author} {\bibfnamefont {B.~F.}\ \bibnamefont {Madore}},
  \bibinfo {author} {\bibfnamefont {F.~C.}\ \bibnamefont {van~den Bosch}}, \
  and\ \bibinfo {author} {\bibfnamefont {M.}~\bibnamefont {Balcells}},\ }\href
  {\doibase 10.1086/345426} {\bibfield  {journal} {\bibinfo  {journal}
  {Astrophys. J.}\ }\textbf {\bibinfo {volume} {583}},\ \bibinfo {pages} {732}
  (\bibinfo {year} {2003})},\ \Eprint {http://arxiv.org/abs/astro-ph/0210152}
  {arXiv:astro-ph/0210152 [astro-ph]} \BibitemShut {NoStop}%
\bibitem [{\citenamefont {McGaugh}\ \emph {et~al.}(2001)\citenamefont
  {McGaugh}, \citenamefont {Rubin},\ and\ \citenamefont
  {de~Blok}}]{McGaugh:2001yc}%
  \BibitemOpen
  \bibfield  {author} {\bibinfo {author} {\bibfnamefont {S.~S.}\ \bibnamefont
  {McGaugh}}, \bibinfo {author} {\bibfnamefont {V.~C.}\ \bibnamefont {Rubin}},
  \ and\ \bibinfo {author} {\bibfnamefont {W.~J.~G.}\ \bibnamefont {de~Blok}},\
  }\href {\doibase 10.1086/323448} {\bibfield  {journal} {\bibinfo  {journal}
  {Astron. J.}\ }\textbf {\bibinfo {volume} {122}},\ \bibinfo {pages} {2381}
  (\bibinfo {year} {2001})},\ \Eprint {http://arxiv.org/abs/astro-ph/0107326}
  {arXiv:astro-ph/0107326 [astro-ph]} \BibitemShut {NoStop}%
\bibitem [{\citenamefont {Kummer}\ \emph {et~al.}(2019)\citenamefont {Kummer},
  \citenamefont {Br{\"u}ggen}, \citenamefont {Dolag}, \citenamefont
  {Kahlhoefer},\ and\ \citenamefont {Schmidt-Hoberg}}]{Kummer:2019yrb}%
  \BibitemOpen
  \bibfield  {author} {\bibinfo {author} {\bibfnamefont {J.}~\bibnamefont
  {Kummer}}, \bibinfo {author} {\bibfnamefont {M.}~\bibnamefont {Br{\"u}ggen}},
  \bibinfo {author} {\bibfnamefont {K.}~\bibnamefont {Dolag}}, \bibinfo
  {author} {\bibfnamefont {F.}~\bibnamefont {Kahlhoefer}}, \ and\ \bibinfo
  {author} {\bibfnamefont {K.}~\bibnamefont {Schmidt-Hoberg}},\ }\href@noop {}
  {\  (\bibinfo {year} {2019})},\ \Eprint {http://arxiv.org/abs/1902.02330}
  {arXiv:1902.02330 [astro-ph.CO]} \BibitemShut {NoStop}%
\bibitem [{\citenamefont {Nishikawa}\ \emph {et~al.}(2019)\citenamefont
  {Nishikawa}, \citenamefont {Boddy},\ and\ \citenamefont
  {Kaplinghat}}]{Nishikawa:2019lsc}%
  \BibitemOpen
  \bibfield  {author} {\bibinfo {author} {\bibfnamefont {H.}~\bibnamefont
  {Nishikawa}}, \bibinfo {author} {\bibfnamefont {K.~K.}\ \bibnamefont
  {Boddy}}, \ and\ \bibinfo {author} {\bibfnamefont {M.}~\bibnamefont
  {Kaplinghat}},\ }\href@noop {} {\  (\bibinfo {year} {2019})},\ \Eprint
  {http://arxiv.org/abs/1901.00499} {arXiv:1901.00499 [astro-ph.GA]}
  \BibitemShut {NoStop}%
\bibitem [{\citenamefont {Tulin}\ \emph {et~al.}(2013)\citenamefont {Tulin},
  \citenamefont {Yu},\ and\ \citenamefont {Zurek}}]{Tulin:2013teo}%
  \BibitemOpen
  \bibfield  {author} {\bibinfo {author} {\bibfnamefont {S.}~\bibnamefont
  {Tulin}}, \bibinfo {author} {\bibfnamefont {H.-B.}\ \bibnamefont {Yu}}, \
  and\ \bibinfo {author} {\bibfnamefont {K.~M.}\ \bibnamefont {Zurek}},\ }\href
  {\doibase 10.1103/PhysRevD.87.115007} {\bibfield  {journal} {\bibinfo
  {journal} {Phys. Rev.}\ }\textbf {\bibinfo {volume} {D87}},\ \bibinfo {pages}
  {115007} (\bibinfo {year} {2013})},\ \Eprint {http://arxiv.org/abs/1302.3898}
  {arXiv:1302.3898 [hep-ph]} \BibitemShut {NoStop}%
\bibitem [{\citenamefont {Feng}\ \emph {et~al.}(2010)\citenamefont {Feng},
  \citenamefont {Kaplinghat},\ and\ \citenamefont {Yu}}]{Feng:2009hw}%
  \BibitemOpen
  \bibfield  {author} {\bibinfo {author} {\bibfnamefont {J.~L.}\ \bibnamefont
  {Feng}}, \bibinfo {author} {\bibfnamefont {M.}~\bibnamefont {Kaplinghat}}, \
  and\ \bibinfo {author} {\bibfnamefont {H.-B.}\ \bibnamefont {Yu}},\ }\href
  {\doibase 10.1103/PhysRevLett.104.151301} {\bibfield  {journal} {\bibinfo
  {journal} {Phys. Rev. Lett.}\ }\textbf {\bibinfo {volume} {104}},\ \bibinfo
  {pages} {151301} (\bibinfo {year} {2010})},\ \Eprint
  {http://arxiv.org/abs/0911.0422} {arXiv:0911.0422 [hep-ph]} \BibitemShut
  {NoStop}%
\bibitem [{\citenamefont {Kesden}\ and\ \citenamefont
  {Kamionkowski}(2006{\natexlab{a}})}]{Kesden:2006zb}%
  \BibitemOpen
  \bibfield  {author} {\bibinfo {author} {\bibfnamefont {M.}~\bibnamefont
  {Kesden}}\ and\ \bibinfo {author} {\bibfnamefont {M.}~\bibnamefont
  {Kamionkowski}},\ }\href {\doibase 10.1103/PhysRevLett.97.131303} {\bibfield
  {journal} {\bibinfo  {journal} {Phys. Rev. Lett.}\ }\textbf {\bibinfo
  {volume} {97}},\ \bibinfo {pages} {131303} (\bibinfo {year}
  {2006}{\natexlab{a}})},\ \Eprint {http://arxiv.org/abs/astro-ph/0606566}
  {arXiv:astro-ph/0606566 [astro-ph]} \BibitemShut {NoStop}%
\bibitem [{\citenamefont {Kesden}\ and\ \citenamefont
  {Kamionkowski}(2006{\natexlab{b}})}]{Kesden:2006vz}%
  \BibitemOpen
  \bibfield  {author} {\bibinfo {author} {\bibfnamefont {M.}~\bibnamefont
  {Kesden}}\ and\ \bibinfo {author} {\bibfnamefont {M.}~\bibnamefont
  {Kamionkowski}},\ }\href {\doibase 10.1103/PhysRevD.74.083007} {\bibfield
  {journal} {\bibinfo  {journal} {Phys. Rev.}\ }\textbf {\bibinfo {volume}
  {D74}},\ \bibinfo {pages} {083007} (\bibinfo {year} {2006}{\natexlab{b}})},\
  \Eprint {http://arxiv.org/abs/astro-ph/0608095} {arXiv:astro-ph/0608095
  [astro-ph]} \BibitemShut {NoStop}%
\bibitem [{\citenamefont {Frieman}\ and\ \citenamefont
  {Gradwohl}(1991)}]{Friedman:1991dj}%
  \BibitemOpen
  \bibfield  {author} {\bibinfo {author} {\bibfnamefont {J.~A.}\ \bibnamefont
  {Frieman}}\ and\ \bibinfo {author} {\bibfnamefont {B.-A.}\ \bibnamefont
  {Gradwohl}},\ }\href {\doibase 10.1103/PhysRevLett.67.2926} {\bibfield
  {journal} {\bibinfo  {journal} {Phys. Rev. Lett.}\ }\textbf {\bibinfo
  {volume} {67}},\ \bibinfo {pages} {2926} (\bibinfo {year}
  {1991})}\BibitemShut {NoStop}%
\bibitem [{\citenamefont {Bean}(2001)}]{Bean:2001ys}%
  \BibitemOpen
  \bibfield  {author} {\bibinfo {author} {\bibfnamefont {R.}~\bibnamefont
  {Bean}},\ }\href {\doibase 10.1103/PhysRevD.64.123516} {\bibfield  {journal}
  {\bibinfo  {journal} {Phys. Rev.}\ }\textbf {\bibinfo {volume} {D64}},\
  \bibinfo {pages} {123516} (\bibinfo {year} {2001})},\ \Eprint
  {http://arxiv.org/abs/astro-ph/0104464} {arXiv:astro-ph/0104464 [astro-ph]}
  \BibitemShut {NoStop}%
\bibitem [{\citenamefont {Gubser}\ and\ \citenamefont
  {Peebles}(2004)}]{Gubser:2004uh}%
  \BibitemOpen
  \bibfield  {author} {\bibinfo {author} {\bibfnamefont {S.~S.}\ \bibnamefont
  {Gubser}}\ and\ \bibinfo {author} {\bibfnamefont {P.~J.~E.}\ \bibnamefont
  {Peebles}},\ }\href {\doibase 10.1103/PhysRevD.70.123510} {\bibfield
  {journal} {\bibinfo  {journal} {Phys. Rev.}\ }\textbf {\bibinfo {volume}
  {D70}},\ \bibinfo {pages} {123510} (\bibinfo {year} {2004})},\ \Eprint
  {http://arxiv.org/abs/hep-th/0402225} {arXiv:hep-th/0402225 [hep-th]}
  \BibitemShut {NoStop}%
\bibitem [{\citenamefont {Nusser}\ \emph {et~al.}(2005)\citenamefont {Nusser},
  \citenamefont {Gubser},\ and\ \citenamefont {Peebles}}]{Nusser:2004qu}%
  \BibitemOpen
  \bibfield  {author} {\bibinfo {author} {\bibfnamefont {A.}~\bibnamefont
  {Nusser}}, \bibinfo {author} {\bibfnamefont {S.~S.}\ \bibnamefont {Gubser}},
  \ and\ \bibinfo {author} {\bibfnamefont {P.~J.~E.}\ \bibnamefont {Peebles}},\
  }\href {\doibase 10.1103/PhysRevD.71.083505} {\bibfield  {journal} {\bibinfo
  {journal} {Phys. Rev.}\ }\textbf {\bibinfo {volume} {D71}},\ \bibinfo {pages}
  {083505} (\bibinfo {year} {2005})},\ \Eprint
  {http://arxiv.org/abs/astro-ph/0412586} {arXiv:astro-ph/0412586 [astro-ph]}
  \BibitemShut {NoStop}%
\bibitem [{\citenamefont {Bean}\ \emph {et~al.}(2008)\citenamefont {Bean},
  \citenamefont {Flanagan}, \citenamefont {Laszlo},\ and\ \citenamefont
  {Trodden}}]{Bean:2008ac}%
  \BibitemOpen
  \bibfield  {author} {\bibinfo {author} {\bibfnamefont {R.}~\bibnamefont
  {Bean}}, \bibinfo {author} {\bibfnamefont {E.~E.}\ \bibnamefont {Flanagan}},
  \bibinfo {author} {\bibfnamefont {I.}~\bibnamefont {Laszlo}}, \ and\ \bibinfo
  {author} {\bibfnamefont {M.}~\bibnamefont {Trodden}},\ }\href {\doibase
  10.1103/PhysRevD.78.123514} {\bibfield  {journal} {\bibinfo  {journal} {Phys.
  Rev.}\ }\textbf {\bibinfo {volume} {D78}},\ \bibinfo {pages} {123514}
  (\bibinfo {year} {2008})},\ \Eprint {http://arxiv.org/abs/0808.1105}
  {arXiv:0808.1105 [astro-ph]} \BibitemShut {NoStop}%
\bibitem [{\citenamefont {Bai}\ \emph {et~al.}(2015)\citenamefont {Bai},
  \citenamefont {Salvado},\ and\ \citenamefont {Stefanek}}]{Bai:2015vca}%
  \BibitemOpen
  \bibfield  {author} {\bibinfo {author} {\bibfnamefont {Y.}~\bibnamefont
  {Bai}}, \bibinfo {author} {\bibfnamefont {J.}~\bibnamefont {Salvado}}, \ and\
  \bibinfo {author} {\bibfnamefont {B.~A.}\ \bibnamefont {Stefanek}},\ }\href
  {\doibase 10.1088/1475-7516/2015/10/029} {\bibfield  {journal} {\bibinfo
  {journal} {JCAP}\ }\textbf {\bibinfo {volume} {1510}},\ \bibinfo {pages}
  {029} (\bibinfo {year} {2015})},\ \Eprint {http://arxiv.org/abs/1505.04789}
  {arXiv:1505.04789 [hep-ph]} \BibitemShut {NoStop}%
\bibitem [{\citenamefont {Baldes}\ and\ \citenamefont
  {Petraki}(2017)}]{Baldes:2017gzw}%
  \BibitemOpen
  \bibfield  {author} {\bibinfo {author} {\bibfnamefont {I.}~\bibnamefont
  {Baldes}}\ and\ \bibinfo {author} {\bibfnamefont {K.}~\bibnamefont
  {Petraki}},\ }\href {\doibase 10.1088/1475-7516/2017/09/028} {\bibfield
  {journal} {\bibinfo  {journal} {JCAP}\ }\textbf {\bibinfo {volume} {1709}},\
  \bibinfo {pages} {028} (\bibinfo {year} {2017})},\ \Eprint
  {http://arxiv.org/abs/1703.00478} {arXiv:1703.00478 [hep-ph]} \BibitemShut
  {NoStop}%
\bibitem [{\citenamefont {Hufnagel}\ \emph
  {et~al.}(2018{\natexlab{a}})\citenamefont {Hufnagel}, \citenamefont
  {Schmidt-Hoberg},\ and\ \citenamefont {Wild}}]{Hufnagel:2017dgo}%
  \BibitemOpen
  \bibfield  {author} {\bibinfo {author} {\bibfnamefont {M.}~\bibnamefont
  {Hufnagel}}, \bibinfo {author} {\bibfnamefont {K.}~\bibnamefont
  {Schmidt-Hoberg}}, \ and\ \bibinfo {author} {\bibfnamefont {S.}~\bibnamefont
  {Wild}},\ }\href {\doibase 10.1088/1475-7516/2018/02/044} {\bibfield
  {journal} {\bibinfo  {journal} {JCAP}\ }\textbf {\bibinfo {volume} {1802}},\
  \bibinfo {pages} {044} (\bibinfo {year} {2018}{\natexlab{a}})},\ \Eprint
  {http://arxiv.org/abs/1712.03972} {arXiv:1712.03972 [hep-ph]} \BibitemShut
  {NoStop}%
\bibitem [{\citenamefont {Hufnagel}\ \emph
  {et~al.}(2018{\natexlab{b}})\citenamefont {Hufnagel}, \citenamefont
  {Schmidt-Hoberg},\ and\ \citenamefont {Wild}}]{Hufnagel:2018bjp}%
  \BibitemOpen
  \bibfield  {author} {\bibinfo {author} {\bibfnamefont {M.}~\bibnamefont
  {Hufnagel}}, \bibinfo {author} {\bibfnamefont {K.}~\bibnamefont
  {Schmidt-Hoberg}}, \ and\ \bibinfo {author} {\bibfnamefont {S.}~\bibnamefont
  {Wild}},\ }\href {\doibase 10.1088/1475-7516/2018/11/032} {\bibfield
  {journal} {\bibinfo  {journal} {JCAP}\ }\textbf {\bibinfo {volume} {1811}},\
  \bibinfo {pages} {032} (\bibinfo {year} {2018}{\natexlab{b}})},\ \Eprint
  {http://arxiv.org/abs/1808.09324} {arXiv:1808.09324 [hep-ph]} \BibitemShut
  {NoStop}%
\bibitem [{\citenamefont {Forestell}\ \emph {et~al.}(2019)\citenamefont
  {Forestell}, \citenamefont {Morrissey},\ and\ \citenamefont
  {White}}]{Forestell:2018txr}%
  \BibitemOpen
  \bibfield  {author} {\bibinfo {author} {\bibfnamefont {L.}~\bibnamefont
  {Forestell}}, \bibinfo {author} {\bibfnamefont {D.~E.}\ \bibnamefont
  {Morrissey}}, \ and\ \bibinfo {author} {\bibfnamefont {G.}~\bibnamefont
  {White}},\ }\href {\doibase 10.1007/JHEP01(2019)074} {\bibfield  {journal}
  {\bibinfo  {journal} {JHEP}\ }\textbf {\bibinfo {volume} {01}},\ \bibinfo
  {pages} {074} (\bibinfo {year} {2019})},\ \Eprint
  {http://arxiv.org/abs/1809.01179} {arXiv:1809.01179 [hep-ph]} \BibitemShut
  {NoStop}%
\bibitem [{\citenamefont {Boddy}\ \emph {et~al.}(2014)\citenamefont {Boddy},
  \citenamefont {Feng}, \citenamefont {Kaplinghat},\ and\ \citenamefont
  {Tait}}]{Boddy:2014yra}%
  \BibitemOpen
  \bibfield  {author} {\bibinfo {author} {\bibfnamefont {K.~K.}\ \bibnamefont
  {Boddy}}, \bibinfo {author} {\bibfnamefont {J.~L.}\ \bibnamefont {Feng}},
  \bibinfo {author} {\bibfnamefont {M.}~\bibnamefont {Kaplinghat}}, \ and\
  \bibinfo {author} {\bibfnamefont {T.~M.~P.}\ \bibnamefont {Tait}},\ }\href
  {\doibase 10.1103/PhysRevD.89.115017} {\bibfield  {journal} {\bibinfo
  {journal} {Phys. Rev.}\ }\textbf {\bibinfo {volume} {D89}},\ \bibinfo {pages}
  {115017} (\bibinfo {year} {2014})},\ \Eprint {http://arxiv.org/abs/1402.3629}
  {arXiv:1402.3629 [hep-ph]} \BibitemShut {NoStop}%
\bibitem [{\citenamefont {Feldman}\ \emph {et~al.}(2007)\citenamefont
  {Feldman}, \citenamefont {Kors},\ and\ \citenamefont
  {Nath}}]{Feldman:2006wd}%
  \BibitemOpen
  \bibfield  {author} {\bibinfo {author} {\bibfnamefont {D.}~\bibnamefont
  {Feldman}}, \bibinfo {author} {\bibfnamefont {B.}~\bibnamefont {Kors}}, \
  and\ \bibinfo {author} {\bibfnamefont {P.}~\bibnamefont {Nath}},\ }\href
  {\doibase 10.1103/PhysRevD.75.023503} {\bibfield  {journal} {\bibinfo
  {journal} {Phys. Rev.}\ }\textbf {\bibinfo {volume} {D75}},\ \bibinfo {pages}
  {023503} (\bibinfo {year} {2007})},\ \Eprint
  {http://arxiv.org/abs/hep-ph/0610133} {arXiv:hep-ph/0610133 [hep-ph]}
  \BibitemShut {NoStop}%
\bibitem [{\citenamefont {Kang}\ \emph {et~al.}(2011)\citenamefont {Kang},
  \citenamefont {Li}, \citenamefont {Liu}, \citenamefont {Tong},\ and\
  \citenamefont {Yang}}]{Kang:2010mh}%
  \BibitemOpen
  \bibfield  {author} {\bibinfo {author} {\bibfnamefont {Z.}~\bibnamefont
  {Kang}}, \bibinfo {author} {\bibfnamefont {T.}~\bibnamefont {Li}}, \bibinfo
  {author} {\bibfnamefont {T.}~\bibnamefont {Liu}}, \bibinfo {author}
  {\bibfnamefont {C.}~\bibnamefont {Tong}}, \ and\ \bibinfo {author}
  {\bibfnamefont {J.~M.}\ \bibnamefont {Yang}},\ }\href {\doibase
  10.1088/1475-7516/2011/01/028} {\bibfield  {journal} {\bibinfo  {journal}
  {JCAP}\ }\textbf {\bibinfo {volume} {1101}},\ \bibinfo {pages} {028}
  (\bibinfo {year} {2011})},\ \Eprint {http://arxiv.org/abs/1008.5243}
  {arXiv:1008.5243 [hep-ph]} \BibitemShut {NoStop}%
\bibitem [{\citenamefont {Chu}\ \emph {et~al.}(2014)\citenamefont {Chu},
  \citenamefont {Mambrini}, \citenamefont {Quevillon},\ and\ \citenamefont
  {Zaldivar}}]{Chu:2013jja}%
  \BibitemOpen
  \bibfield  {author} {\bibinfo {author} {\bibfnamefont {X.}~\bibnamefont
  {Chu}}, \bibinfo {author} {\bibfnamefont {Y.}~\bibnamefont {Mambrini}},
  \bibinfo {author} {\bibfnamefont {J.}~\bibnamefont {Quevillon}}, \ and\
  \bibinfo {author} {\bibfnamefont {B.}~\bibnamefont {Zaldivar}},\ }\href
  {\doibase 10.1088/1475-7516/2014/01/034} {\bibfield  {journal} {\bibinfo
  {journal} {JCAP}\ }\textbf {\bibinfo {volume} {1401}},\ \bibinfo {pages}
  {034} (\bibinfo {year} {2014})},\ \Eprint {http://arxiv.org/abs/1306.4677}
  {arXiv:1306.4677 [hep-ph]} \BibitemShut {NoStop}%
\bibitem [{\citenamefont {Mambrini}(2011)}]{Mambrini:2011dw}%
  \BibitemOpen
  \bibfield  {author} {\bibinfo {author} {\bibfnamefont {Y.}~\bibnamefont
  {Mambrini}},\ }\href {\doibase 10.1088/1475-7516/2011/07/009} {\bibfield
  {journal} {\bibinfo  {journal} {JCAP}\ }\textbf {\bibinfo {volume} {1107}},\
  \bibinfo {pages} {009} (\bibinfo {year} {2011})},\ \Eprint
  {http://arxiv.org/abs/1104.4799} {arXiv:1104.4799 [hep-ph]} \BibitemShut
  {NoStop}%
\bibitem [{\citenamefont {Holdom}(1986)}]{Holdom:1985ag}%
  \BibitemOpen
  \bibfield  {author} {\bibinfo {author} {\bibfnamefont {B.}~\bibnamefont
  {Holdom}},\ }\href {\doibase 10.1016/0370-2693(86)91377-8} {\bibfield
  {journal} {\bibinfo  {journal} {Phys. Lett.}\ }\textbf {\bibinfo {volume}
  {166B}},\ \bibinfo {pages} {196} (\bibinfo {year} {1986})}\BibitemShut
  {NoStop}%
\bibitem [{\citenamefont {Dienes}\ \emph {et~al.}(1997)\citenamefont {Dienes},
  \citenamefont {Kolda},\ and\ \citenamefont {March-Russell}}]{Dienes:1996zr}%
  \BibitemOpen
  \bibfield  {author} {\bibinfo {author} {\bibfnamefont {K.~R.}\ \bibnamefont
  {Dienes}}, \bibinfo {author} {\bibfnamefont {C.~F.}\ \bibnamefont {Kolda}}, \
  and\ \bibinfo {author} {\bibfnamefont {J.}~\bibnamefont {March-Russell}},\
  }\href {\doibase 10.1016/S0550-3213(97)80028-4,
  10.1016/S0550-3213(97)00173-9} {\bibfield  {journal} {\bibinfo  {journal}
  {Nucl. Phys.}\ }\textbf {\bibinfo {volume} {B492}},\ \bibinfo {pages} {104}
  (\bibinfo {year} {1997})},\ \Eprint {http://arxiv.org/abs/hep-ph/9610479}
  {arXiv:hep-ph/9610479 [hep-ph]} \BibitemShut {NoStop}%
\bibitem [{\citenamefont {Abel}\ \emph {et~al.}(2008)\citenamefont {Abel},
  \citenamefont {Goodsell}, \citenamefont {Jaeckel}, \citenamefont {Khoze},\
  and\ \citenamefont {Ringwald}}]{Abel:2008ai}%
  \BibitemOpen
  \bibfield  {author} {\bibinfo {author} {\bibfnamefont {S.~A.}\ \bibnamefont
  {Abel}}, \bibinfo {author} {\bibfnamefont {M.~D.}\ \bibnamefont {Goodsell}},
  \bibinfo {author} {\bibfnamefont {J.}~\bibnamefont {Jaeckel}}, \bibinfo
  {author} {\bibfnamefont {V.~V.}\ \bibnamefont {Khoze}}, \ and\ \bibinfo
  {author} {\bibfnamefont {A.}~\bibnamefont {Ringwald}},\ }\href {\doibase
  10.1088/1126-6708/2008/07/124} {\bibfield  {journal} {\bibinfo  {journal}
  {JHEP}\ }\textbf {\bibinfo {volume} {07}},\ \bibinfo {pages} {124} (\bibinfo
  {year} {2008})},\ \Eprint {http://arxiv.org/abs/0803.1449} {arXiv:0803.1449
  [hep-ph]} \BibitemShut {NoStop}%
\bibitem [{\citenamefont {Hall}\ \emph {et~al.}(2010)\citenamefont {Hall},
  \citenamefont {Jedamzik}, \citenamefont {March-Russell},\ and\ \citenamefont
  {West}}]{Hall:2009bx}%
  \BibitemOpen
  \bibfield  {author} {\bibinfo {author} {\bibfnamefont {L.~J.}\ \bibnamefont
  {Hall}}, \bibinfo {author} {\bibfnamefont {K.}~\bibnamefont {Jedamzik}},
  \bibinfo {author} {\bibfnamefont {J.}~\bibnamefont {March-Russell}}, \ and\
  \bibinfo {author} {\bibfnamefont {S.~M.}\ \bibnamefont {West}},\ }\href
  {\doibase 10.1007/JHEP03(2010)080} {\bibfield  {journal} {\bibinfo  {journal}
  {JHEP}\ }\textbf {\bibinfo {volume} {03}},\ \bibinfo {pages} {080} (\bibinfo
  {year} {2010})},\ \Eprint {http://arxiv.org/abs/0911.1120} {arXiv:0911.1120
  [hep-ph]} \BibitemShut {NoStop}%
\bibitem [{\citenamefont {Chu}\ \emph {et~al.}(2012)\citenamefont {Chu},
  \citenamefont {Hambye},\ and\ \citenamefont {Tytgat}}]{Chu:2011be}%
  \BibitemOpen
  \bibfield  {author} {\bibinfo {author} {\bibfnamefont {X.}~\bibnamefont
  {Chu}}, \bibinfo {author} {\bibfnamefont {T.}~\bibnamefont {Hambye}}, \ and\
  \bibinfo {author} {\bibfnamefont {M.~H.~G.}\ \bibnamefont {Tytgat}},\ }\href
  {\doibase 10.1088/1475-7516/2012/05/034} {\bibfield  {journal} {\bibinfo
  {journal} {JCAP}\ }\textbf {\bibinfo {volume} {1205}},\ \bibinfo {pages}
  {034} (\bibinfo {year} {2012})},\ \Eprint {http://arxiv.org/abs/1112.0493}
  {arXiv:1112.0493 [hep-ph]} \BibitemShut {NoStop}%
\bibitem [{\citenamefont {Baldes}\ \emph {et~al.}(2018)\citenamefont {Baldes},
  \citenamefont {Cirelli}, \citenamefont {Panci}, \citenamefont {Petraki},
  \citenamefont {Sala},\ and\ \citenamefont {Taoso}}]{Baldes:2017gzu}%
  \BibitemOpen
  \bibfield  {author} {\bibinfo {author} {\bibfnamefont {I.}~\bibnamefont
  {Baldes}}, \bibinfo {author} {\bibfnamefont {M.}~\bibnamefont {Cirelli}},
  \bibinfo {author} {\bibfnamefont {P.}~\bibnamefont {Panci}}, \bibinfo
  {author} {\bibfnamefont {K.}~\bibnamefont {Petraki}}, \bibinfo {author}
  {\bibfnamefont {F.}~\bibnamefont {Sala}}, \ and\ \bibinfo {author}
  {\bibfnamefont {M.}~\bibnamefont {Taoso}},\ }\href {\doibase
  10.21468/SciPostPhys.4.6.041} {\bibfield  {journal} {\bibinfo  {journal}
  {SciPost Phys.}\ }\textbf {\bibinfo {volume} {4}},\ \bibinfo {pages} {041}
  (\bibinfo {year} {2018})},\ \Eprint {http://arxiv.org/abs/1712.07489}
  {arXiv:1712.07489 [hep-ph]} \BibitemShut {NoStop}%
\bibitem [{\citenamefont {Hambye}\ \emph {et~al.}(2018)\citenamefont {Hambye},
  \citenamefont {Tytgat}, \citenamefont {Vandecasteele},\ and\ \citenamefont
  {Vanderheyden}}]{Hambye:2018dpi}%
  \BibitemOpen
  \bibfield  {author} {\bibinfo {author} {\bibfnamefont {T.}~\bibnamefont
  {Hambye}}, \bibinfo {author} {\bibfnamefont {M.~H.~G.}\ \bibnamefont
  {Tytgat}}, \bibinfo {author} {\bibfnamefont {J.}~\bibnamefont
  {Vandecasteele}}, \ and\ \bibinfo {author} {\bibfnamefont {L.}~\bibnamefont
  {Vanderheyden}},\ }\href {\doibase 10.1103/PhysRevD.98.075017} {\bibfield
  {journal} {\bibinfo  {journal} {Phys. Rev.}\ }\textbf {\bibinfo {volume}
  {D98}},\ \bibinfo {pages} {075017} (\bibinfo {year} {2018})},\ \Eprint
  {http://arxiv.org/abs/1807.05022} {arXiv:1807.05022 [hep-ph]} \BibitemShut
  {NoStop}%
\bibitem [{\citenamefont {Han}\ \emph {et~al.}(2018)\citenamefont {Han},
  \citenamefont {Pi},\ and\ \citenamefont {Sasaki}}]{Han:2018yrk}%
  \BibitemOpen
  \bibfield  {author} {\bibinfo {author} {\bibfnamefont {C.}~\bibnamefont
  {Han}}, \bibinfo {author} {\bibfnamefont {S.}~\bibnamefont {Pi}}, \ and\
  \bibinfo {author} {\bibfnamefont {M.}~\bibnamefont {Sasaki}},\ }\href
  {\doibase 10.1016/j.physletb.2019.02.037} {\  (\bibinfo {year} {2018}),\
  10.1016/j.physletb.2019.02.037},\ \Eprint {http://arxiv.org/abs/1809.05507}
  {arXiv:1809.05507 [hep-ph]} \BibitemShut {NoStop}%
\bibitem [{\citenamefont {Carrington}(1992)}]{Carrington:1991hz}%
  \BibitemOpen
  \bibfield  {author} {\bibinfo {author} {\bibfnamefont {M.~E.}\ \bibnamefont
  {Carrington}},\ }\href {\doibase 10.1103/PhysRevD.45.2933} {\bibfield
  {journal} {\bibinfo  {journal} {Phys. Rev.}\ }\textbf {\bibinfo {volume}
  {D45}},\ \bibinfo {pages} {2933} (\bibinfo {year} {1992})}\BibitemShut
  {NoStop}%
\bibitem [{\citenamefont {Quiros}(1999)}]{Quiros:1999jp}%
  \BibitemOpen
  \bibfield  {author} {\bibinfo {author} {\bibfnamefont {M.}~\bibnamefont
  {Quiros}},\ }in\ \href@noop {} {\emph {\bibinfo {booktitle} {{High energy
  physics and cosmology. Proceedings, Summer School, Trieste, Italy, June
  29-July 17, 1998}}}}\ (\bibinfo {year} {1999})\ pp.\ \bibinfo {pages}
  {187--259},\ \Eprint {http://arxiv.org/abs/hep-ph/9901312}
  {arXiv:hep-ph/9901312 [hep-ph]} \BibitemShut {NoStop}%
\bibitem [{\citenamefont {D'Onofrio}\ and\ \citenamefont
  {Rummukainen}(2016)}]{DOnofrio:2015gop}%
  \BibitemOpen
  \bibfield  {author} {\bibinfo {author} {\bibfnamefont {M.}~\bibnamefont
  {D'Onofrio}}\ and\ \bibinfo {author} {\bibfnamefont {K.}~\bibnamefont
  {Rummukainen}},\ }\href {\doibase 10.1103/PhysRevD.93.025003} {\bibfield
  {journal} {\bibinfo  {journal} {Phys. Rev.}\ }\textbf {\bibinfo {volume}
  {D93}},\ \bibinfo {pages} {025003} (\bibinfo {year} {2016})},\ \Eprint
  {http://arxiv.org/abs/1508.07161} {arXiv:1508.07161 [hep-ph]} \BibitemShut
  {NoStop}%
\bibitem [{\citenamefont {Csikor}\ \emph {et~al.}(1999)\citenamefont {Csikor},
  \citenamefont {Fodor},\ and\ \citenamefont {Heitger}}]{Csikor:1998eu}%
  \BibitemOpen
  \bibfield  {author} {\bibinfo {author} {\bibfnamefont {F.}~\bibnamefont
  {Csikor}}, \bibinfo {author} {\bibfnamefont {Z.}~\bibnamefont {Fodor}}, \
  and\ \bibinfo {author} {\bibfnamefont {J.}~\bibnamefont {Heitger}},\ }\href
  {\doibase 10.1103/PhysRevLett.82.21} {\bibfield  {journal} {\bibinfo
  {journal} {Phys. Rev. Lett.}\ }\textbf {\bibinfo {volume} {82}},\ \bibinfo
  {pages} {21} (\bibinfo {year} {1999})},\ \Eprint
  {http://arxiv.org/abs/hep-ph/9809291} {arXiv:hep-ph/9809291 [hep-ph]}
  \BibitemShut {NoStop}%
\bibitem [{\citenamefont {Fukugita}\ and\ \citenamefont
  {Yanagida}(1986)}]{Fukugita:1986hr}%
  \BibitemOpen
  \bibfield  {author} {\bibinfo {author} {\bibfnamefont {M.}~\bibnamefont
  {Fukugita}}\ and\ \bibinfo {author} {\bibfnamefont {T.}~\bibnamefont
  {Yanagida}},\ }\href {\doibase 10.1016/0370-2693(86)91126-3} {\bibfield
  {journal} {\bibinfo  {journal} {Phys. Lett.}\ }\textbf {\bibinfo {volume}
  {B174}},\ \bibinfo {pages} {45} (\bibinfo {year} {1986})}\BibitemShut
  {NoStop}%
\bibitem [{\citenamefont {Davidson}\ \emph {et~al.}(2008)\citenamefont
  {Davidson}, \citenamefont {Nardi},\ and\ \citenamefont
  {Nir}}]{Davidson:2008bu}%
  \BibitemOpen
  \bibfield  {author} {\bibinfo {author} {\bibfnamefont {S.}~\bibnamefont
  {Davidson}}, \bibinfo {author} {\bibfnamefont {E.}~\bibnamefont {Nardi}}, \
  and\ \bibinfo {author} {\bibfnamefont {Y.}~\bibnamefont {Nir}},\ }\href
  {\doibase 10.1016/j.physrep.2008.06.002} {\bibfield  {journal} {\bibinfo
  {journal} {Phys. Rept.}\ }\textbf {\bibinfo {volume} {466}},\ \bibinfo
  {pages} {105} (\bibinfo {year} {2008})},\ \Eprint
  {http://arxiv.org/abs/0802.2962} {arXiv:0802.2962 [hep-ph]} \BibitemShut
  {NoStop}%
\bibitem [{\citenamefont {Shaposhnikov}(1987)}]{Shaposhnikov:1987tw}%
  \BibitemOpen
  \bibfield  {author} {\bibinfo {author} {\bibfnamefont {M.~E.}\ \bibnamefont
  {Shaposhnikov}},\ }\href {\doibase 10.1016/0550-3213(87)90127-1} {\bibfield
  {journal} {\bibinfo  {journal} {Nucl. Phys.}\ }\textbf {\bibinfo {volume}
  {B287}},\ \bibinfo {pages} {757} (\bibinfo {year} {1987})}\BibitemShut
  {NoStop}%
\bibitem [{\citenamefont {Trodden}(1999)}]{Trodden:1998ym}%
  \BibitemOpen
  \bibfield  {author} {\bibinfo {author} {\bibfnamefont {M.}~\bibnamefont
  {Trodden}},\ }\href {\doibase 10.1103/RevModPhys.71.1463} {\bibfield
  {journal} {\bibinfo  {journal} {Rev. Mod. Phys.}\ }\textbf {\bibinfo {volume}
  {71}},\ \bibinfo {pages} {1463} (\bibinfo {year} {1999})},\ \Eprint
  {http://arxiv.org/abs/hep-ph/9803479} {arXiv:hep-ph/9803479 [hep-ph]}
  \BibitemShut {NoStop}%
\bibitem [{\citenamefont {Konstandin}(2013)}]{Konstandin:2013caa}%
  \BibitemOpen
  \bibfield  {author} {\bibinfo {author} {\bibfnamefont {T.}~\bibnamefont
  {Konstandin}},\ }\href {\doibase 10.3367/UFNe.0183.201308a.0785} {\bibfield
  {journal} {\bibinfo  {journal} {Phys. Usp.}\ }\textbf {\bibinfo {volume}
  {56}},\ \bibinfo {pages} {747} (\bibinfo {year} {2013})},\ \bibinfo {note}
  {[Usp. Fiz. Nauk183,785(2013)]},\ \Eprint {http://arxiv.org/abs/1302.6713}
  {arXiv:1302.6713 [hep-ph]} \BibitemShut {NoStop}%
\bibitem [{\citenamefont {Berkooz}\ \emph {et~al.}(2004)\citenamefont
  {Berkooz}, \citenamefont {Nir},\ and\ \citenamefont
  {Volansky}}]{Berkooz:2004kx}%
  \BibitemOpen
  \bibfield  {author} {\bibinfo {author} {\bibfnamefont {M.}~\bibnamefont
  {Berkooz}}, \bibinfo {author} {\bibfnamefont {Y.}~\bibnamefont {Nir}}, \ and\
  \bibinfo {author} {\bibfnamefont {T.}~\bibnamefont {Volansky}},\ }\href
  {\doibase 10.1103/PhysRevLett.93.051301} {\bibfield  {journal} {\bibinfo
  {journal} {Phys. Rev. Lett.}\ }\textbf {\bibinfo {volume} {93}},\ \bibinfo
  {pages} {051301} (\bibinfo {year} {2004})},\ \Eprint
  {http://arxiv.org/abs/hep-ph/0401012} {arXiv:hep-ph/0401012 [hep-ph]}
  \BibitemShut {NoStop}%
\bibitem [{\citenamefont {Baldes}\ \emph {et~al.}(2016)\citenamefont {Baldes},
  \citenamefont {Konstandin},\ and\ \citenamefont {Servant}}]{Baldes:2016gaf}%
  \BibitemOpen
  \bibfield  {author} {\bibinfo {author} {\bibfnamefont {I.}~\bibnamefont
  {Baldes}}, \bibinfo {author} {\bibfnamefont {T.}~\bibnamefont {Konstandin}},
  \ and\ \bibinfo {author} {\bibfnamefont {G.}~\bibnamefont {Servant}},\ }\href
  {\doibase 10.1007/JHEP12(2016)073} {\bibfield  {journal} {\bibinfo  {journal}
  {JHEP}\ }\textbf {\bibinfo {volume} {12}},\ \bibinfo {pages} {073} (\bibinfo
  {year} {2016})},\ \Eprint {http://arxiv.org/abs/1608.03254} {arXiv:1608.03254
  [hep-ph]} \BibitemShut {NoStop}%
\bibitem [{\citenamefont {Ellis}\ \emph {et~al.}(2019)\citenamefont {Ellis},
  \citenamefont {Ipek},\ and\ \citenamefont {White}}]{Ellis:2019flb}%
  \BibitemOpen
  \bibfield  {author} {\bibinfo {author} {\bibfnamefont {S.~A.~R.}\
  \bibnamefont {Ellis}}, \bibinfo {author} {\bibfnamefont {S.}~\bibnamefont
  {Ipek}}, \ and\ \bibinfo {author} {\bibfnamefont {G.}~\bibnamefont {White}},\
  }\href@noop {} {\  (\bibinfo {year} {2019})},\ \Eprint
  {http://arxiv.org/abs/1905.11994} {arXiv:1905.11994 [hep-ph]} \BibitemShut
  {NoStop}%
\bibitem [{\citenamefont {Frampton}(1976)}]{Frampton:1976kf}%
  \BibitemOpen
  \bibfield  {author} {\bibinfo {author} {\bibfnamefont {P.~H.}\ \bibnamefont
  {Frampton}},\ }\href {\doibase 10.1103/PhysRevLett.37.1378,
  10.1103/PhysRevLett.37.1716.2} {\bibfield  {journal} {\bibinfo  {journal}
  {Phys. Rev. Lett.}\ }\textbf {\bibinfo {volume} {37}},\ \bibinfo {pages}
  {1378} (\bibinfo {year} {1976})},\ \bibinfo {note} {[Erratum: Phys. Rev.
  Lett.37,1716(1976)]}\BibitemShut {NoStop}%
\bibitem [{\citenamefont {Coleman}(1977)}]{Coleman:1977py}%
  \BibitemOpen
  \bibfield  {author} {\bibinfo {author} {\bibfnamefont {S.~R.}\ \bibnamefont
  {Coleman}},\ }\href {\doibase 10.1103/PhysRevD.15.2929,
  10.1103/PhysRevD.16.1248} {\bibfield  {journal} {\bibinfo  {journal} {Phys.
  Rev.}\ }\textbf {\bibinfo {volume} {D15}},\ \bibinfo {pages} {2929} (\bibinfo
  {year} {1977})},\ \bibinfo {note} {[Erratum: Phys.
  Rev.D16,1248(1977)]}\BibitemShut {NoStop}%
\bibitem [{\citenamefont {Callan}\ and\ \citenamefont
  {Coleman}(1977)}]{Callan:1977pt}%
  \BibitemOpen
  \bibfield  {author} {\bibinfo {author} {\bibfnamefont {C.~G.}\ \bibnamefont
  {Callan}, \bibfnamefont {Jr.}}\ and\ \bibinfo {author} {\bibfnamefont
  {S.~R.}\ \bibnamefont {Coleman}},\ }\href {\doibase 10.1103/PhysRevD.16.1762}
  {\bibfield  {journal} {\bibinfo  {journal} {Phys. Rev.}\ }\textbf {\bibinfo
  {volume} {D16}},\ \bibinfo {pages} {1762} (\bibinfo {year}
  {1977})}\BibitemShut {NoStop}%
\bibitem [{\citenamefont {Branchina}\ and\ \citenamefont
  {Messina}(2013)}]{Branchina:2013jra}%
  \BibitemOpen
  \bibfield  {author} {\bibinfo {author} {\bibfnamefont {V.}~\bibnamefont
  {Branchina}}\ and\ \bibinfo {author} {\bibfnamefont {E.}~\bibnamefont
  {Messina}},\ }\href {\doibase 10.1103/PhysRevLett.111.241801} {\bibfield
  {journal} {\bibinfo  {journal} {Phys. Rev. Lett.}\ }\textbf {\bibinfo
  {volume} {111}},\ \bibinfo {pages} {241801} (\bibinfo {year} {2013})},\
  \Eprint {http://arxiv.org/abs/1307.5193} {arXiv:1307.5193 [hep-ph]}
  \BibitemShut {NoStop}%
\bibitem [{\citenamefont {Branchina}\ \emph {et~al.}(2015)\citenamefont
  {Branchina}, \citenamefont {Messina},\ and\ \citenamefont
  {Sher}}]{Branchina:2014rva}%
  \BibitemOpen
  \bibfield  {author} {\bibinfo {author} {\bibfnamefont {V.}~\bibnamefont
  {Branchina}}, \bibinfo {author} {\bibfnamefont {E.}~\bibnamefont {Messina}},
  \ and\ \bibinfo {author} {\bibfnamefont {M.}~\bibnamefont {Sher}},\ }\href
  {\doibase 10.1103/PhysRevD.91.013003} {\bibfield  {journal} {\bibinfo
  {journal} {Phys. Rev.}\ }\textbf {\bibinfo {volume} {D91}},\ \bibinfo {pages}
  {013003} (\bibinfo {year} {2015})},\ \Eprint {http://arxiv.org/abs/1408.5302}
  {arXiv:1408.5302 [hep-ph]} \BibitemShut {NoStop}%
\end{thebibliography}%
\end{document}